\documentclass{emulateapj} 

\shorttitle{Nuclear IR SED of Type II AGNs}
\shortauthors{Videla et al.}

\begin{document}

\title{Nuclear Infrared Spectral Energy Distribution \\ of Type II
  Active Galactic Nuclei}

\author{Liza Videla, Paulina Lira}
\affil{Departamento de Astronom\'ia, Universidad de Chile.}

\author{Heather Andrews} 
\affil{Leiden Observatory, Leiden University, NL-2300 RA Leiden, The Netherlands}

\author{Almudena Alonso-Herrero} 
\affil{Instituto de F\'isica de Catabria, CSIC-UC, 39005 Santander, Spain}

\author{David M. Alexander and Martin Ward}
\affil{Department of Physics, Durham University.}

\begin{abstract}

We present near and mid--IR observations of a sample of Seyfert II galaxies
drawn from the $12\mu$m Galaxy Sample. The sample was observed in the J, H, K,
L, M and N bands. Galaxy Surface Brightness Profiles are modeled using
nuclear, bulge, bar (when necessary) and disk components. To check the
reliability of our findings the procedure was tested using {\em Spitzer\/}
observations of M\,31. Nuclear Spectral Energy Distributions (SEDs) are
determined for 34 objects, and optical spectra are presented for 38, including
analysis of their stellar populations using the STARLIGHT spectral synthesis
code. Emission line diagnostic-diagrams are used to discriminate between
genuine AGN and HII nuclei. Combining our observations with those found in the
literature, we have a total of 40 SEDs. It is found that about 40\%\ of the
SEDs are characterized by an upturn in the near-IR, which we have quantified
as a NIR slope $\alpha < 1$ for an SED characterized as $\lambda f_{\lambda}
\propto \lambda^{\alpha}$. Three objects with an HII nucleus and two Seyfert
nuclei with strong contamination from a circumnuclear starburst, also show an
upturn. For genuine AGN this component could be explained as emission from the
accretion disk, a jet, or from a very hot dust component leaking from the
central region through a clumpy obscuring structure. The presence of a very
compact nuclear starburst as the origin for this NIR excess emission is not
favored by our spectroscopic data for these objects.

\end{abstract}

\keywords{}

\section{Introduction}

\subsection{Unified Model}

The presence of well--collimated radio jets in radio loud Active Galactic
Nuclei (AGN), and the expectation of an axis of symmetry defined by the
geometrically thin accretion disk, argue for strong anisotropies in the
central engines of AGN. This implies that the appearance of an AGN will depend
on the observer's position relative to the system's axis. Following this
reasoning, and to explain a wealth of observational evidence, the Unified
Model postulates that Type I and Type II AGN are the same kind of object, but
that they appear distinct because of the difference in angle between the
central engine's axis and our line of sight.

Support for this model has been particularly strong in the case of Seyfert
galaxies. Optical spectra of Seyfert I objects show Broad Emission Line
Regions (BLR); those of Seyfert II objects do not.  Explaining this as an
orientation effect, with obscuration of the central source, goes back to at
least \citet{osterbrock78}, who suggested that both types are physically the
same, but that in Seyfert II galaxies the BLR is blocked from our view,
probably by dust in a toroidal structure surrounding the central source.
Below we summarize some of the work that followed and supported this thesis.

Evidence for obscuration of Type II objects comes from the excess absorption
in X-ray, UV and optical wavelengths, when observations are compared with Type
I nuclei
\citep{lawrence82,mass-hesse93,turner97,malkan98,risaliti99}. Recently,
\citet{shi06} have shown a clear correlation between the column densities
derived from X-ray data and the strength of the $9.7\mu$m silicate feature:
Type II systems with large hydrogen columns show large absorption features,
while objects classified as Type I often show the feature in emission.

In those AGN that are radio-loud sources, the radio structure appears to be
aligned with the torus axis (Wilson \& Tsvetanov, 1994; Nagar et al.~1999;
Barbosa et al.~2009). This anisotropy is also observed in the cone--like
structures of the Narrow Line Region, where the torus collimates ionizing
radiation from the central source \citep{pogge88a,pogge88b,falcke98}.

In the IR, detection of broad components from hydrogen recombination lines
such as Pa$\beta$ and Br$\gamma$ in objects optically classified as Seyfert II
galaxies is consistent with extinction along a line--of--sight that grazes the
torus surface, sufficient to suppress the BLR emission at optical wavelengths,
but optically thin in the infrared \citep{ruiz94,veilleux97,veilleux99}.

More direct support for unification comes from the HST, which has imaged an
extended nuclear structure of $\sim\! 90$ pc in NGC\,4261 \citep{jaffe93},
while interferometry observations are starting to shed light on the inner
obscuring torus (Wittkowski et al.~1998, 2004; Weinberger et al.~1999; Swain
et al.~2003; Jaffe et al.~2004; Weigelt et al.~2004; Meisenheimer et al.~2007;
Tristram et al.~2007, 2009; Beckert et al.~2008; Kishimoto et al.~2009, 2011;
Raban et al.~2009).

There is also evidence for structure in the obscuring material.  Changes in
broad line profiles \citep{penston84} and inferred column densities imply the
presence of clumpy structures, leading to variable obscuration of the central
engine as they move across the line of sight
\citep{risaliti11,risaliti09,puccetti07,elvis04} (although the column density
variations are more likely to correspond to obscuration by BLR clouds than the
torus).

Finally, the most famous, and arguably most convincing evidence in support of
the Unified Model comes from optical spectropolarimetry of Type II objects.
\citet{antonucci85} found scattered emission from the BLR in polarized flux
from NGC\,1068, strongly supporting the idea of unification by orientation,
and so implying the existence of an obscuring structure.  Their work was
extended by \citep{young95,young96,moran00}.

\subsection{Spectral Energy Distribution}

The Spectral Energy Distribution (SED) is an important tool for studying the
central engine in AGN. In most luminous and intermediate luminosity AGN it can
be characterized by two bumps --- one peaking in the UV and the other in the
mid-IR --- and an inflection point between, at $\sim\! 1.5~\mu$m
\citep{elvis94}. The UV feature (the ``Big Blue Bump'') is normally identified
with emission from the central accretion disk, while the IR flux is assumed to
originate in the dusty obscuring structure, the torus.  The latter has a broad
SED, with a peak somewhere between $5-30 \mu$m, decreasing slowly towards the
far-IR.  The emission mechanism is believed to be thermal reprocessing of
central, optical-UV emission by dust at a wide range of temperatures --- the
highest ($\sim\! 1000-1500$ K) corresponding to dust closest to the central
engine.

Obtaining the SED of the torus in AGN is difficult for a number of reasons.

First, the SED inflection point matches the peak in starlight emission from
the host galaxy.  In intermediate luminosity Seyfert galaxies the host galaxy
contribution is often comparable to the emission from the nucleus, flattening
the optical-IR region of the SED and lowering the relative strength of the UV
and IR bumps. To study the {\emph{nuclear}} SED, galactic starlight must be
removed.

Second, until recently, only small size IR arrays have been available, making
it impossible to image the complete host galaxy for nearby sources,
compromising the modeling of the surface brightness distribution.

Third, observations at $\lambda \gtrsim 10 \mu$m can only be achieved from
space, which are characterized by lower angular resolution.

Fourth, and finally, observations obtained at different epochs can be affected
by variability.  Fortunately, the torus region is large, damping most
variations of the central source, so this is not expected to be significant.

The number of compiled AGN SEDs in the literature increases constantly, but
many of them incorporate data gathered by different groups using different
instruments; this often implies data of different resolutions, analyzed in
different ways.  To properly study the emission from the dusty torus in AGN,
and to accurately infer physical and geometrical characteristics, this paper
describes a more homogenous dataset.  We have constructed nuclear IR SEDs of a
sample of 48 Type II Seyfert galaxies using near-- and mid--IR
high--resolution ground--based images (described here), and modeled them in
clumpy using the approach of \citet{nenkova08a,nenkova08b,nenkova02}
(described in a companion paper, \citep{lira12}). The main contribution of
this work is the number of objects we study, and the detailed treatment of the
{\emph{nuclear}} SED construction.

In \S\ref{data} we present the sample, observations and data reduction,
including the optical spectroscopy and the stellar population
analysis. \S\ref{construction} describes, in detail, the path from IR images
to the final nuclear IR SEDs of our sample, including the analysis of M\,31 to
validate our methodology. Section \S\ref{SEDs} discusses the results and
possible interpretations. A summary is presented in \S\ref{summary}

\section{Observations}\label{data}

The sample used in this work contains all Seyfert II galaxies from the
Extended 12$\mu$m Galaxy Sample \citep{rush93} that are located in the
southern hemisphere.  The 48 objects are described in Table~\ref{fechas}.

\begin{deluxetable*}{llccllccc}
\tabletypesize{\footnotesize}
\tablecolumns{9}

\tablecaption{Observed sample of 48 Seyfert 2 galaxies, indicating the
  detector used in each infrared observation.\label{fechas}}

\tablehead{ 
\colhead{Galaxy} & \colhead{Hubble Type}  & \colhead{Redshift} & \colhead{Inc} &  \colhead{RA} & \colhead{DEC} & \colhead{JHK bands} & \colhead{LM bands} &\colhead{N band} \\
\colhead{(1)} & \colhead{(2)}  & \colhead{(3)} & \colhead{(4)} &  \colhead{(5)} & \colhead{(6)} & \colhead{(7)} & \colhead{(8)} &\colhead{(9)} 
}

\startdata
    NGC\,34, MRK\,938	&$\wr$\,Sc	&0.0198	&0.14	& 00 11 06.55	& -- 12 06 26.32 &ISA-2      &ISA71-B	&--	\\ 
    F\,00198-7926 	&$\wr$\,Pec	&0.0728	&0.31	& 00 21 52.90	& -- 79 10 08.00 &ISA-2      &ISA71-B	&--	\\ 
    F\,00521-7054 	&E-S0		&0.0689	&0.22	& 00 53 56.15	& -- 70 38 04.16 &ISA-2      &ISA71-B	&--	\\ 
    ESO\,541-IG12 	&Mult		&0.0564	&0.29	& 01 02 17.55	& -- 19 40 08.67 &ISA-2	     &ISA71-B	&G04	\\ 
    NGC\,424, TOL0109	&(R)SB(r)0/a	&0.0117	&$lost$ & 01 11 27.51	& -- 38 05 01.08 &ISA-2	     &ISA71-B	&--	\\ 
    F\,01475-0740 	&E-S0		&0.0177	&0.17	& 01 50 02.70	& -- 07 25 48.48 &ISA-2	     &ISA71-B	&--	\\ 
    NGC\,1068 		&$\wr$\,(R)SA(rs)b&0.0038&0.07	& 02 42 40.71   & -- 00 00 47.81 &AH03	     &AH03	&AH03 	\\ 
    NGC\,1097 		&(R'\_1)SB(r'l)b&0.0042	&$lost$ & 02 46 18.99	& -- 30 16 28.68 &ISA-2	     &ISA71-B	&--	\\ 
    NGC\,1125 		&SAB0		&0.0109	& --    & 02 51 40.27	& -- 16 39 03.69 & --	     &ISA71-B	&--	\\ 
    NGC\,1144 	        &S0 pec		&0.0288	&0.08	& 02 55 10.85	& -- 00 10 47.20 &ISA-2	     &ISA71-B	&NED	\\ 
    MCG\,-2-8-39, F\,02581-1136&SAB(rs)a&0.0299	&0.64	& 03 00 30.69	& -- 11 24 54.06 &ISA-2	     &ISA71-B	&NED	\\ 
    NGC\,1194 		&SA0+		&0.0136	&0.46	& 03 03 49.11	& -- 01 06 13.39 &ISA-2	     &--	&G04	\\ 
    NGC\,1241 		&SB(rs)b	&0.0135	&$lost$ & 03 11 14.90	& -- 08 55 20.89 &ISA-2	     &--	&--	\\ 
    NGC\,1320, MRK\,607	&$\wr$\,Sa sp	&0.0094	&0.69	& 03 24 48.72	& -- 03 02 31.99 &ISA-2	     &--  	&G04	\\ 
    NGC\,1386 		&SB(s)0+	&0.0029	&$lost$ & 03 36 46.40	& -- 36 00 02.00 &ISA-2	     &-- 	&--	\\ 
    F\,03362-1642	&Pec		&0.0369	&0.68   & 03 38 34.54	& -- 16 32 15.84 &ISA-2	     &--	&G04	\\ 
    F\,04385-0828	&S0		&0.0151	&0.54	& 04 40 54.96	& -- 08 22 22.22 &ISA-2	     &--	&G04	\\ 
    NGC\,1667 		&SAB(r)c	&0.0152	&0.31   & 04 48 37.14	& -- 06 19 11.87 &ISA-2	     &--	&GEMI	\\ 
    ESO\,33-G2 		&SB0		&0.0186	&0.34	& 04 55 58.99	& -- 75 32 28.06 &ISA-2	     &--	&--	\\ 
    F\,05189-2524	&Pec		&0.0426	&0.06	& 05 21 01.47	& -- 25 21 45.38 &ISA-2	     &--	&GEMI	\\ 
    ESO\,253-G3 	&Sa tidal	&0.0425	&$2\,nuc$& 05 25 18.29	& -- 46 00 19.60 &ISA-2	     &--	&GEMI	\\ 
    MCG\,+0-29-23 	&SAB(s)b	&0.0248	&0.07	& 11 21 12.26	& -- 02 59 03.45 &SofI(S)    &ISA70-B	&G04	\\ 
    NGC\,3660, MRK\,1291&SB(r)bc	&0.0123	&0.14	& 11 23 32.24   & -- 08 39 30.21 &SofI(S)    &ISA70-B	&GEMI	\\ 
    NGC\,4388 		&SA(s)b sp	&0.0084	&0.64	& 12 25 46.75   & + 12 39 43.51	 &AH03	     &AH03	&G04	\\ 
    NGC\,4501, M\,88 	&SA(rs)b	&0.0076	&0.47	& 12 31 59.16   & + 14 25 13.60	 &SofI(S)    &ISA70-B	&GEMI-G04\\ 
    TOL\,1238-364, IC\,3639&$\wr$\,SB(rs)bc&0.0109&0.12	& 12 40 52.88   & -- 36 45 21.52 &SofI(S)    &ISA70-B	&GEMI	\\ 
    NGC\,4941 		&R)SAB(r)ab	&0.0037	&0.50	& 13 04 13.06   & -- 05 33  5.79 &SofI(S)    &ISA70-B	&G04	\\ 
    NGC\,4968  		&$\wr$\,(R')SAB0&0.0099	&0.34	& 13 07 05.89   & -- 23 40 39.38 &AH03	     &AH03	&G04	\\ 
    MCG\,-3-34-64, F\,13197-1627&SB?	&0.0165	&0.25	& 13 22 24.46   & -- 16 43 42.90 &SofI(S)    &ISA70-B	&G04	\\ 
    NGC\,5135 		&$\wr$\,SB(l)ab	&0.0137	&0.01	& 13 25 43.97   & -- 29 50 02.26 &SofI(S)    &ISA70-B	&G04	\\ 
    MRK\,463 		&Merger		&0.0503	&$2\,nuc$& 13 56 02.87  & +  18 22 19.48 &SofI(S)    &ISA70-B	&G04	\\ 
    NGC\,5506 		&Sa pec sp	&0.0062	&0.52	& 14 13 14.86   & -- 03 12 26.94 &AH03	     &AH03 	&G04	\\ 
    NGC\,5953, MRK\,9031&SAa pec 	&0.0065	&0.37	& 15 34 33.70   & + 15 11 49.50	 &SofI(S)    &ISA70-B	&GEMI-G04\\ 
    NGC\,5995, MCG\,-2-40-4&S(B)c	&0.0252	&0.19	& 15 48 24.96   & -- 13 45 27.94 &SofI(S)    &ISA70-B	&G04	\\ 
    F\,15480-0344	&$\wr$\,S0	&0.0303	&0.11	& 15 50 41.51   & -- 03 53 18.34 &SofI(S)    &ISA71-B	&G04	\\ 
    F\,19254-7245	&Merger		&0.0617	&$2\,nuc$& 19 31 21.40  & -- 72 39 17.96 &SofI(V)    &ISA71-B	&GEMI	\\ 
    NGC\,6810, F\,19393-5846&SA(s)ab:sp	&0.0068 &0.50	& 19 43 34.16   & -- 58 39 20.54 &SofI(V)    &ISA71-B   &GEMI	\\ 
    NGC\,6890 		&R')SA(r:)ab	&0.0081	&0.15	& 20 18 18.07   & -- 44 48 23.36 &SofI(V)    &ISA71-B	&GEMI	\\ 
    IC\,5063 		&SA(s)0+	&0.0114	&0.2	& 20 52 02.01   & -- 57 04 09.12 &SofI(V)    &ISA71-B	&GEMI	\\ 
    MRK\,897, UGC\,11680&Compact	&0.0264	&0.21	& 21 07 43.70   & + 03 52 19.00	 &SofI(V)    &ISA71-B	&GEMI	\\ 
    NGC\,7130, IC\,5135	&$\wr$\,Sa pec	&0.0162	&0.01	& 21 48 19.49   & -- 34 57 05.98 &SofI(V)    &ISA71-B 	&GEMI	\\ 
    NGC\,7172 		&Sa pec sp	&0.0087	&0.25	& 22 02 01.68   & -- 31 52 18.12 &AH03	     &AH03 	&G04	\\ 
    MCG\,-3-58-7, F\,22469-1932 &(R')SAB(s)0/a&0.0317&0.13& 22 49 37.15 & -- 19 16 26.39 &SofI(V)    &ISA71-B	&G04	\\ 
    NGC\,7496 		&(R')SB(rs)bc	&0.005	&0.11	& 23 09 47.26   & -- 43 25 39.76 &SofI(V)    &ISA71-B   &--     \\ 
    NGC\,7582 		&(R'\_1)SB(s)ab	&0.0053	&0.69	& 23 18 23.50   & -- 42 22 13.98 &SofI(SV)   &ISA71-B	&--	\\ 
    NGC\,7590 		&S(r?)bc	&0.0053	&0.64	& 23 18 54.60   & -- 42 14 21.00 &SofI(SV)   &ISA71-B	&--	\\ 
    NGC\,7674, MRK\,533 &$\wr$\,SA(r)bc pec&0.0289&0.04	& 23 27 56.74   & + 08 46 44.52	 &AH03	     &AH03 	&G04	\\ 
    CGCG\,381-051, F\,23461+0157&SBc	&0.0307	&0.10   & 23 48 41.64   & + 02 14 24.02	 &SofI(V)    &ISA71-B	&G04	
\enddata

\tablecomments{In column (2), $\wr$ indicates Compton Thick (CT) sources. In
  column (4) we give the galaxy inclination obtained from the mean ellipticity
  ($e=1-b/a$) of the last fitted isophotes in the near-IR bands
  (\S\ref{2Dvs1D}); {\emph{2 nuc}} indicates the detection of 2 nuclei in the
  center of the galaxy, close enough to prevent reliable SBP fitting of 2
  separated galaxies; {\emph{lost}} indicates that the observations of the
  galaxy were lost, either because of the size of the galaxy compared to the
  detector (bad sky subtraction), or because of non--photometric observing
  conditions (NGC\,424, \S\ref{NIR_reduc}). In column (7), V indicates
  observations made in visitor mode and S indicates observations made in
  service mode. In column (9), G04 indicates observations made by
  \citet{gorjian04}; AH03 indicates SEDs determined by \citet{ah03}; NED
  indicates values obtained from the NASA/IPAC Extragalactic
  Database\footnote{The NASA/IPAC Extragalactic Database (NED) is operated by
    the Jet Propulsion Laboratory, California Institute of Technology, under
    contract with the National Aeronautics and Space Administration.}}
\end{deluxetable*}

The 12$\mu$m Galaxy Sample has several advantages over other catalogs.  Most
importantly, it is selected in the mid-infrared (MIR), giving a more
representative population of nearby (z$\leq$0.07) AGNs.  In comparison,
optically--selected samples can miss obscured systems and are more susceptible
to the effects of host galaxy reddening (e.g., Huchra \& Burg, 1992; Maiolino
\& Rieke, 1995).  This is important because a wide range of nuclear
obscuration (hydrogen columns of $\sim\! 10^{22}-10^{25}$ cm$^{-2}$) can
provide more general tests of nuclear emission models.  In addition, the
12$\mu$m Sample includes elliptical, lenticular and spiral galaxies (which
helps avoid systematic errors in the decomposition process of the surface
brightness profiles) and a wide range of galaxy inclinations (which allows the
importance of galactic, as well as nuclear, absorption to be assessed).

However, the sample may be biased towards Seyfert II sources with strong star
formation (Buchanan et al.~2006) --- the stellar IR emission can compensate
for a weak AGN component, increasing the likelihood of inclusion in a flux
limited sample. This might explain the prevalence of star formation in Seyfert
II galaxies when compared with Seyfert I nuclei (e.g., Maiolino et al.~1995).

Our analysis of the central region of galaxies is built on extensive studies
of the 12$\mu$m Galaxy Sample at many wavelengths, from radio to X-rays
(Spinoglio et al.~1995, 2002; Rush et al.~1996b; Hunt et al.~1999ab; Bassani et
al.~1999; Thean et al.~2000, 2001; Imanishi, 2003, 2004; Gorjian et al.~2004;
Strong et al.~2004; Buchanan et al.~2006; Tommasin et al.~2008, 2010; Wu et
al.~2009; Gallimore et al.~2010; Baum et al.~2010), including
spectropolarimetric observations (Tran 2001).

The initial classification of objects in the sample, by Rush et al.~(1993),
was based on existing catalogs of active galaxies
\citep{veron-cetty91,hewitt91}.  Since then, as better optical spectra became
available, some objects have been re-classified.  To verify the
classifications, and to study stellar populations, we obtained high quality
optical spectroscopy for 38 objects. The analysis of these data is presented
next.

\subsection{Optical Spectroscopy and Analysis}

\begin{figure}
  \begin{center}
    \includegraphics[scale=0.45,trim=50 100 0 100]{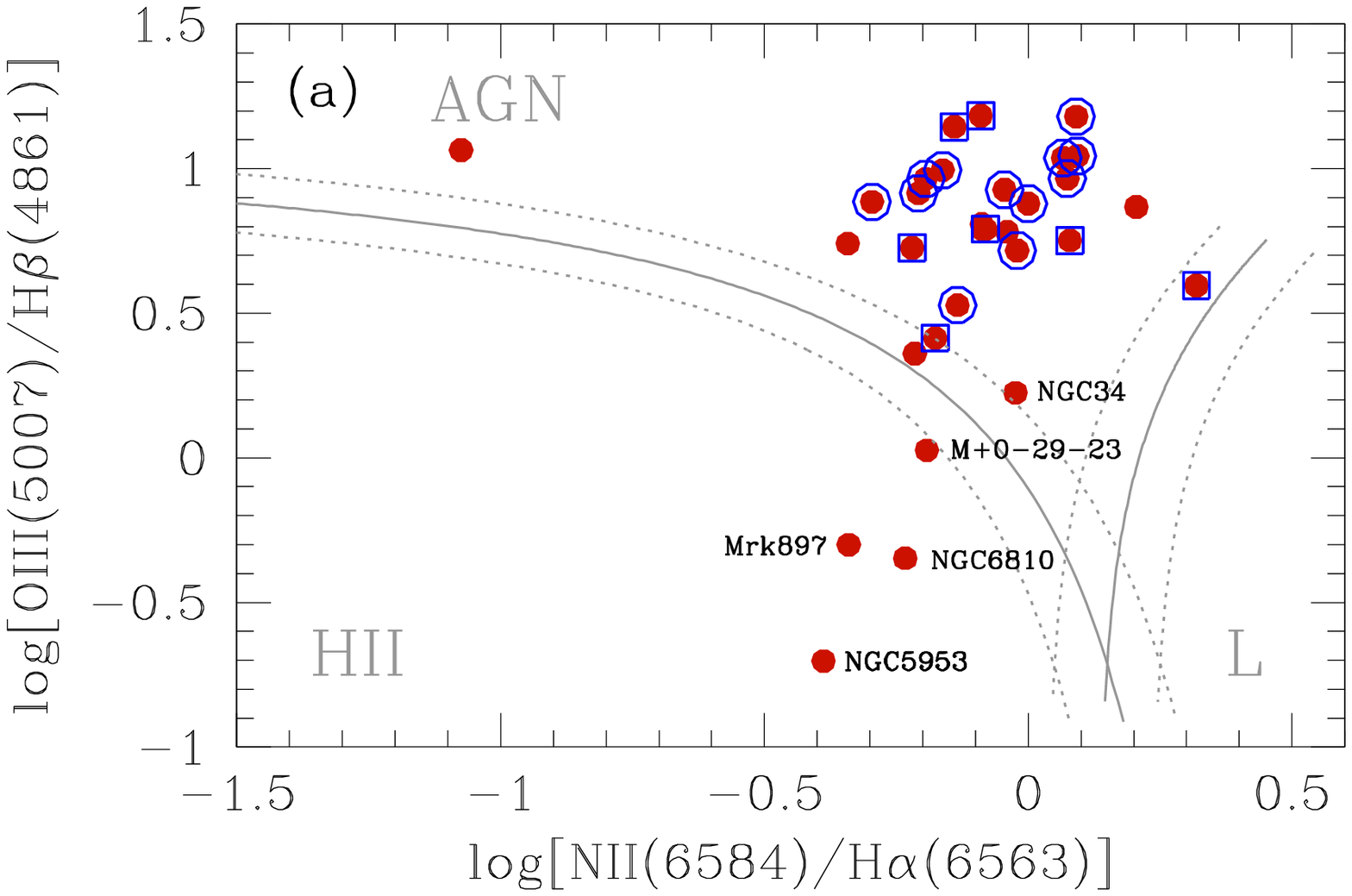}\\
    \includegraphics[scale=0.45,trim=50 100 0 100]{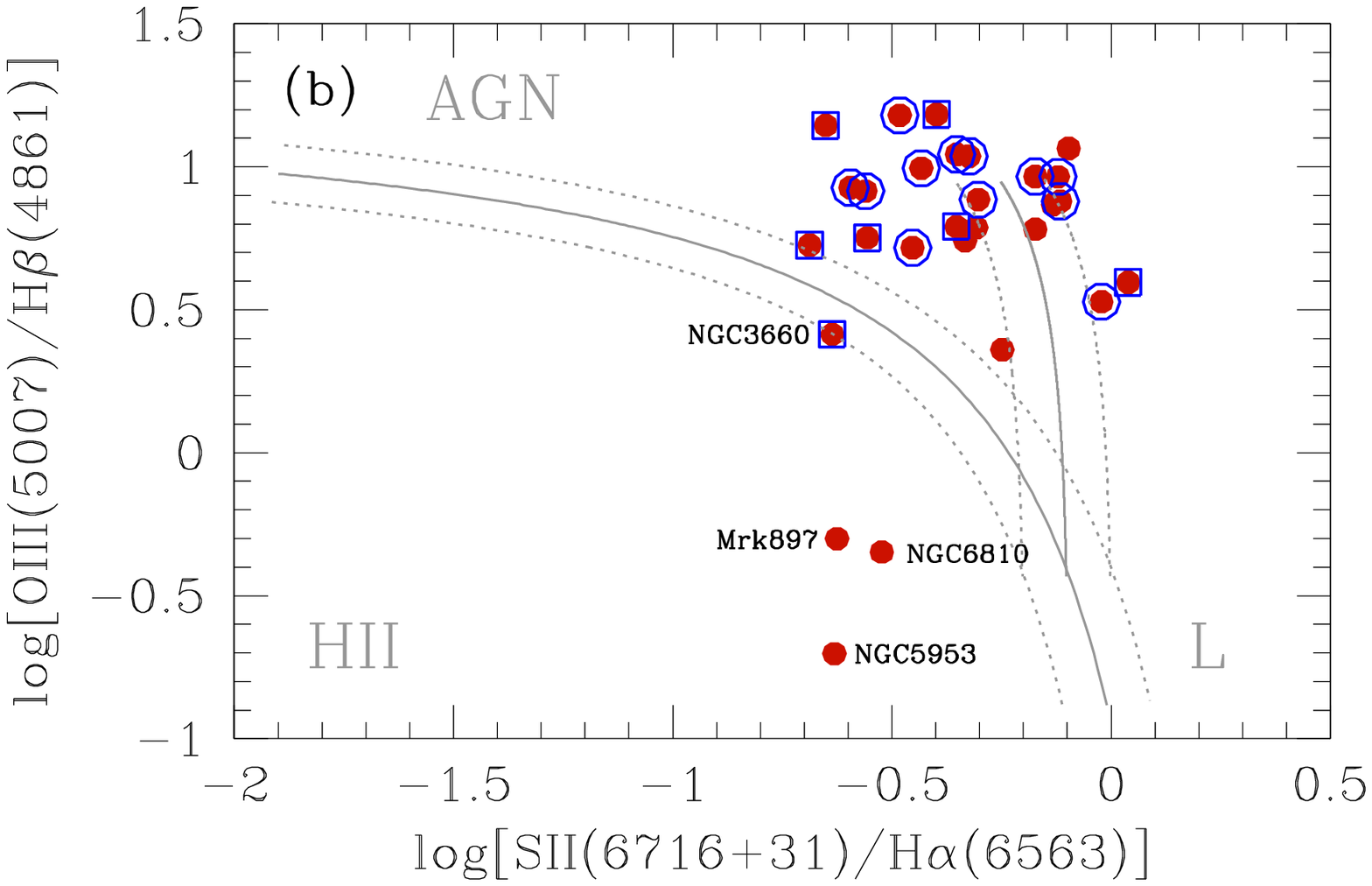}\\
    \includegraphics[scale=0.45,trim=50 0 0 100]{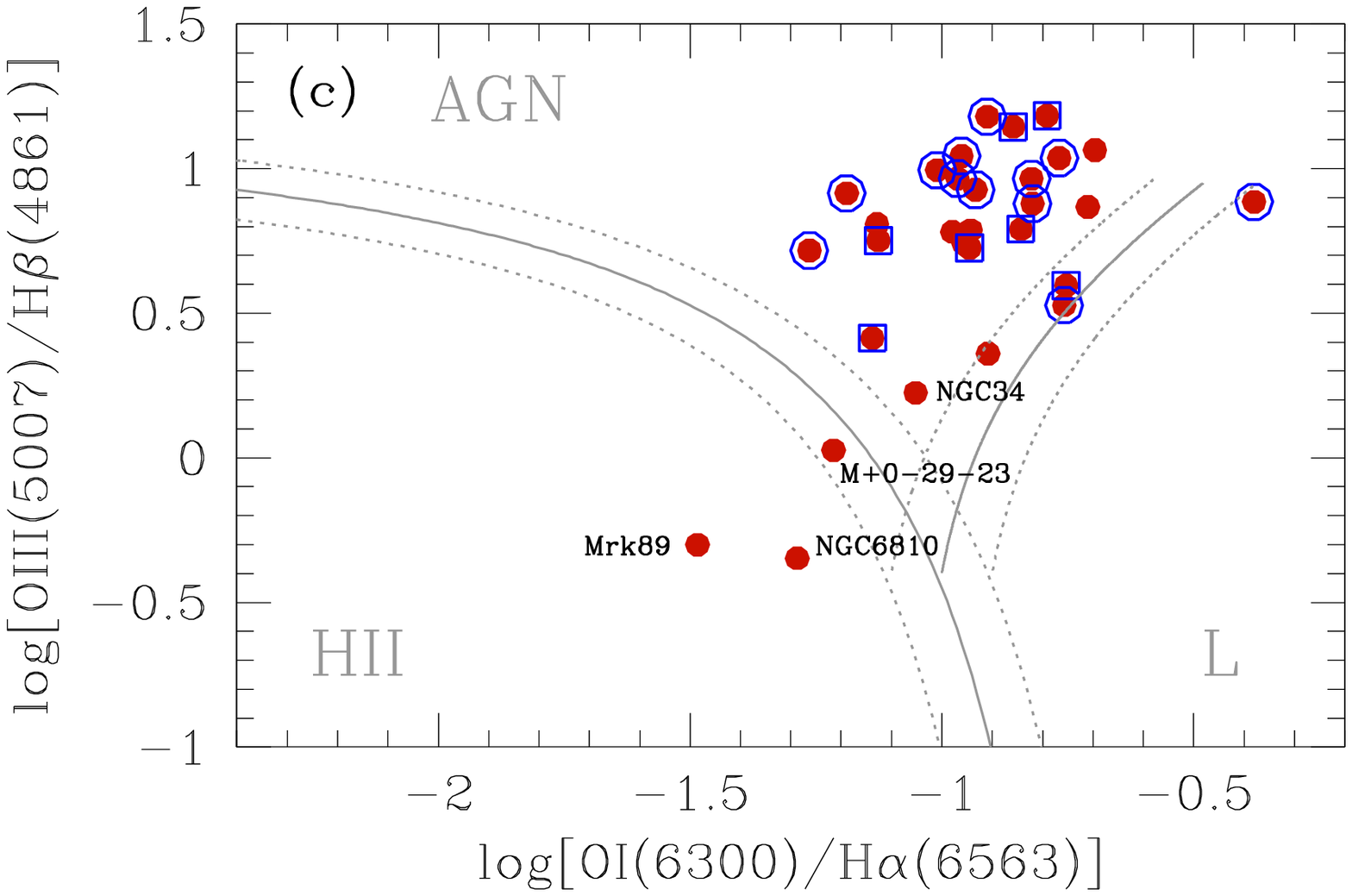}

    \caption{Diagnostic diagrams for emission line spectra.  The theoretical
      divisions between the star-forming, AGN and LINER classes, as
      parameterized by \citet{kewley01}, are shown with thick gray lines in
      panels (a), (b), and (c). Dotted lines indicate $\pm 0.1$ dex. Nuclei
      that fall in the HII region are labeled. Also, AGN nuclei are shown with
      an extra circle or square if they have a NIR slope grater or smaller
      than 1, respectively (see \S 4). \label{diag_diag}}
  \end{center} 
\end{figure}

Long--slit optical spectroscopy was obtained with the R--C Spectrograph on the
Blanco 4m telescope at CTIO, in August 2007 and February 2008. We used the
KPGL3-1 grating and a slit width of $1\arcsec$. The slit was positioned at the
parallactic angle to minimize light losses. In both runs the seeing varied
between $\sim\! 0.8$ and $\sim\! 1.2$ arc-sec. The February 2008 run was
photometric, while clouds were presented during the August 2007 run. However,
absolute flux calibration is not necessary to characterize the nebular and
stellar components of the nuclear spectra.

The data were reduced using IRAF\footnote{IRAF is distributed by the National
  Optical Astronomy Observatories, which are operated by the Association of
  Universities for Research in Astronomy, Inc., under cooperative agreement
  with the National Science Foundation.} software.  Standard reduction
procedures (bias subtraction, flat-fielding, slit illumination correction)
were followed.  Extraction of the galactic spectra used a window of width
$\sim\! 2.2\arcsec$.  Wavelength calibration used comparison copper-neon and
copper-argon lamps.  The spectra were corrected for the foreground extinction
tabulated by \citet{schlegel98} using the empirical extinction function of
\citet{cardelli89}.

We determined the stellar component of the spectra using the STARLIGHT
spectral synthesis code developed by \citet{cid04,cid05}. A total of 45
different stellar population templates were used, covering an age range of
$\sim\!  10^{6} - 10^{10}$ years and a metalicity range of $0.2 - 2.5 \times
Z_{\odot}$. The best fit spectrum was found by minimizing $\chi^{2}$ after
masking all emission lines.

Following \citet{cid04,cid05} we grouped output from the population synthesis
into 3 age groups: a Young component (Y: $t\!  <\! 10^{8}$ yr); an
Intermediate component (I: $10^{8}\! <\! t\! <\! 10^{9}$ yr); and an Old
component (O: $t\! >\! 2.5\times10^{9}$ yr). The results are presented in
Table~\ref{table_spec}.

\begin{deluxetable*}{lrrrrrrrrrr}
\tabletypesize{\footnotesize}
\tablecolumns{11}

\tablecaption{Population synthesis analysis and measured emission line
  fluxes and flux ratios. \label{table_spec}}

\tablehead{ 
\colhead{Name} & \colhead{Y(\%)} & \colhead{I(\%)} & \colhead{O(\%)} & \colhead{$f$(4861)} & \colhead{$f$(5007)} & \colhead{(5007/4861)} & \colhead{(6583/6563)} & \colhead{(6300/6563)} & \colhead{(6716/6563)} & \colhead{(6563/4861)} \\
& \colhead{(2)} & \colhead{(3)} & \colhead{(4)} & \colhead{(5)} & \colhead{(6)}& \colhead{(7)} & \colhead{(8)} & \colhead{(9)} & \colhead{(10)} & \colhead{(11)} }
\startdata
    NGC\,34	   & 46.8  &34.6  &18.6    &0.9    &1.6:  &0.2:     &-0.02     &-1.1	  &---	  &13.7  \\ 
    F\,00198-7926  &  0.0  & 0.0  &100.0   &0.5    &3.9   &---      &---       &---	  &---	  &---   \\ 
    F\,00521-7054  &  0.0  &10.6  &89.4    &0.8    &6.8   &0.9      &-0.05     &-0.9	  &-0.6	  &8.3:  \\ 
    ESO\,541-IG12  & 53.6  & 0.0  &46.4    &0.7    &5.8   &0.9      &-0.2      &-1.2	  &-0.6	  &6.1:  \\ 
    NGC\,424       & 23.9  & 0.0  &76.1    &10.0   &54.5  & ---     &---       &---	  &---	  &---	 \\ 
    F\,01475-0740  & 53.6  & 0.0  &46.4    &1.4    &7.5   &0.7      &-0.2      &-0.9	  &-0.7	  &7.1	 \\ 
    NGC\,1068	   & ---   & ---  & ---    & ---   & ---  & ---     &---       &---	  &---    &---	 \\ 
    NGC\,1097 	   & ---   & ---  & ---    & ---   & ---  & ---     &---       &---	  &---    &---	 \\ 
    NGC\,1125 	   & 50.3  & 3.9  &45.8    &2.4    &14.7  &0.8      &-0.09     &-0.9	  &-0.3	  &4.8	 \\ 
    NGC\,1144      &  0.0  & 0.0  &100.0   &0.6    &6.4   &---      &---       &---	  &---	  &---	 \\ 
    MCG\,-2-8-39   & 21.8  & 0.0  &78.2    &1.8    &26.8  &1.2      &-0.09     &-0.8	  &-0.4	  &4.2   \\ 
    NGC\,1194 	   & 26.7  & 0.0  &73.3    &0.3    &2.6   &0.9      &-0.3      &-0.4	  &-0.3	  &6.6	 \\ 
    NGC\,1241 	   & ---   & ---  & ---    & ---   & ---  & ---     &---       &---	  &---    &---	 \\ 
    NGC\,1320      & 4.80  &15.9  &79.3    &2.6    &25.5  &1.0      &-0.2      &-1.0	  &-0.4	  &3.8	 \\ 
    NGC\,1386	   & ---   & ---  & ---    &---    & ---  & ---     &---       &---	  &---    &---	 \\ 
    F\,03362-1642  & 51.6  & 0.0  &48.4    &0.7    &4.5   &0.8      &-0.09     &-1.1	  &---	  &5.3	 \\ 
    F\,04385-0828  & 45.8  & 0.0  &54.2    &0.5    &1.8   &0.5      &-0.1      &-0.8	  &-0.02  &4.0:	 \\ 
    NGC\,1667	   & 25.2  & 0.0  &74.8    &1.1    &7.9   &0.9      &0.2       &-0.7	  &-0.1	  &5.7:  \\ 
    ESO\,33-G2 	   & 32.3  & 1.2  &66.5    &1.3    &12.4  &1.0      &0.07      &-1.0	  &-0.2	  &5.0:  \\ 
    F\,05189-2524  & 1.9   &26.7  &71.4    &1.9    &11.0  & ---     &---       &---	  &---	  &---	 \\ 
    ESO\,253-G3    & 75.6  & 0.0  &24.4    &2.7    &15.1  &0.7      &-0.3      &-1.0	  &-0.3	  &3.6:  \\ 
    MCG\,+0-29-23  & 28.2  &10.7  &61.1    &0.8    &0.9:  &0.02:    &-0.2      &-1.2	  &---	  &8.1   \\ 
    NGC\,3660 	   & 3.3   & 0.0  &96.7    &1.3    &3.5   &0.4      &-0.2      &-1.1	  &-0.6	  &3.3:	 \\ 
    NGC\,4388 	   & 35.2  & 0.0  &64.8    &4.9    &45.2  &1.0      &-0.2      &-0.8	  &-0.1	  &4.0   \\ 
    NGC\,4501 	   & 7.8   & 0.0  &92.2    &0.9    &3.6:  &0.6:     &0.3       &-0.8	  &0.04	  &3.7:	 \\ 
    TOL\,1238-364  & 26.2  &20.5  &53.3    &8.1    &49.8  &0.8      &-0.08     &-0.8	  &-0.4	  &4.0   \\ 
    NGC\,4941 	   & 0.0   & 0.0  &100.0   &2.8    &31.9  &1.0      &-1.1      &-0.7	  &-0.1	  &4.6	 \\ 
    NGC\,4968      & 28.7  & 6.2  &65.1    &2.4    &26.0  &1.0      &0.07      &-0.8	  &-0.3	  &4.3	 \\ 
    MCG\,-3-34-64  & 0.0   &22.0  &78.0    &16.3   & ---  & ---     &---       &---	  &---	  &---	 \\ 
    NGC\,5135	   & 53.9  &16.9  &29.2    &4.5    &23.3  &0.7      &-0.02     &-1.3	  &-0.5	  &5.6   \\ 
    MRK\,463a *    & 60.1  &23.7  &16.2    &1.1    &2.5   &0.4      &-0.2      &-0.9	  &-0.2	  &3.2	 \\ 
    MRK\,463b *	   & 63.6  &19.1  &17.3    &9.2    &67.0  &---      &---       &---	  &---	  &---	 \\ 
    NGC\,5506 	   & 53.1  & 0.0  &46.9    &9.2    &59.7  &0.9      &0.0       &-0.8	  &-0.1	  &4.4	 \\ 
    NGC\,5953 	   & 66.6  &12.2  &21.2    &8.0    &0.3:  &-0.7:    &-0.4      &---	  &-0.6	  &4.2	 \\ 
    MCG\,-2-40-4   & 27.3  & 4.2  &68.5    &1.6    &8.8   & ---     &---       &---	  &---	  &---	 \\ 
    F\,15480-0344  & 11.2  &13.8  &75.0    &1.7    &65.0  &1.1      &-0.1      &-0.9	  &-0.7	  &4.6:  \\ 
    F\,19254-7245  & 18.9  & 0.0  &81.1    &4.7    &4.3   & ---     &---       &---	  &---	  &---	 \\ 
    NGC\,6810 	   & 44.2  & 2.1  &53.7    &0.6    &1.1:  &-0.3:    &-0.2      &-1.3	  &-0.5	  &7.4:  \\ 
    NGC\,6890 	   & 19.2  & 6.2  &74.5    &2.5    &15.0  &1.2      &0.09      &-0.9	  &-0.5	  &5.0	 \\ 
    IC\,5063 	   & ---   & ---  & ---    &1.0    & ---  & ---     &---       &---	  &---    &---	 \\ 
    MRK\,897       & 34.3  &38.2  &27.5    & ---   &1.5   &-0.3     &-0.3      &-1.5	  &-0.6	  &4.3	 \\ 
    NGC\,7130      & 65.8  &12.1  &22.1    &3.0    &27.1  &0.8      &0.08      &-1.1	  &-0.6	  &7.9	 \\ 
    NGC\,7172 	   & ---   & ---  & ---    &4.8    & ---  & ---     &---       &---	  &---    &---	 \\ 
    MCG\,-3-58-7   & 33.6  &27.8  &38.7    & ---   &18.6  &1.0      &0.09      &-1.0	  &-0.4	  &4.7   \\ 
    NGC\,7582 	   & ---   & ---  & ---    &5.9    & ---  & ---     &---       &---	  &---    &---	 \\ 
    NGC\,7590 	   & 23.0  & 8.0  &69.0    & ---   &3.3   &0.8      &-0.04     &-1.0	  &-0.2	  &5.9	 \\ 
    NGC\,7674      & ---   & ---  & ---    &0.5    & ---  & ---     &---       &---	  &---    &---	 \\ 
    CGCG\,381-051  & ---   & ---  & ---    & ---   & ---  & ---     &---       &---	  &---    &---	 \\      
\enddata                                                  

\tablecomments{Columns (5) and (6) are in units of $10^{-14}$
  ergs/s/cm$^2$. Unreliable flux measurements (blending and calibration
  issues) are labeled with ':'.  Columns (7) to (10) report the logarithm of
  the flux ratios. Column (11) is the Balmer decrement. *: Spectra for both
  nuclei in MRK\,463 were obtained.}
\end{deluxetable*}

The best--fit stellar models were subtracted from the observed data to give
pure emission line spectra. This reveals weak emission lines present in
targets with significant intermediate age stellar populations. Table
\ref{table_spec} contains the measured emission line flux ratios.

Diagnostic diagrams were constructed using emission line fluxes and are shown
in Fig.~\ref{diag_diag}. The diagrams allow us to distinguish different
ionizing mechanisms of the emission nebula using line ratios that are
sensitive to changes in the spectral shape of the ionizing radiation
\citep{baldwin81,veilleux87,kewley01}.  In this way we can test the
identification of the sources as Seyfert nuclei.

We will discuss possible contaminants in Fig.~\ref{diag_diag} by type.  First,
we argue that there is little evidence for LINERs in our sample.

In panel (a), there are no data points in the LINER region.  The border
between the HII and LINER regions is defined to include all star--forming
models (it is calculated from the highest possible line ratios), but the
[NII]\,$\lambda$6584/H$\alpha$ ratio for star--forming regions and LINERs can
overlap \citep{kewley01}, so LINERs can be located to the left of the border.
Also, theoretical models of shock-excitation give line ratios in the LINER
region only when no precursor is observed \citep{allen08}.  For both these
reasons panel (a) does not discriminate these two classes of object well.

In panel (b), variations in density can introduce uncertainties in the
[SII]\,$\lambda\lambda$6717,6730/H$\alpha$ ratio, so objects should not be
classified as a LINER based purely on this diagnostic \citep{kewley01}.

Given the above, and the single LINER candidate in panel (c) which gives the
most reliable test, none of our sources can be firmly classified as a LINER
nucleus.

Second, four sources systematically appear in the HII section of the diagrams,
consistent with the classification of Gallimore et al.~(2010). These are:
NGC\,5953 (but see next), MCG\,+0-29-23, NGC\,6810, and MRK\,897. Also,
NGC\,34 appears systematically close to the starburst region.

NGC\,5953 was classified as an AGN by \citet{rafanelli90} after making a
careful identification of the nuclear source, which can be confused with a
bright star--forming region located $3\arcsec$ to the West. Unfortunately we
centered our slit on the star--formation knot and the slit orientation did not
include the nucleus, so our optical spectrum is of the wrong object. The IR
images, however, are dominated by the nuclear region --- the star--forming
region can also be seen, but is much less prominent. The determined SED,
however, suggests that contamination by this circumnuclear starburst or a
compact nuclear starburst is present (\S 4.1).

One of the most reliable signatures of an active nucleus is the detection of
luminous, hard X-ray emission. We checked the literature for X-ray data on
MCG\,+0-29-23, NGC\,6810, and MRK\,897, but found little support for
classification as Seyferts. \citet{strickland07} presents the discovery of a
starburst driven wind in NGC\,6810 from XMM-Newton observations, and explains
that the claim of a hard $L_{\rm x}\! \sim\!  10^{42}$ ergs/s source in this
galaxy was due to the misidentification of a HEAO1 source.  Upper limits of
$L_{\rm x}\!  <\! 1 \times 10^{42}$ and $6 \times 10^{41}$ ergs/s in the soft
band were reported by \citet{rush96a} for MRK\,897 and MCG\,+0-29-23.

Given the above, NGC\,5953 seems to harbor a genuine Seyfert II nucleus, but
MCG\,+0-29-23, NGC\,6810, and MRK\,897 do not. Contamination by a starburst
component in NGC\,34 and NGC\,5953 seems likely. We will present data for all
objects in this work, but exclude the three canonical HII nuclei from analysis
in paper II.

\subsection{Imaging and data reduction}

The following galaxies in our sample have existing high quality SEDs in
\citet{ah03}: NGC\,1068, NGC\,4388, NGC\,4968, NGC\,5506, NGC\,7172 and
NGC\,7674. The remaining 42 Type II Seyfert galaxies were observed in the
following infrared bands: near-Infrared (NIR) J (1.25$\mu$m), H (1.65$\mu$m),
and K (2.2$\mu$m) bands; and mid-IR (MIR) L (3.56$\mu$m), M (4.66$\mu$m), and
N (10$\mu$m) bands.  Details of the observing runs are listed in
Table~\ref{runs}; typical seeing conditions (for a H$_0$=70 km/s/Mpc) are
summarized in Table~\ref{seeing}.

\begin{deluxetable}{ccccc}
\tabletypesize{\footnotesize}
\tablecolumns{5}

\tablecaption{Summary of infrared observing runs. \label{runs}}

\tablehead{
\colhead{Period} & \colhead{Telescope} &
\colhead{Detector}&	\colhead{Mode} & \colhead{Bands}
}

\startdata
2002B	&  ANTU (UT1)	& ISAAC	& Visitor& JHK	\\
2002B	&  ANTU (UT1)	& ISAAC	& Service& LM	\\
2003A	&  ANTU (UT1)	& ISAAC & Service& LM	\\
2003B	&  NTT		& SofI	& Service& JHK	\\
2004A	&  Gemini S	& T-ReCS& Service& N	\\
2004B	&  NTT		& SofI	& Visitor& JHK	
\enddata
\end{deluxetable}

\begin{deluxetable}{ccccccc}
\tabletypesize{\footnotesize}
\tablecolumns{7}
\tablecaption{Summary of the seeing conditions. \label{seeing}}

\tablehead{
\colhead{Band} & \colhead{Mean} & \colhead{$\sigma$} & \colhead{Median} & \colhead{Mean} & \colhead{$\sigma$} & \colhead{Median}\\
\colhead{} & \colhead{[$\arcsec$]}  & \colhead{[$\arcsec$]} & \colhead{[$\arcsec$]} & \colhead{[pc]}& \colhead{[pc]} & \colhead{[pc]}
}

\startdata
J &0.7	  &0.1	&0.7	&325	  &317  &224	\\
H &0.7	  &0.1  &0.6	&318	  &319  &203	\\
K &0.6	  &0.1  &0.6	&296	  &283  &208	\\
L &0.4	  &0.1	&0.4	&206	  &179  &127	\\
M &0.4	  &0.1  &0.4	&227	  &178  &162	\\
N &0.4	  &0.1  &0.3	&105	  &103	&69	\\
\enddata
\end{deluxetable}

\subsubsection{NIR observations} \label{NIR_reduc}

Observations in the near--IR bands followed the usual strategy for this
wavelength regime: several images of short integration time (3--20 seconds)
are obtained, to avoid saturating the detector and to minimize fluctuations in
sky brightness; the location of the object on the detector is changed for each
image, to subtract the sky efficiently and to avoid bad pixels.

The data were reduced, in part, with the XMOSAIC task from the IRAF XDIMSUM
package.  Darks were subtracted and flat fields applied.  Sky images were
calculated from a masked median of the object images or from the mean of
sky--only images.

Sky images were subtracted and results were registered and shifted before
co--adding.  The standard deviation per pixel was calculated from an exposure
map.

The observations with ISAAC in visiting mode were done under non--photometric
conditions.  Flux calibration was achieved using field stars found in the
2MASS catalog\footnote{This publication makes use of data products from the
  Two Micron All Sky Survey, which is a joint project of the University of
  Massachusetts and the Infrared Processing and Analysis Center/California
  Institute of Technology, funded by the National Aeronautics and Space
  Administration and the National Science Foundation.} as standard flux
calibrators. This strategy did not work for NGC\,424, which had no field
stars in the galaxy frame, and those observations were lost.

\subsubsection{MIR observations} \label{MIR_reduc}

At these wavelengths the atmosphere is very bright, and temporarily and
spatially variable; the telescope and surroundings also emit and reflect
radiation.  So observations were made using standard chopping and nodding
techniques.

Mid--IR detectors are small, and the probability of finding a star in the same
frame as the galaxy is slim, so each mid--IR galaxy observation was followed
by observation of a star to determine the Point Spread Function (PSF).  Known
standards were used for flux calibration.

The ISAAC mid--IR observations were reduced with the ECLIPSE ``jitter''
pipeline. In a number of cases observing conditions degraded at the end of the
integration, so some of the nodding series were discarded by hand.

The Gemini N band data were reduced using the MIDIR task from the IRAF GEMINI
package.  When pipeline results were compared with a reduction by hand the
images were very similar.

\section{Constructing the SEDs}\label{construction}

The SEDs contain nuclear magnitudes extracted from near-- and mid--IR images.
In \S\ref{2Dvs1D} we describe how 1-D brightness profiles are calculated from
2-D images; the modeling of these profiles, to separate the nuclear
component, is described in \S\ref{modeling}.

\subsection{Determining the Surface Brightness Profiles}\label{2Dvs1D}

Deep IR images of a sample of Type II Seyfert galaxies were obtained. These
images cover the entire galaxy, but only the nuclear emission is needed for
the SED.  To deconvolve the galaxy into its different components --- bulge,
disk, bar, and nucleus --- we first determine the radial Surface Brightness
Profile (SBP) of the galaxy
\citep{kotilainen92,zitelli93,ah96,prieto01,prieto02}. The main advantage of
this 1-D approach is that smaller structures (spiral arms, HII Regions, rings)
are diluted relative to the more fundamental structures (bulge, disk, nucleus
and bar) and the quantity of data that must be fitted is reduced from $\sim\!
500\! \times\! 500$ pixels to $\sim\! 200$ points.

The task ELLIPSE from the IRAF STSDAS package is used on the sky subtracted
images to obtain the SBPs. A manual masking removes contamination from
different sources that might distort the isophote fitting, like foreground
bright stars or other galaxies. ELLIPSE fits ellipses to the galaxy isophotes;
these are described by the center coordinates, the position angle and the
ellipticity. The center coordinates were held fixed while the other parameters
were allowed to vary. A linear radial sampling was used, with steps of 0.5--1
pixel, depending on the size of the galaxy. In cases where the edge of the
galaxy was too faint and the ellipse fitting process did not converge, the
fitted isophotes observed in other images were adopted. In order to take into
account the inclination of the galaxy, the SBP was corrected using the mean
ellipticity of the last fitted isophotes.

\subsection{Modeling the Surface Brightness Profiles}\label{modeling}

Most spiral galaxies can be described using a disk, bulge, optional bar, and,
sometimes, an active nucleus.  Historically each component has been modeled
with a specific analytical function.  These are described below.

\paragraph{Nucleus}
The nuclear point source is modeled using Dirac's Delta:
\begin{eqnarray}
\int_{-a}^a\delta(t)dt &=& 1
\end{eqnarray}
So the nuclear SBP is described by:
\begin{eqnarray}\label{Sigma_nucleus}
\Sigma_n(r) &=& \sigma_n\delta(r),
\end{eqnarray}
where $\sigma_n$ is the nuclear surface brightness.

\paragraph{Disk}
The disk is modeled using an exponential law:
\begin{eqnarray}
\Sigma_d(r) &=& \sigma_d \exp\{-r/r_d\},
\end{eqnarray}
where $\sigma_d$ is the surface brightness of the disk at $r=0$,
$\sigma_d=\Sigma_d(r=0)$, and $r_d$ is the disk scale length.

\paragraph{Bulge}\label{bulge}
The bulge is modeled using a general S\'ersic's law:
\begin{eqnarray}
\Sigma_B(r) &=& \sigma_B \exp\{-b_n \cdot (r/r_B)^{1/n}\},
\end{eqnarray}
where $n$ is a free parameter and $b_n$ is defined by:
\begin{eqnarray}
\Gamma(2n) &=& 2\gamma(2n,b_n),
\end{eqnarray}
where $\Gamma$ is the (complete) gamma function, and $\gamma(2n,x)$ is the
incomplete gamma function defined by
\begin{eqnarray}
\gamma(2n,x) &=& \int_0^x e^{-t}t^{2n-1}dt.
\end{eqnarray}
Common values of $b_n$ are $b_1=1.678$, corresponding to an exponential SBP,
and $b_4=7.6692$, corresponding to a de Vaucouleur's profile. The most usual
analytical expressions for $b_n$ are $b_n=1.9992 \cdot n - 0.3271$ for $0.5\!
<\! n\! <\! 10$, and $b_n=2n-1/3~$ for $n \gtrsim 8$
\citep{graham01,graham05}. In this work we adopt the first approximation,
because the typical values of $n$ are between 0.5 and 4 in the IR bands
\citep{balcells03,hunt04,grosbol04,kormendy04}.

\paragraph{Bar}
Many galaxies in the sampled are classified as barred, and approximately 2/3
of all spiral galaxies classified as non-barred look barred in the IR
(\citet{kormendy04}, and references therein). In order to model the bar, a
second general S\'ersic's law was included in the model, which introduced 3
new free parameters.

\paragraph{Final Model}
The final model is the sum of the 3 or 4 components convolved with the image
PSF in order to introduce the effect of the seeing:
\begin{eqnarray}\label{final_model}
\Sigma (r) &=& [ \sigma_n\delta(r) + \sigma_d e^{-\frac{r}{r_d}} +
  \sigma_B e^{-b_{n_B} (\frac{r}{r_B})^{1/n_B}} \nonumber\\ & & + { } \sigma_b
  e^{-b_{n_b} (\frac{r}{r_b})^{1/n_b}} ] \otimes PSF(r)
\end{eqnarray}

\subsubsection{The Point Spread Function}\label{moff_psf}

\begin{figure}
  \begin{center}
    \includegraphics[scale=0.4,angle=270]{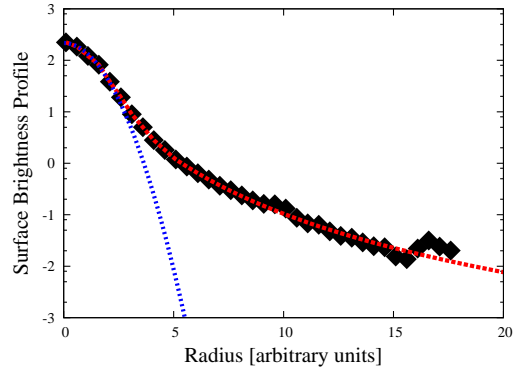}
    \caption{Example of the Moffat fit of a PSF star. The observed data are
      the black diamonds and the Moffat profile is represented by the red
      line. A Gaussian profile is shown for comparison
      (blue).\label{moff-gauss}}
  \end{center} 
\end{figure}

To determine the correct parameters for the galaxy profile we need a precise
representation of the PSF describing the seeing and instrument response of
each observation. For our NIR imaging, the PSF is very accurately determined
from a star found in the galaxy frame, which guarantees the same distortion
for galaxy and star. In the MIR imaging the star is observed before and/or
after the galaxy and, as the integration is quite long at these wavelengths,
seeing conditions can change significantly.  This means that the MIR stellar
PSF may not accurately match the galaxy.  In such cases the bulge profile is
taken from the NIR fit, but scaled to match the external part of the observed
profiles, and the nuclear flux was obtained assuming $F_{nuc} =
F_{gal}-F_{bul}$. See Appendix~\ref{apendiceA} for details.

The PSF radial profile was obtained using the same technique as the galaxy
profile (Section~\ref{2Dvs1D}) and then modeled with a Moffat function to
allow for interpolation / extrapolation when fitting the galaxy profile. The
Moffat function is similar to a Gaussian, but has more extended tails which
better fit the effects of seeing, as shown in Fig.~\ref{moff-gauss}. It is
given by:
\begin{eqnarray}
M\!o\!f\!\!f(r) &=& \frac{peak}{(1 + (r/\alpha)^2)^\beta}~.
\end{eqnarray}

\subsubsection{The Fitting Routine}\label{fit_rou}

To measure the nuclear flux --- i.e., to deconvolve the nuclear emission from
that of the galaxy --- the radial SBP was modeled using the expression
described above (Eqn.~\ref{final_model}).  The fitting routine uses the
``Downhill Simplex Method in Multidimensions'' first proposed by Nelder \&
Mead (1965): the minimization process starts with an initial guess from the
user and then takes a series of steps exploring the available parameter space
of the problem until it achieves the termination criterium which is,
typically, an insignificant variation between one step and the next.

As the algorithm does not guarantee that the minimum found is global, it was
started from 4 or 5 extreme points of the parameter space: the first initial
guess includes all the galactic components of the model with comparable
contributions to the total flux; other initial points each zero one of the
components. These first guesses are not completely blind, but instead are
based on the morphology seen in the images. An iterative process that randomly
perturbs the initial point is also included.  Tests showed that 200 iterations
were sufficient to reliably find the global minimum, which corresponds to the
model that minimizes $\chi^2$; the percentage contributions of the galactic
components are obtained from this model.

\begin{figure*}
  \begin{center}
    \epsscale{0.36}
    \plotone{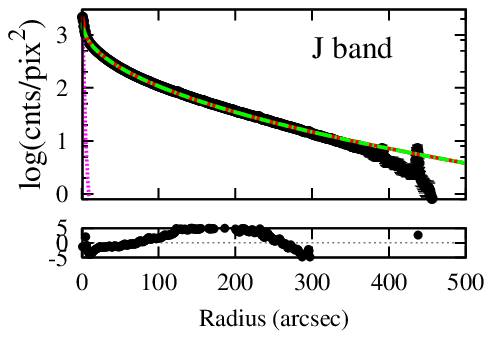}
    \plotone{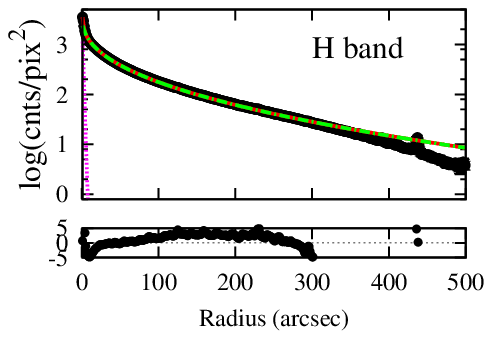}
    \plotone{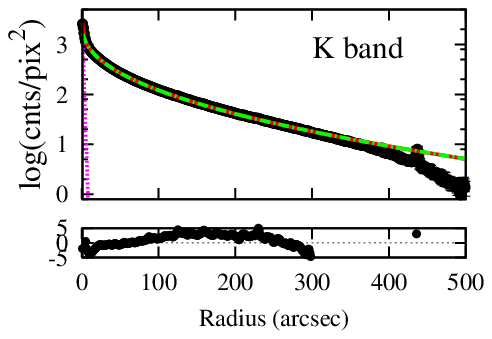}
    \plotone{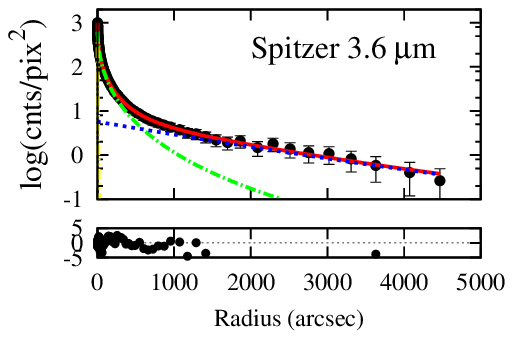}
    \plotone{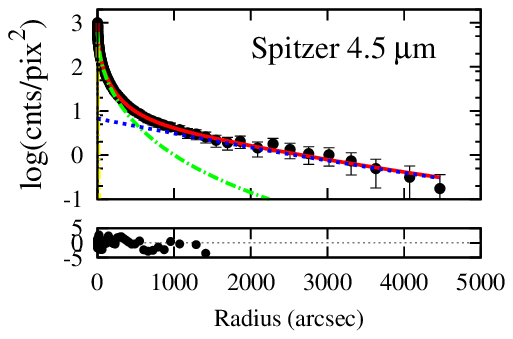}
    \plotone{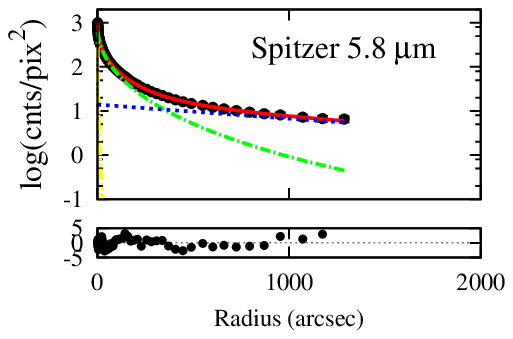}
    \plotone{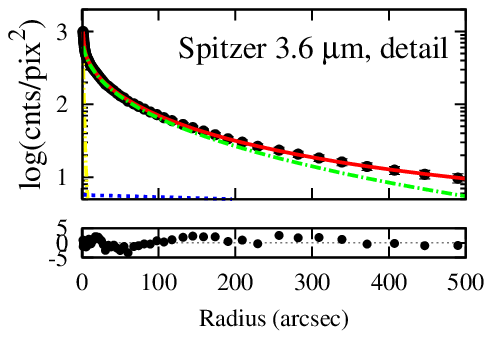}
    \plotone{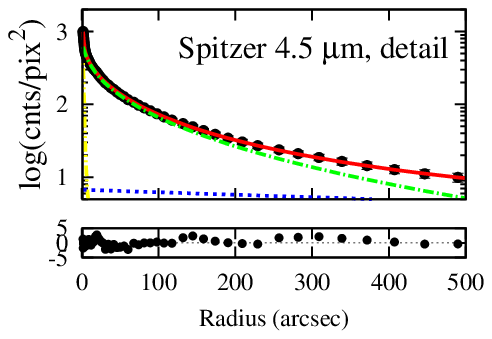}
    \plotone{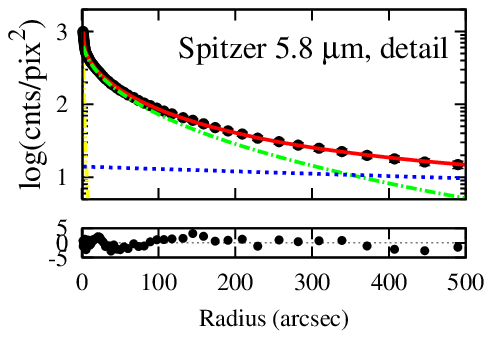}
    \caption{Radial surface brightness profile fits of J, H, and K bands (top
      panels), and $3.6\mu$m, $4.5\mu$m, and $5.8\mu$m (middle and lower
      panels) of galaxy Messier~31.  The abscissas show Log(counts/pix$^2$)
      and the ordinates show the distance from the center in arc-seconds. The
      percentage residues are shown in the bottom panels. The model (red thick
      continuous line) is composed by a nuclear PSF (pink dotted-line), a
      bulge (green dashed-dotted line), a disk (blue dashed line), and a bar
      (yellow dashed-double-dotted). The black points correspond to the
      observation. The brightness excess in the central part of the galaxy is
      modeled as a bar in the IRAC bands and as a nuclear component in the
      2MASS bands. See text for details. \label{fit_m31}}
  \end{center}
\end{figure*}

\subsubsection{Errors in the fitted components}\label{error_seds}

The flux of each fitted galactic component has two main sources of
uncertainty: the first from the flux calibration (the zero points, obtained by
the standard calibration procedure) and the second from the fitting
process. The confidence limits for the fitted parameters, determined using the
fitting routine described previously, are calculated using Monte Carlo
simulations: the best--fit parameters are perturbed (one at a time) to
construct many synthetic data sets. Each of these synthetic data sets has a
particular $\chi_{syn}^2$ value and the first confidence level (corresponding
to 1$\sigma$ error), which occurs 68.3\% of the times, corresponds to the
cases where $\Delta\chi^2 = \chi_{min}^2 - \chi_{syn}^2 \leq 1$.  The final
errors are calculated as the quadratic sum of the individual errors for the
two sources of uncertainty.

\subsubsection{Synthetic Data}

Synthetic data were used to test the fitting routines. SBPs of different
``kinds'' of galaxies were constructed: including all components equally; all
components with very small or very large contributions; omitting one or more
components; with wrong PSFs; and with different sampling steps (for the galaxy
and the PSF star).

The first test recovered the parameters of the synthetic SBP. The routine
gives excellent estimates of the parameters for galaxies composed of 3 or 4
structural parameters with non-negligible contributions to the total
emission. In cases when one or more components were not present, the results
were also good, corresponding to observed examples like the non-barred
galaxies, or a couple of cases where nor disk nor bar is observed.  Including
all the 4 components might give a better $\chi^2$, because there will be more
free parameters, but if the image does not show a bar, for example, it is
obvious that it should not be included in the fit.

Next, changes in the PSF were used to test the effects on the results (i.e.,
using a wrong PSF to describe the seeing conditions). We found that the
results could be recovered (this is, at least 2 of the initial starting points
could reach the actual solution) whenever the FWHM was narrower or broader by
up to 50\%\ of its real value. These tests apply especially to the mid--IR
observations: as the integration time is long, the probability of a change in
the seeing conditions between the observation of the PSF star and galaxy is
greater.

The truncation of the galaxy profile severely affects the performance of the
fitting routine. Fortunately there are very few such cases in the
observations, and then, typically, only one of the broad--band images is
affected, which means that results can be checked with the unaffected profiles
and parameters constrained appropriately.

Finally, it is found that the sampling step is not particularly important,
providing that the PSF is sampled by a finer discretization than the sampling
of the galaxy.

\subsection{Messier 31}\label{m31}

\begin{deluxetable}{lcccccc}
\tabletypesize{\footnotesize}
\tablecolumns{6} \tablecaption{Values for the relevant fitted
  parameters of M 31 deconvolution process. \label{m31_param}}

\tablehead{
\colhead{Component} & \colhead{1.25$\mu$m} & \colhead{1.65$\mu$m}  & \colhead{2.2$\mu$m} & \colhead{3.6$\mu$m} & \colhead{4.5$\mu$m}  & \colhead{5.8$\mu$m} 
}

\startdata
Bulge $r_{e\!f\!f}$	&185    	&228    	&207    	&301    	&259    	&209    \\
Bulge $n$		&2.0      	&2.3      	&2.2      	&2.5      	&2.4      	&2.2      \\
         		&         	&         	&         	&         	&         	&         \\
Disk $r$		&-- 	  	&--   	  	&--   	  	&1165     	&991    	&996    \\
         		&         	&         	&         	&         	&         	&         \\
Bar $r_{e\!f\!f}$		&-- 	  	&--   	  	&--   	  	&1.2   	  	&1.2   	  	&0.6   	  \\
Bar $n$		        &-- 	        &--      	&--      	&0.6            &0.5      	&1.2      \\
\enddata

\tablecomments{The scale lengths are expressed in arc-seconds. The bright
  excess in the central part of the galaxy profiles is modeled as a bar in the
  IRAC bands and as a nuclear component in the 2MASS bands (due to the
  differences in pixel scale). The disk is not included in the 2MASS fits
  because of the smaller image coverage. See text for details.}
\end{deluxetable}

To test the SBP fitting procedure, Messier\,31, the Andromeda galaxy, was
studied in the same way as the objects in our sample.

Images of M31 were taken by IRAC as part of the {\emph{Spitzer}} General
Observer program, ID 3126, in August 2005. \citet{barmby06} measured the SBPs
by fitting elliptical isophotes to the images using the ELLIPSE task. Profiles
in the three longer wavelength bands (at 4.5, 5.8, and 8.0 $\mu$m) were
measured using the ellipses fitted to the $3.6~\mu$m image, so that colors
could be measured at the same spatial locations from all 4 bands. These
profiles were kindly provided by P. Barmby \citep{barmby06}.

Images in the J, H, and K bands were obtained from the 2MASS public catalog.
The profiles were constructed using ELLIPSE. However, M31 is so close that the
2MASS images do not cover the full extent of the galaxy: while
{\emph{Spitzer}}-IRAC covers $\gtrsim 4000\arcsec$ in radius, the images of
2MASS cover only $\sim\! 500\arcsec$. This compromises the results for the
disk component in the 2MASS data.

M31 is one of the nearest galaxies, and all galactic components can be
detected in all the IR bands, which is not the case for the galaxies in our
sample. The SBPs of M31 were fitted from $1.25\mu$m to $5.8\mu$m
(Fig.~\ref{fit_m31}). The $8.0\mu$m SBP was not included because it is
dominated by a 10 Kpc star-forming ring and individual stellar features. The
$5.8\mu$m SBP was truncated at $\sim\! 1300''$ to avoid wiggles produced by
the 10 Kpc ring. The relevant fitted parameters are listed in
Table~\ref{m31_param}.

It is interesting to note the bright excess in the central part of the galaxy
profiles. This excess is modeled as a bar in the IRAC bands and as a nucleus
in the 2MASS bands (due to the difference in pixel scale). \citet{peng02b}
studied the central region of M31 in detail using the {\emph Hubble Space
  Telescope} (HST). He modeled the center of the galaxy using 6 components to
account for 3 different peaks of emission in UV and optical images. However
the UV peak disappears in the near-IR images and the 2 V-band peaks diminish,
so here we assume that the nucleus/bar fitted by the process is real, but the
pixel scale does not allow a more detailed modeling.

As the disk could not be fitted in the 2MASS profiles only the fitted radius
of the bulge is compared: 700 pc in the J band, 862 pc in the H band, 787 pc
in the K band, 1140 pc at $3.6~\mu$m, 983 pc at $4.5~\mu$m, and 795 pc at
$5.8~\mu$m. The dispersion of these values about the mean is between 20 and
30\% and they show no significant systematic change moving toward longer
wavelengths.

The results corroborate some na\"ive expectations: unless there are strong
color gradients within each of the structural components (the bulge, disk,
and/or bar) the scale lengths should remain fairly constant for the galaxies
in our sample, while the brightness scale factor should decrease from the NIR
to the MIR as the stellar populations become dimmer. Also, since the bulge has
older stars and is redder than the disk and the bar, it is expected to be the
brightest component in the mid-IR.

Given these findings for M\,31, we decided, when fitting the SBPs of the
galaxies in our sample, to constrain the scale lengths in the MIR using the
values fitted in the NIR, since the very dim stellar components
under--determine the fitting at these wavelengths. See
Appendix~\ref{apendiceA} for details.

\begin{deluxetable}{lccccccc}
\tabletypesize{\footnotesize} 
\tablecolumns{8} 

\tablecaption{Mean, standard deviation, skewness, median and mode percentage
  contribution of each galactic component\label{aju_contri}}

\tablehead{
\colhead{} & \colhead{Band} & \colhead{Number} & \colhead{Mean} & \colhead{$\sigma$} & \colhead{Skew} & \colhead{Median} & \colhead{Mode}\\ 
           &                &                  & \colhead{(\%)} & \colhead{(\%)}     & \colhead{(\%)} & \colhead{(\%)}   & \colhead{(\%)}\\
}
\startdata
Nucleus		&M	&20/28	&84     &27     &-1.5	&100	&100\\
		&L	&25/28	&42 	&28 	& 0.2	&38 	&44 \\
		&K	&32/34	&13 	&18 	& 2.1   &3.3	&13 \\
		&H	&31/34	&8.4	&14 	& 2.4   &1.8	&0.8\\\smallskip
		&J	&32/34	&5.9	&12 	& 3.9   &1.6	&0.8\\ 
Bulge		&M	&7/28	&44 	&31 	& 0.3	&30 	&44 \\
		&L	&25/28	&52 	&28 	& 0.1	&53 	&57 \\
		&K	&34/34	&35 	&25 	& 0.9	&29 	&35 \\
		&H	&34/34	&33 	&22 	& 1.0   &27 	&33 \\\smallskip
		&J	&34/34	&34 	&23 	& 0.9	&27 	&34 \\
Disk		&M	&0/28   &--     &--     &--     &--     &-- \\
		&L	&4/28	&36 	&21 	&-0.2	&37 	&36 \\
		&K	&31/34	&47 	&25 	& 0.1	&42 	&47 \\
		&H	&31/34	&52 	&25 	&-0.2	&51 	&52 \\\smallskip
		&J	&31/34	&55 	&24 	&-0.1	&48 	&55 \\
Bar		&M	&0/28   &--     &--     &--     &--     &-- \\
		&L	&1/28   &17     &--     &--     &17     &17 \\
		&K	&12/34	&28 	&18 	& 0.5	&26 	&28 \\
		&H	&12/34	&30 	&17 	&-0.3	&31 	&30 \\
		&J	&12/34	&30 	&17 	& 0.1	&31 	&30 \\ 

\enddata \tablecomments{In column 'Number' we indicate the number of fits
  which include that particular component over the total number of images
  available in a particular band. Results obtained by AH03 are not included in
  this Table.}
\end{deluxetable}

\begin{turnpage}
\begin{deluxetable*}{lccccccccccc}
\tabletypesize{\footnotesize}
\tablecolumns{11}

\tablecaption{Fitted parameters of the galaxy
deconvolution in each band of observation.\label{fitted_param}}

\tablehead{
\colhead{Galaxy} & \colhead{Band} & \colhead{$\sigma_{nuc}$} & \colhead{$\sigma_{Bul}$} & \colhead{r$_{eff}^{Bul}$} & \colhead{n$_{Bul}$} & \colhead{$\sigma_{dis}$} & \colhead{r$_{dis}$} & \colhead{$\sigma_{bar}$} & \colhead{r$_{eff}^{bar}$} & \colhead{n$_{bar}$} \\
\colhead{} & \colhead{} & \colhead{Watts/m$^2$/$\mu$/[\arcsec]$^2$} & \colhead{Watts/m$^2$/$\mu$/[\arcsec]$^2$} & \colhead{pc} & \colhead{} & \colhead{Watts/m$^2$/$\mu$/[\arcsec]$^2$} & \colhead{pc} & \colhead{Watts/m$^2$/$\mu$/[\arcsec]$^2$} & \colhead{pc} & \colhead{} 
}

\startdata
NGC\,34			&J	& 5.6e-14	& 2.4e-14	&  269	& 1.4	& 1.2e-15	& 1102	& -- 		& -- 		& -- 	 \\
			&H	& 1.2e-13	& 2.9e-14	&  225	& 1.3	& 1.5e-15	& 1071	& -- 		& -- 		& -- 	  \\
			&K	& 4.8e-14	& 3.1e-14	&  292	& 2.2	& 3.3e-16	& 1134	& -- 	        & -- 	 	& -- 	 \\
			&L	& 2.9e-14	& 1.0e-14	&  162	& 1.2	& 9.7e-17	&  980  	& -- 		& -- 		& -- 	 \\
			&M	& 1.5e-14	& 7.5e-15	&  225	& 1.8	& -- 		& --  	& -- 	        & -- 		& -- 	 \\
&&&&&&&&&&\\

IRAS\,00198-7926\,N	&J	& 9.0e-15	& 2.9e-16	& 1397	& 0.5	& 8.8e-17	& 7765	& -- 		& -- 		& -- 	 \\
			&H	& 6.7e-15	& 2.3e-16	& 1722	& 0.5	& 5.0e-17	& 8908	& -- 		& -- 		& -- 	  \\
			&K	& 5.7e-15	& 9.5e-17	& 1559	& 0.3	& 3.5e-17	& 8670	 & -- 	        & -- 	 	& -- 	 \\
			&L	& 6.6e-14	& 6.3e-16	& 1262 	& 0.5	& -- 		& --    	& -- 		& -- 		& -- 	 \\
			&M	& 9.5e-14	& -- 		& --  	& -- 	& -- 		& --  	& -- 	        & -- 		& -- 	 \\
&&&&&&&&&&\\

IRA\,00198-7926\,S	&J	& 3.4e-15	& 5.3e-16	& 2450	& 0.9	& 1.9e-16	& 5939	& -- 	& -- 		& -- 	 \\
			&H	& 8.2e-15	& 2.4e-16	& 2108	& 0.5	& 1.5e-16	& 5671	& -- 	& -- 		& -- 	 \\
			&K	& 2.8e-14	& 6.6e-17	& 2197	& 0.2	& 1.0e-16	& 5538	& -- 	& -- 	 	& -- 	 \\
			&L	& 6.6e-14	& 6.3e-16	& 1262 	& 0.5	& -- 		& --    & -- 	& -- 	 	& -- 	 \\
			&M	& 9.5e-14	& -- 		& --  	& -- 	& -- 		& --  	& -- 	& -- 		& -- 	 \\
&&&&&&&&&&\\
                        	
IRAS\,00521-7054	&J	& 5.7e-14	& 5.0e-16	& 4225	& 1.4	& -- 		& --  	& -- 	& -- 		& -- 	 \\
			&H	& 7.5e-14	& 2.4e-15	& 3004	& 1.9	& -- 	        & --  	& -- 	& -- 		& -- 	 \\
			&K	& 1.2e-13	& 4.8e-16	& 2821	& 1.2	& -- 		& --  	& -- 	& -- 	 	& -- 	 \\
			&L	& 1.7e-16	& 8.0e-16	& 3088 	& 1.9	& -- 		& --    & -- 	& -- 	 	& -- 	 \\
&&&&&&&&&&\\
                        	
ESO\,541-IG12		&J	& 4.3e-14	& 9.9e-16	& 2460	& 1.1	& 7.9e-17	& 5763	& -- 	& -- 		& -- 	 \\
			&H	& 9.5e-14	& 7.6e-16	& 2506	& 1.0	& 4.4e-17	& 6810	& -- 	& -- 		& -- 	 \\
			&K	& 1.3e-13	& 8.7e-16	& 2346	& 1.4	& 1.9e-17	& 6902	& -- 	& -- 	 	& -- 	 \\
			&L	& 4.4e-13	& 3.3e-16	& 2153 	& 1.4	& -- 		& --    & -- 	& -- 	 	& -- 	 \\
			&M	& 2.5e-13	& 3.5e-16	& 2107	& 1.4	& -- 		& --  	& -- 	& -- 		& -- 	 \\
&&&&&&&&&&\\
                        	
IRAS\,01475-0740	&J	& 2.2e-14	& 1.3e-15	&  466	& 0.9	& 1.3e-16	& 1090	& -- 	& -- 		& -- 	 \\
			&H	& 2.2e-14	& 1.4e-15	&  462	& 1.0	& 1.4e-16	&  799	& -- 	& -- 		& --     \\
			&K	& 2.0e-14	& 4.7e-16	&  444	& 0.9	& 4.9e-17	&  852	& -- 	& -- 	 	& -- 	 \\
			&L	& 5.9e-14	& 1.6e-16	&  380  & 1.0	& -- 		& --    & -- 	& -- 	 	& -- 	 \\
&&&&&&&&&&\\
                        	
NGC\,1144		&J	& 1.6e-14	& 4.1e-15	&  404	& 0.6	& 1.6e-15	& 2478	& -- 	& -- 		& -- 	 \\
			&H	& 1.1e-14	& 3.6e-15	&  422	& 0.7	& 1.0e-15	& 2478	& -- 	& -- 		& -- 	 \\
			&K	& 8.0e-15	& 1.4e-15	&  399	& 0.6	& 4.4e-16	& 2253	& -- 	& -- 	 	& -- 	 \\
			&L	& 2.1e-14	& 3.7e-16	&  329  & 0.9	& 1.0e-16	& 2149 	& -- 	& -- 	 	& -- 	 \\
&&&&&&&&&&\\
                        	
MCG\,-2-8-39		&J	& 3.5e-15	& 2.1e-15	& 1762	& 1.6	& 2.6e-17	& 11483	& 2.4e-17	& 3216	& 0.2	 \\
			&H	& 2.1e-15	& 1.5e-17	& 2698	& 0.2	& 1.2e-17	& 17104	& 1.9e-15	& 1971	& 1.8	 \\
			&K	& 1.5e-15	& 1.4e-15	& 2076	& 2.1	& 7.5e-18	& 14443	& 7.4e-18       & 2550 	& 0.2	 \\
			&L	& 1.2e-14	& 2.9e-16	& 1750 	& 2.0	& -- 		& --    & -- 	& -- 	 	& -- 	 \\
&&&&&&&&&&\\
                        	
NGC\,1194		&J	& 1.1e-15	& 2.5e-14	& 3877	& 3.3	& 2.5e-17	& 7256	& -- 	& -- 		& -- 	 \\
			&H	& 4.7e-15	& 2.3e-14	& 3288	& 3.4	& 3.7e-17	& 4109	& -- 	& -- 		& -- 	 \\
			&K	& 1.8e-14	& 1.3e-14	& 3313	& 3.5	& 1.1e-17	& 5935	& -- 	& -- 	 	& -- 	 \\
&&&&&&&&&&\\
                        	
NGC\,1320		&J	& 7.3e-15	& 7.7e-15	& 1597	& 2.3	& 1.2e-16	& 3091	& -- 	& -- 		& -- 	 \\
			&H	& 1.7e-14	& 5.8e-15	& 1357	& 2.1	& 1.0e-16	& 3238	& -- 	& -- 		& -- 	 \\
			&K	& 2.6e-14	& 2.6e-15	& 1818	& 2.3	& 3.2e-17	& 3203  & -- 	& -- 	 	& -- 	 \\
\enddata 
\tablecomments{The remaining of this Table is available in the online version of the Journal.}
\end{deluxetable*}
\end{turnpage}

\section{Galaxy Fitting Results}\label{SEDs}

The fitting of the SBPs for each of the galaxies in our sample was done
assuming the model described in \S\ref{modeling}. Visual inspection of the
images and the Hubble classification of each galaxy were used to check for
coherent results.

\begin{figure}
  \begin{center}
    \includegraphics[scale=0.35,angle=0,trim=100 0 250 0]{LOIII_nuc.eps}%
    \includegraphics[scale=0.35,angle=90]{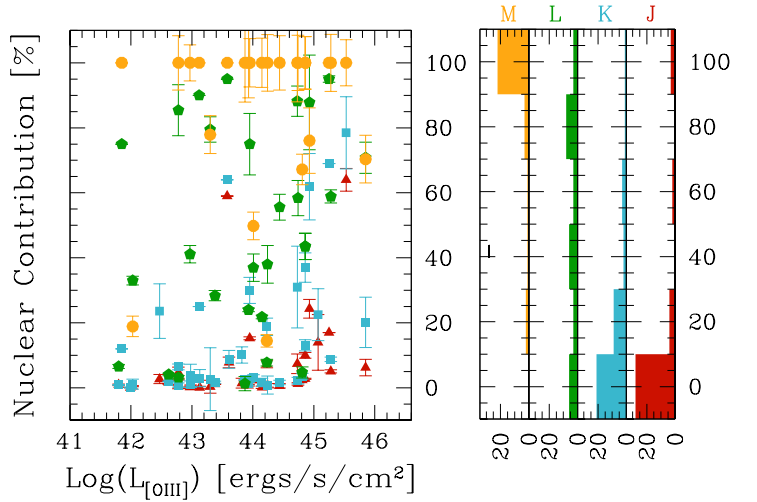}
    \caption{Contribution from the nuclear flux to the total flux as a
      function of band: J, K, L and M measurements are presented with
      triangles, squares, pentagons, and circles, respectively. The
      right--hand side histograms show the projected
      distributions. \label{aju_contri_fig}}
  \end{center}
\end{figure}

\begin{figure*}
  \begin{center}
    \includegraphics[scale=0.25,trim=0   100 0 100]{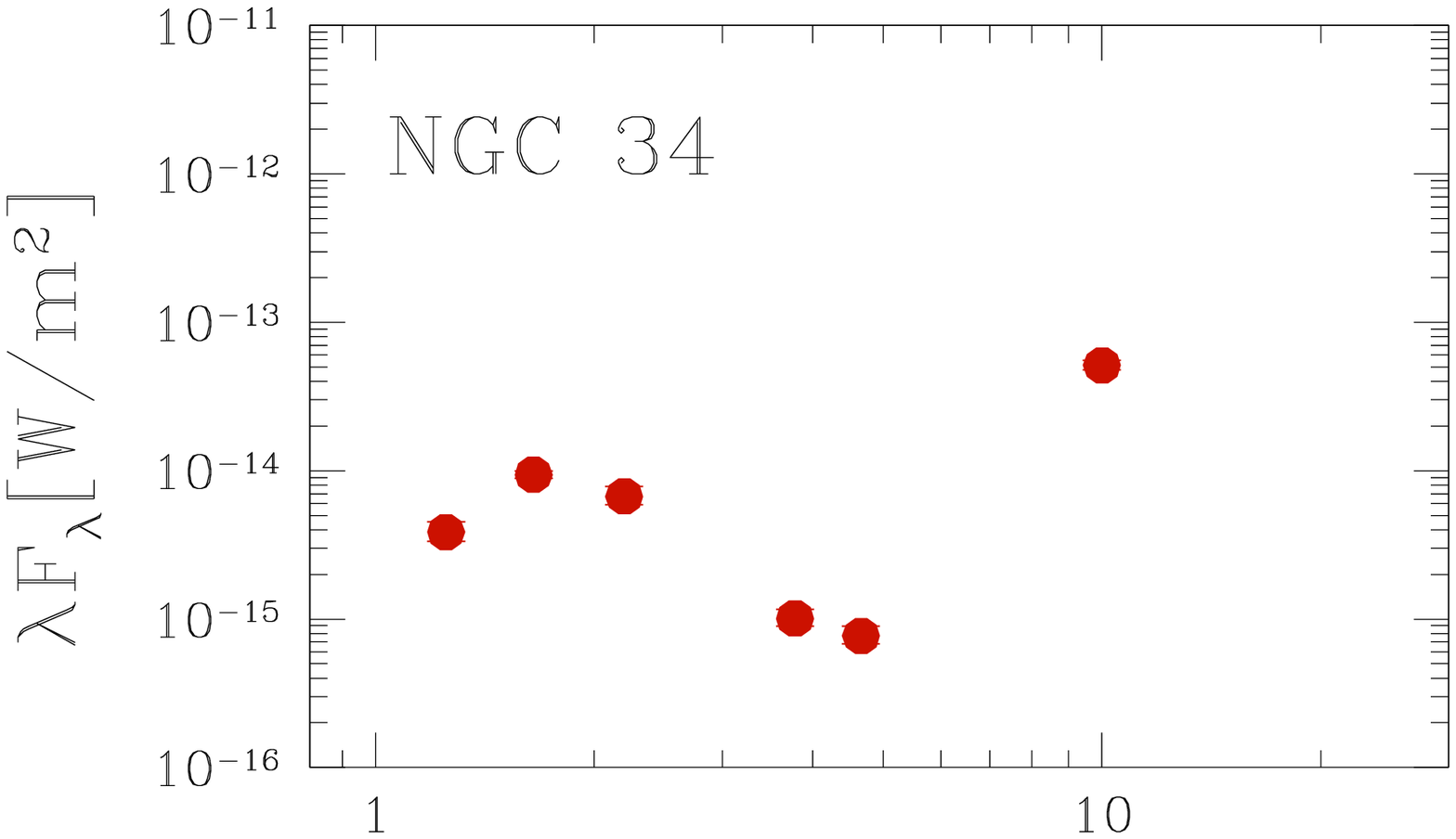}
    \includegraphics[scale=0.25,trim=100 100 0 100]{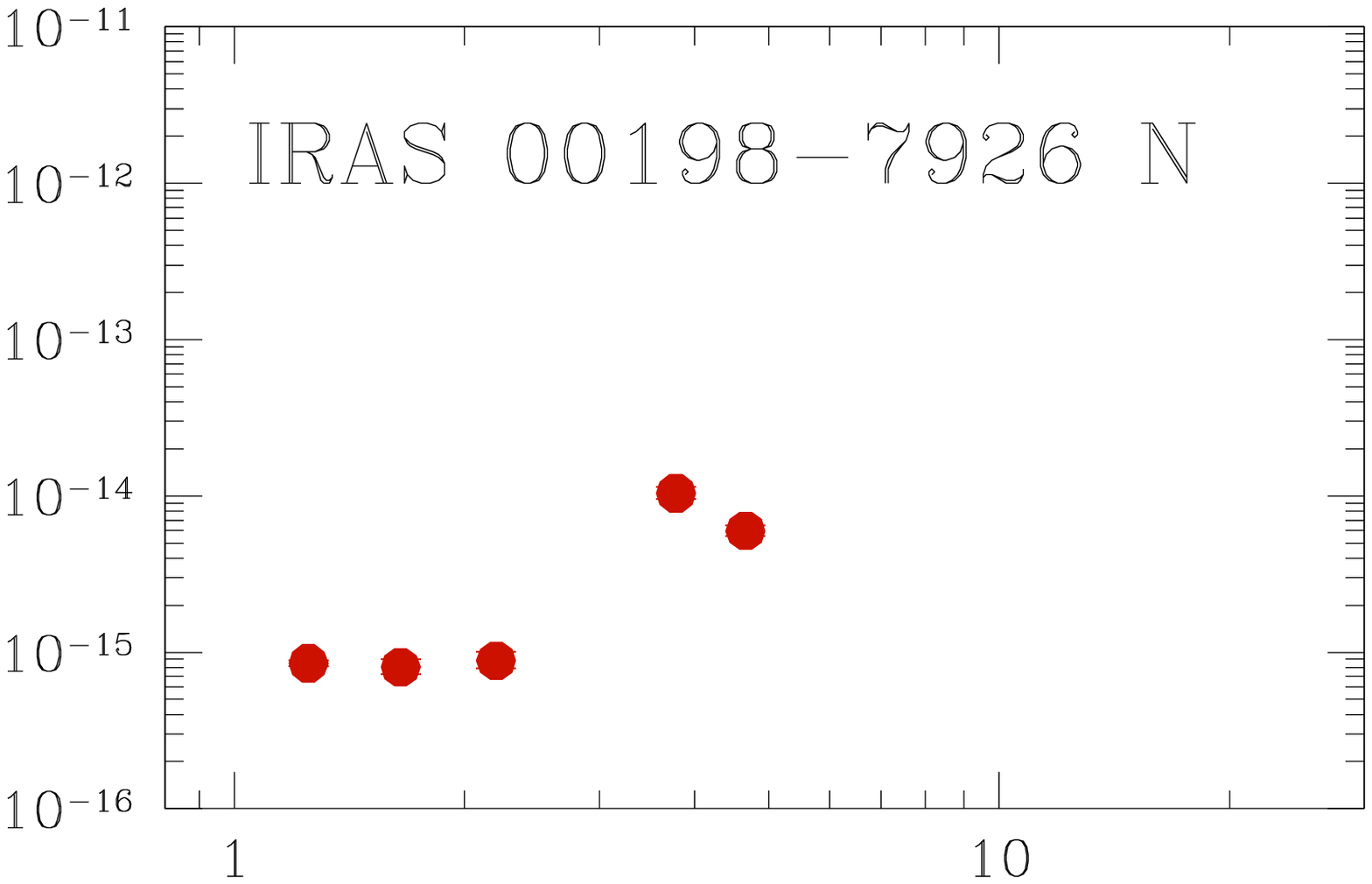}
    \includegraphics[scale=0.25,trim=100 100 0 100]{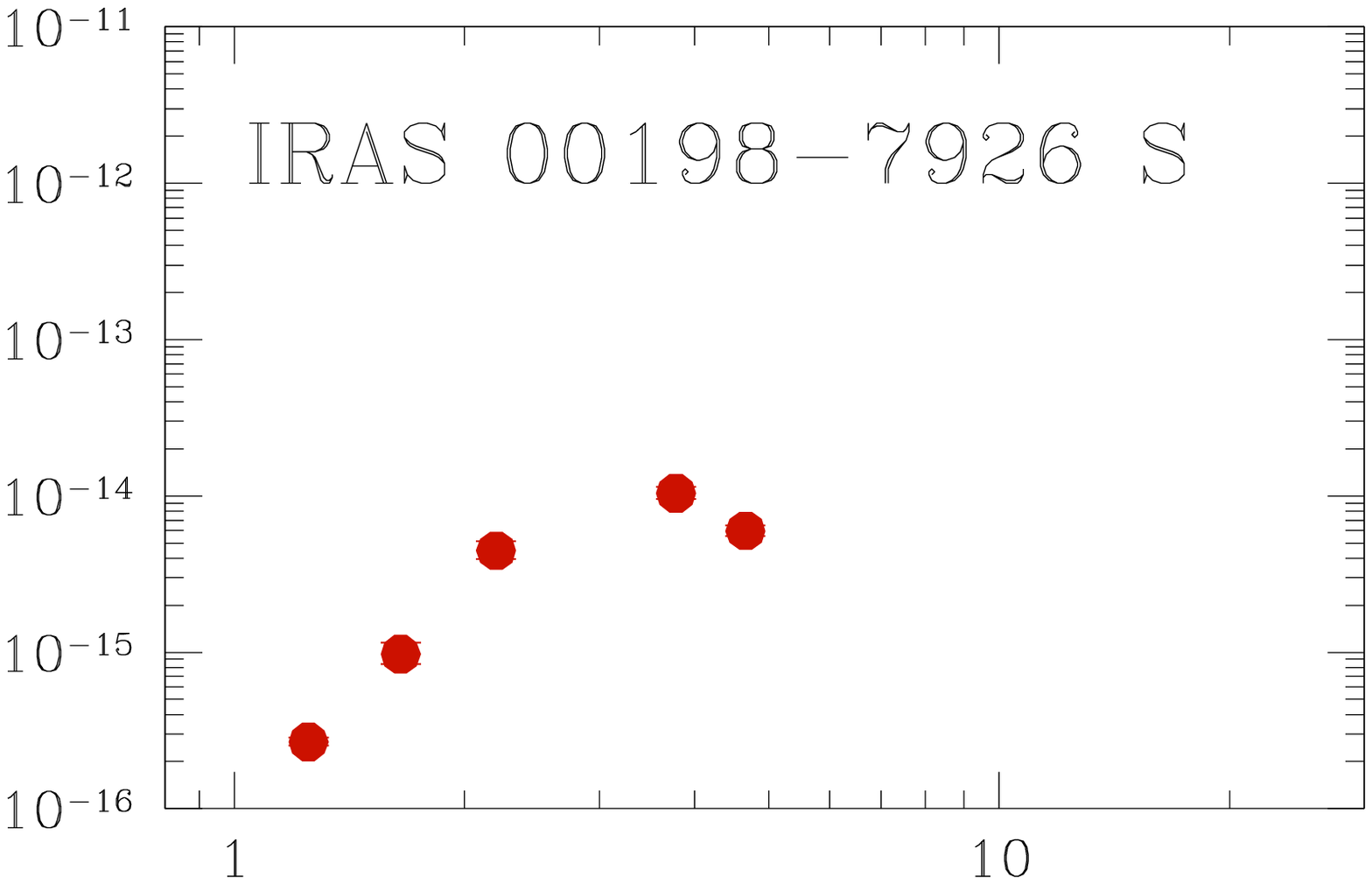}
    \includegraphics[scale=0.25,trim=100 100 0 100]{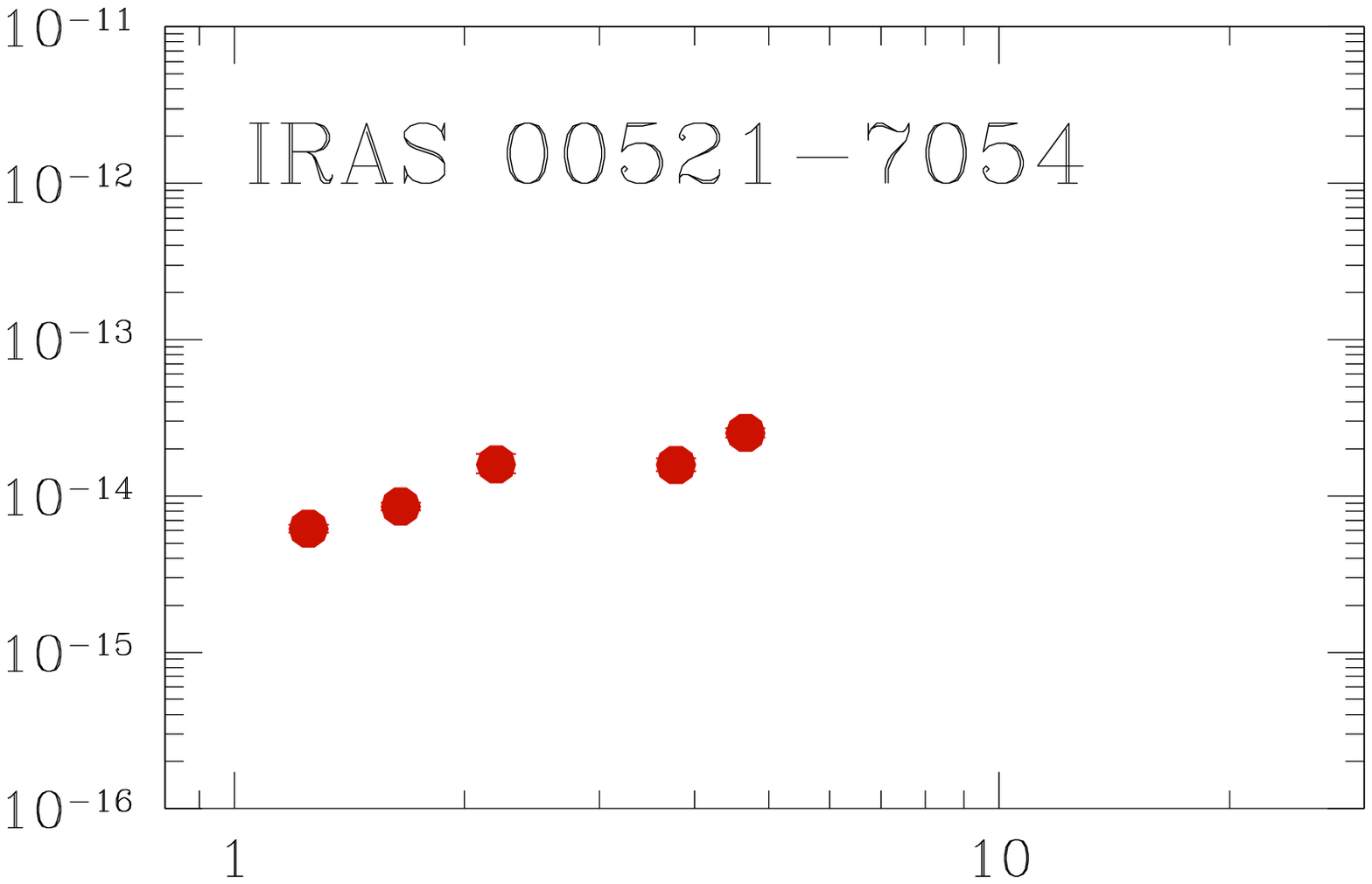}\\
    \includegraphics[scale=0.25,trim=0   100 0 100]{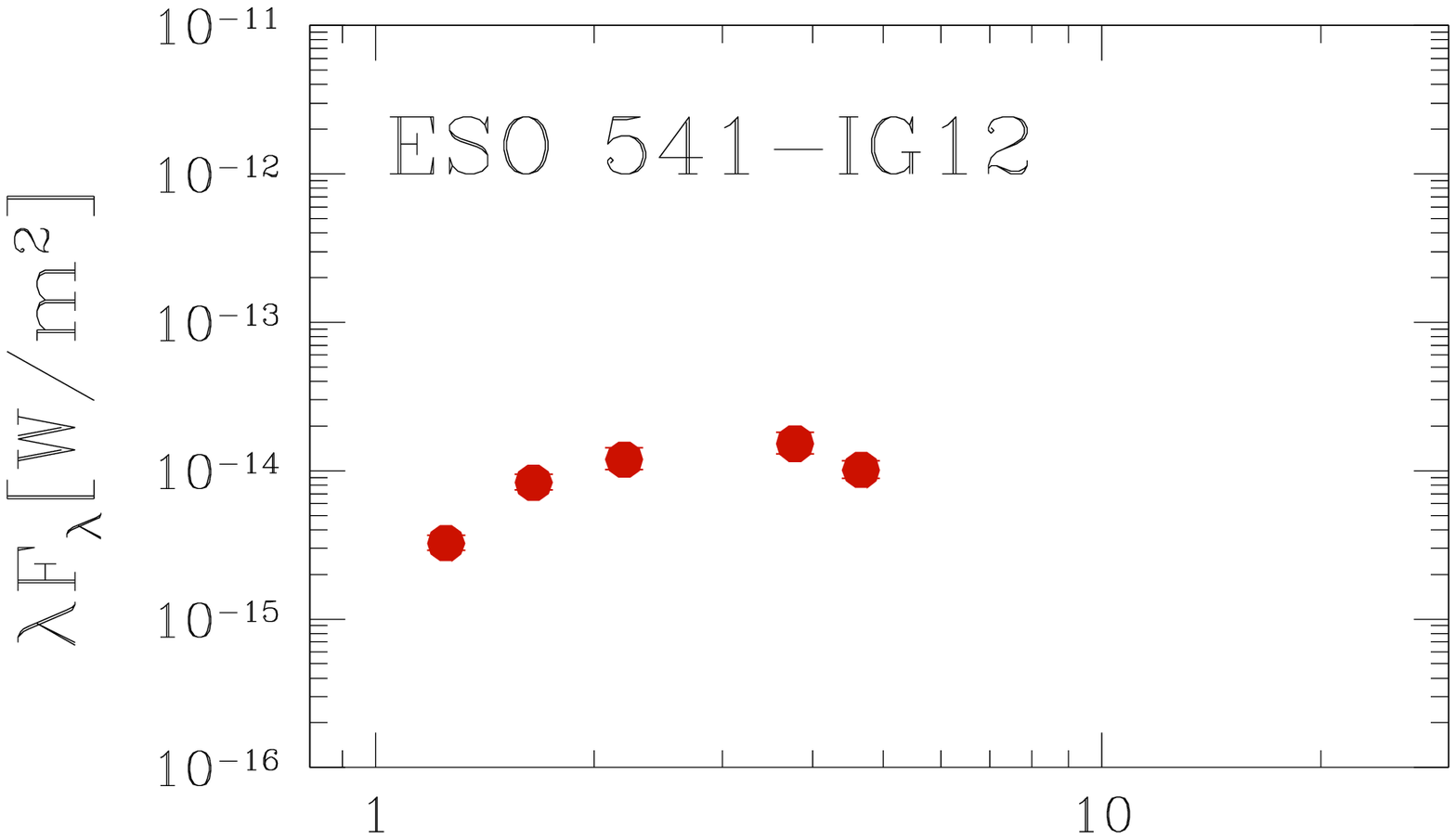}
    \includegraphics[scale=0.25,trim=100 100 0 100]{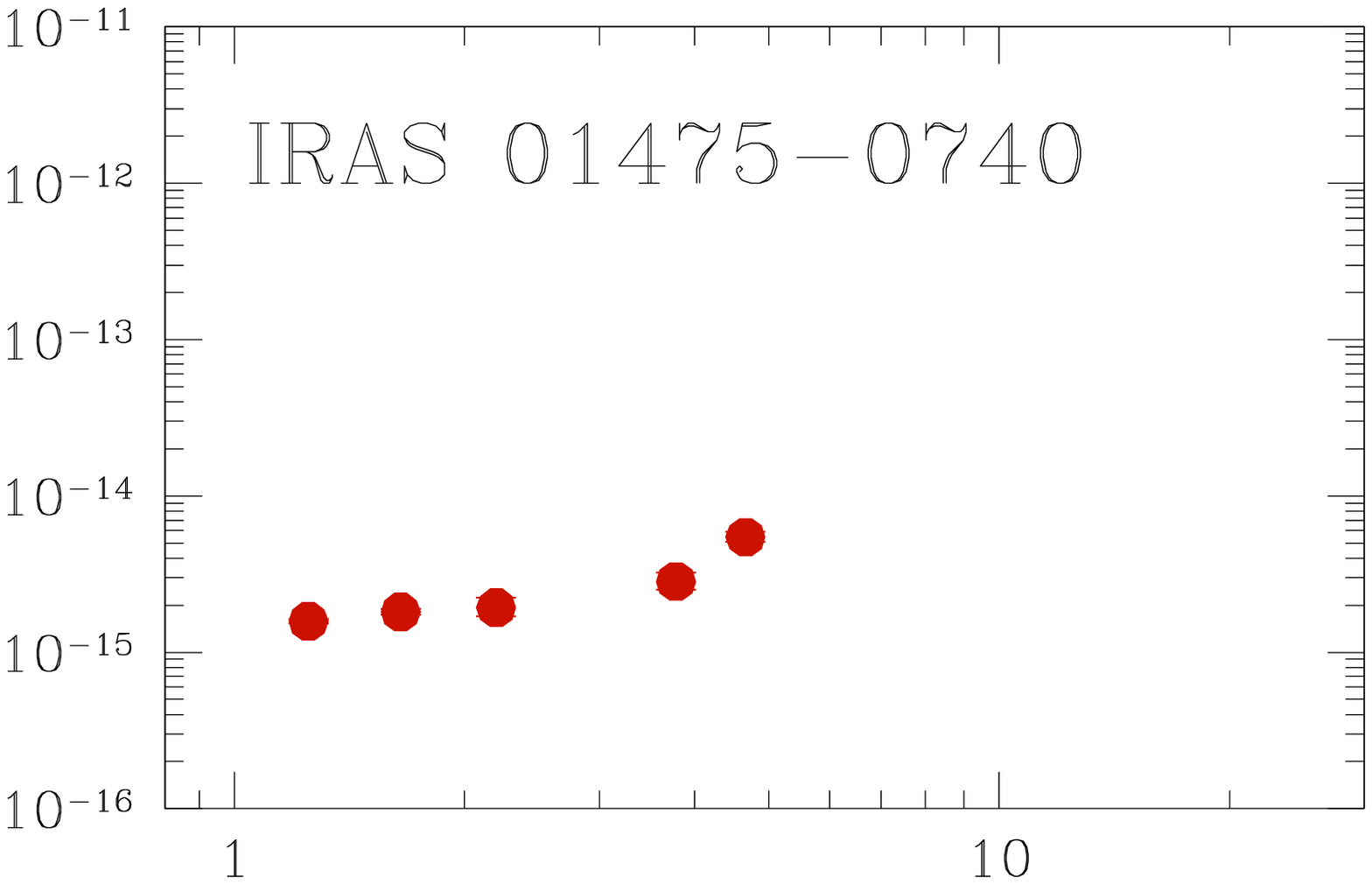}
    \includegraphics[scale=0.25,trim=100 100 0 100]{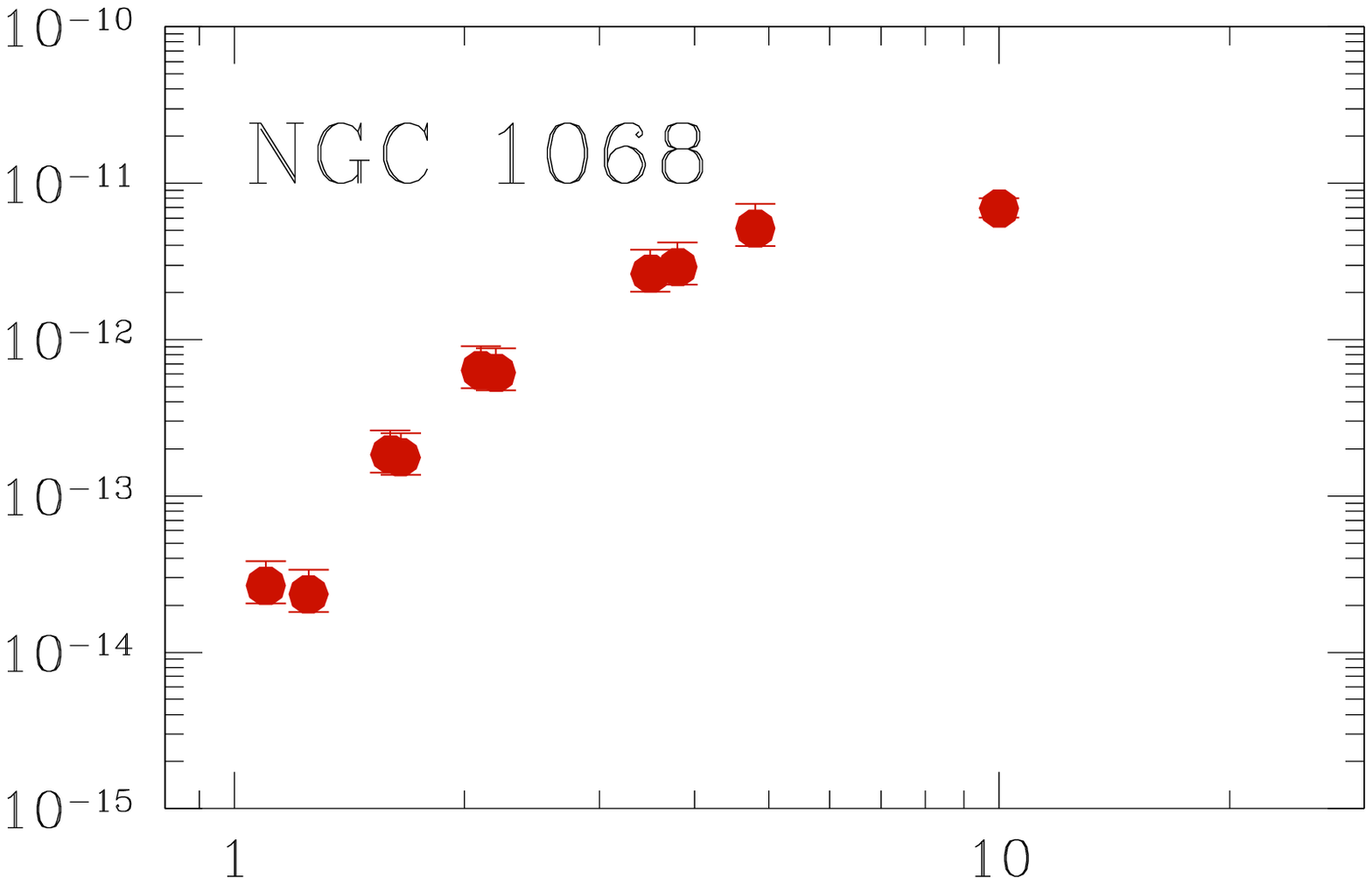}
    \includegraphics[scale=0.25,trim=100 100 0 100]{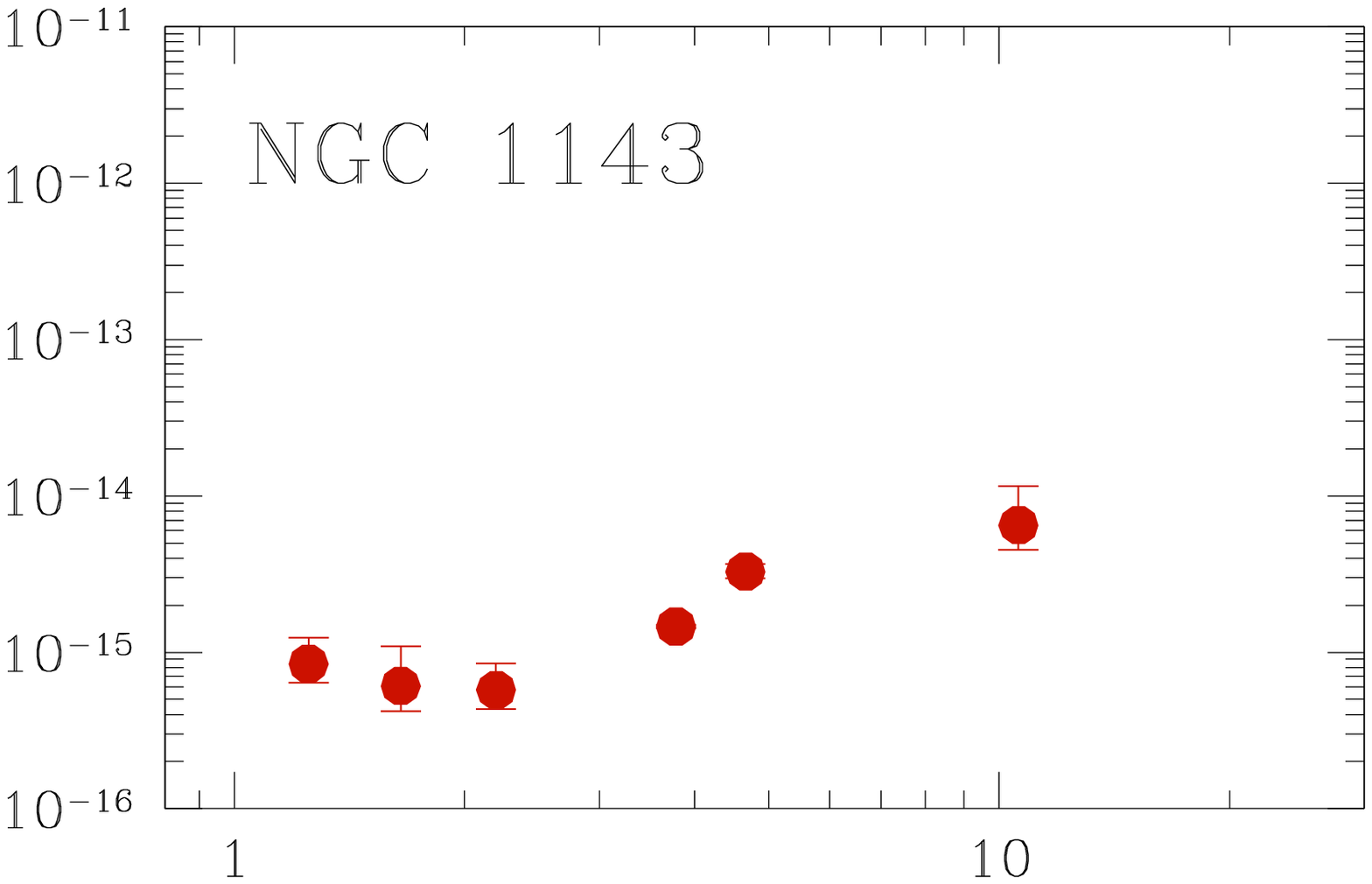}\\
    \includegraphics[scale=0.25,trim=0   100 0 100]{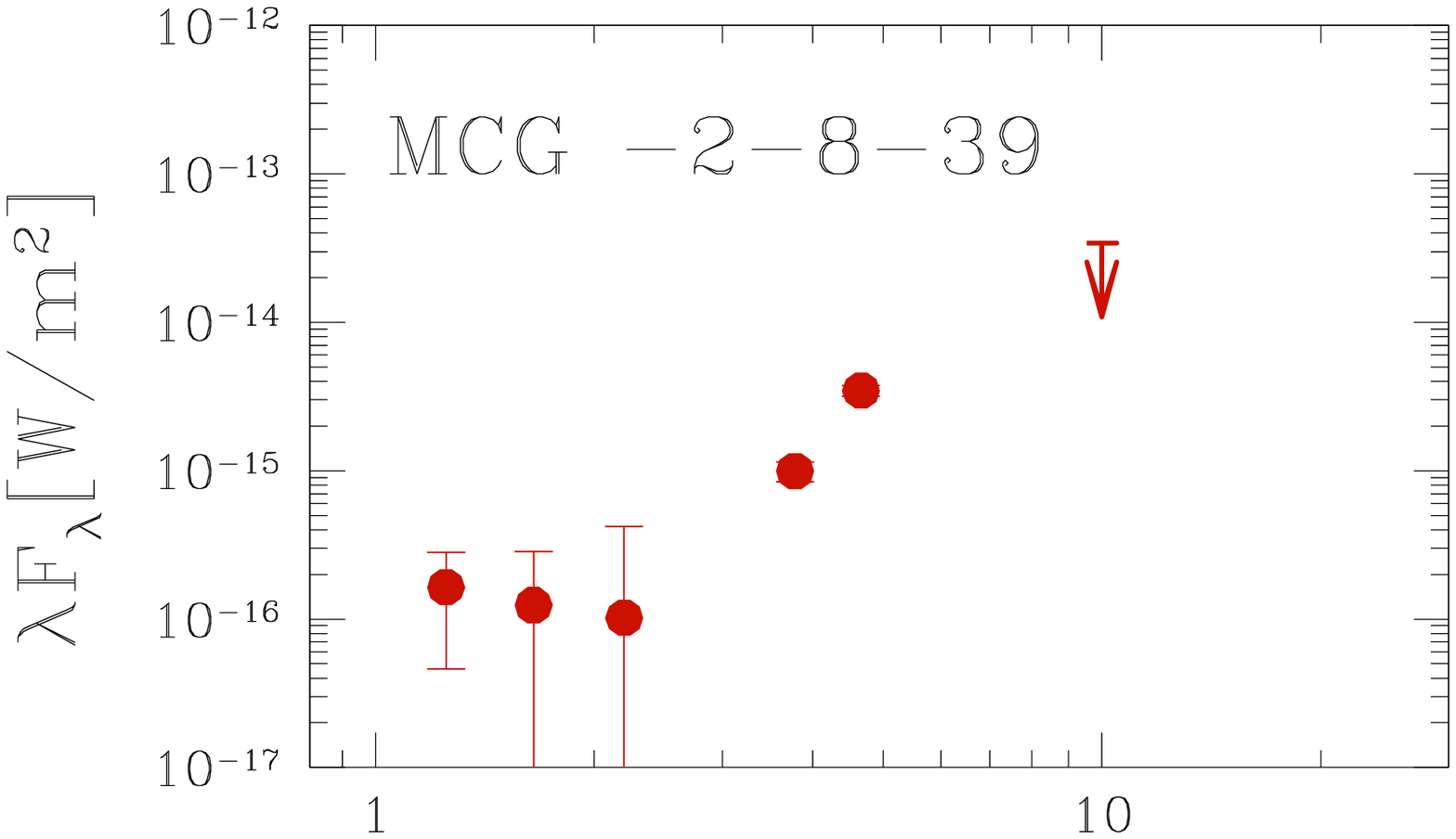}
    \includegraphics[scale=0.25,trim=100 100 0 100]{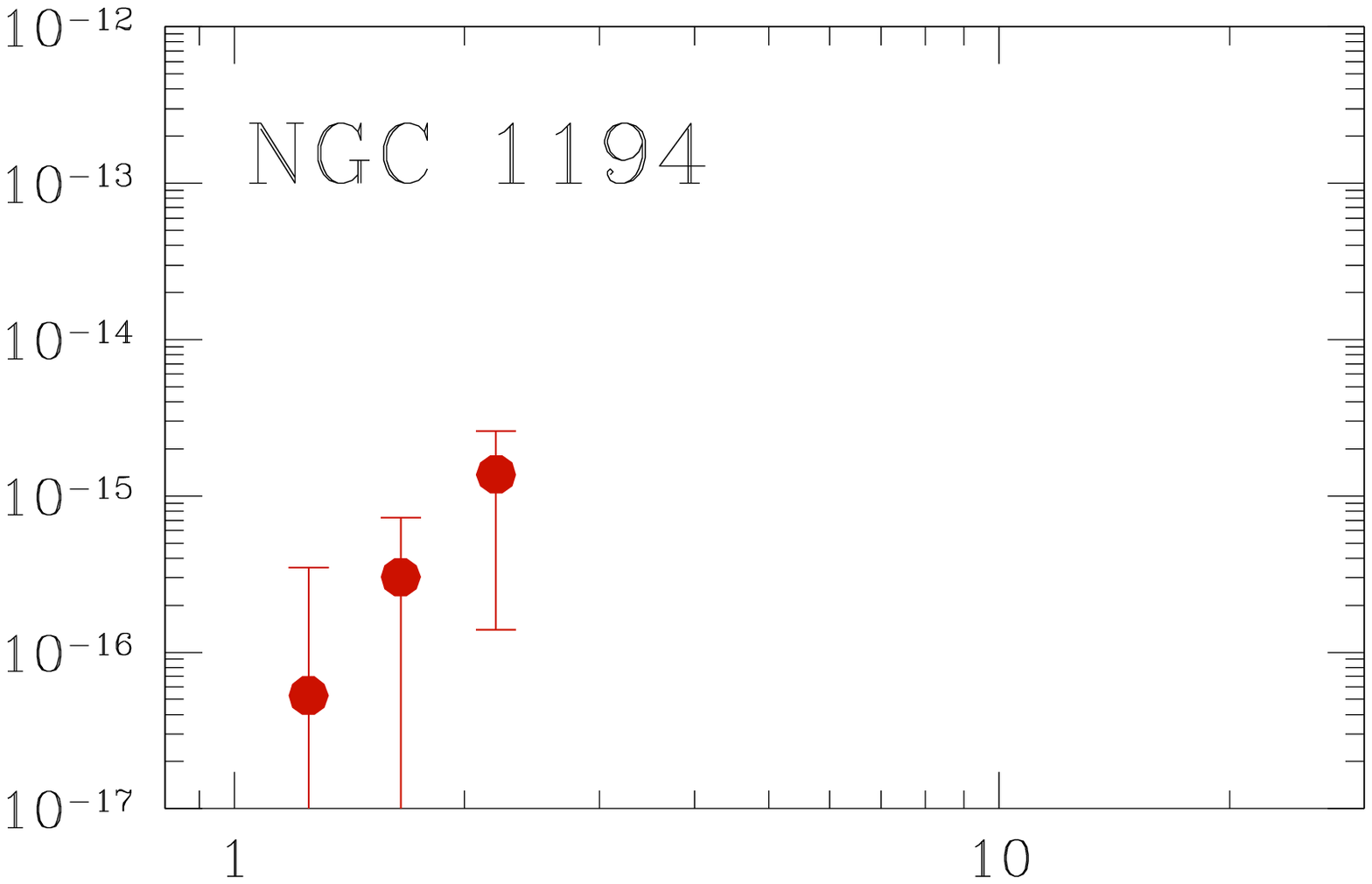}
    \includegraphics[scale=0.25,trim=100 100 0 100]{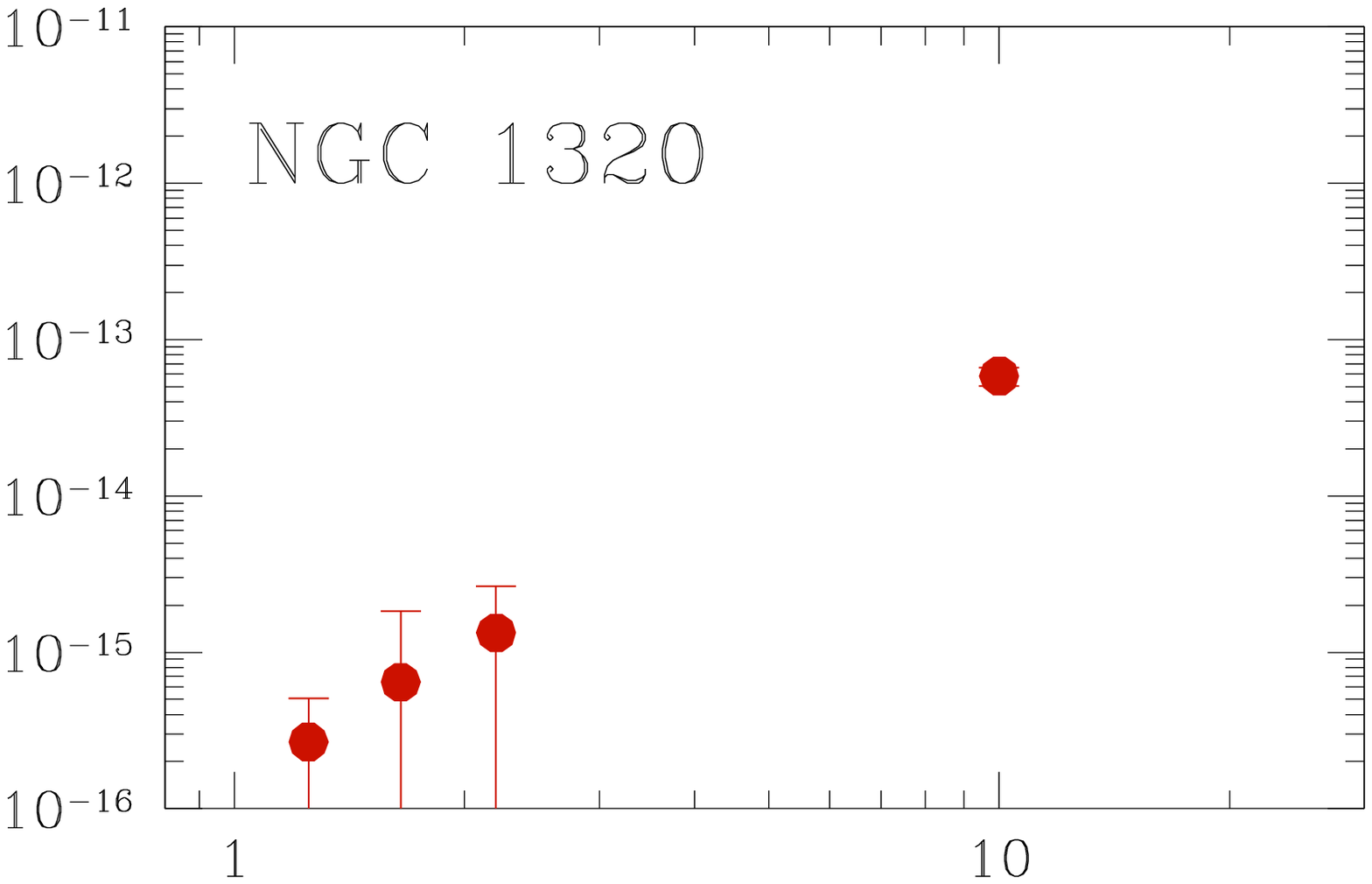}
    \includegraphics[scale=0.25,trim=100 100 0 100]{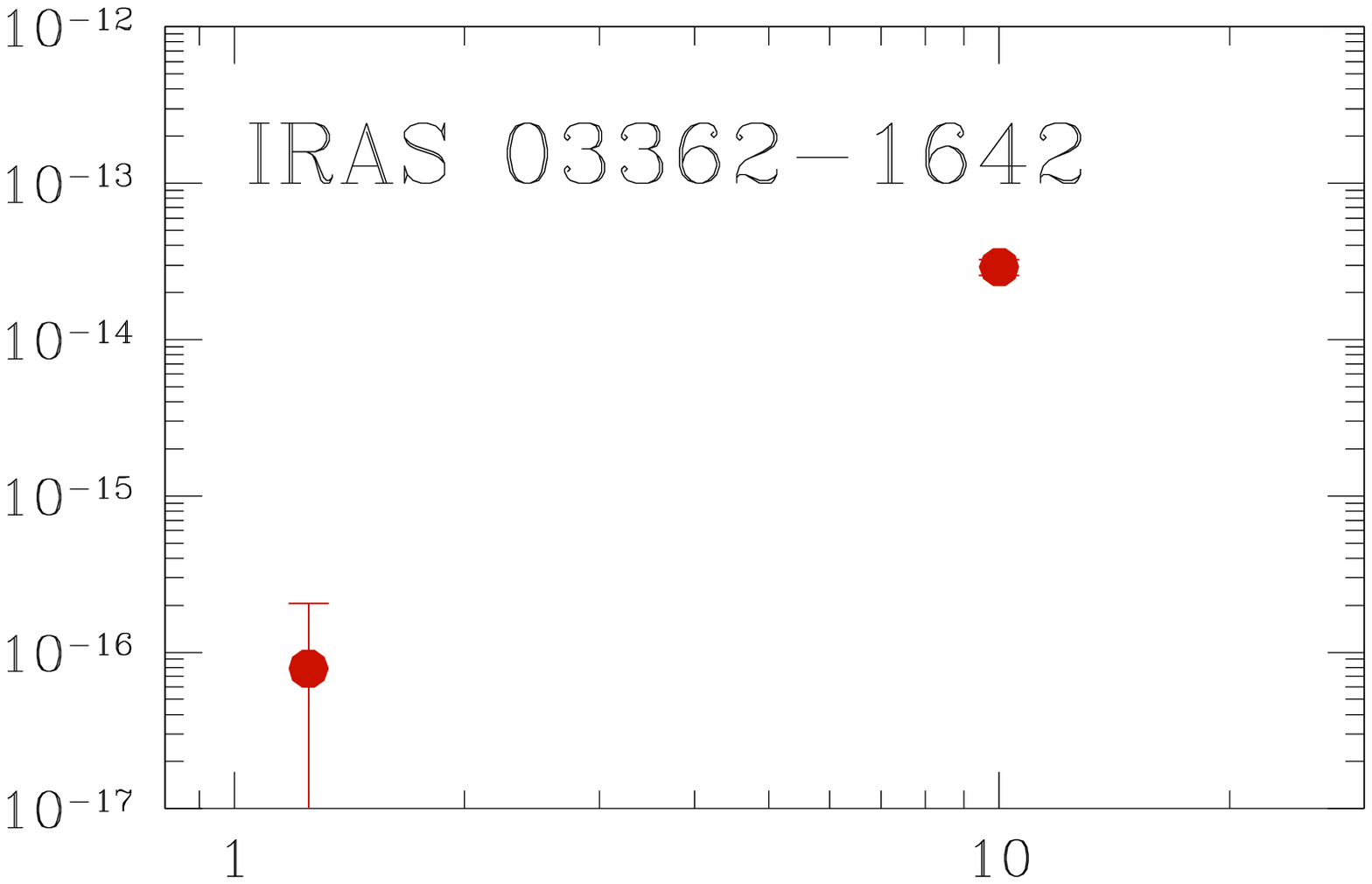}\\
    \includegraphics[scale=0.25,trim=0   100 0 100]{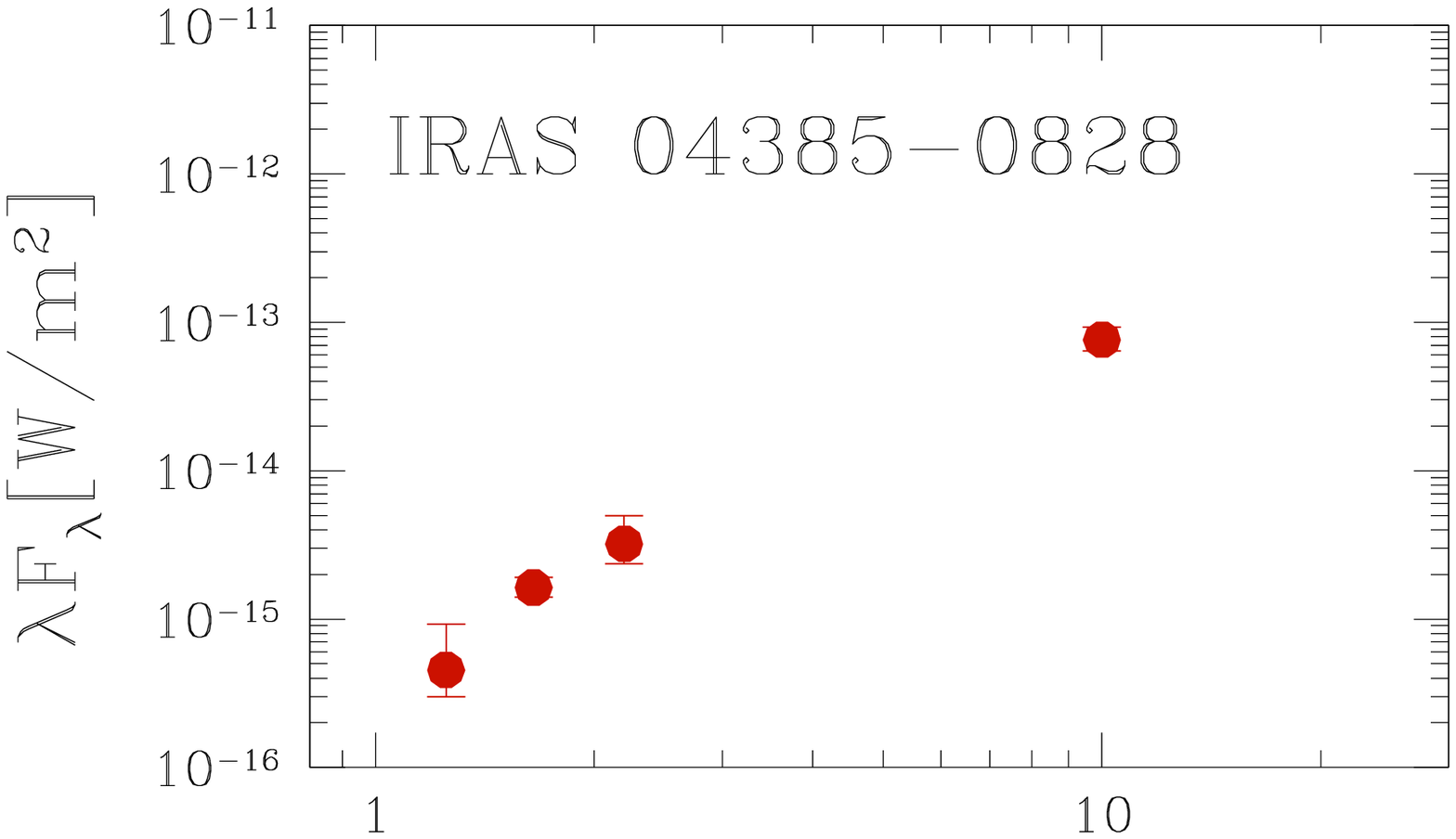}  
    \includegraphics[scale=0.25,trim=100 100 0 100]{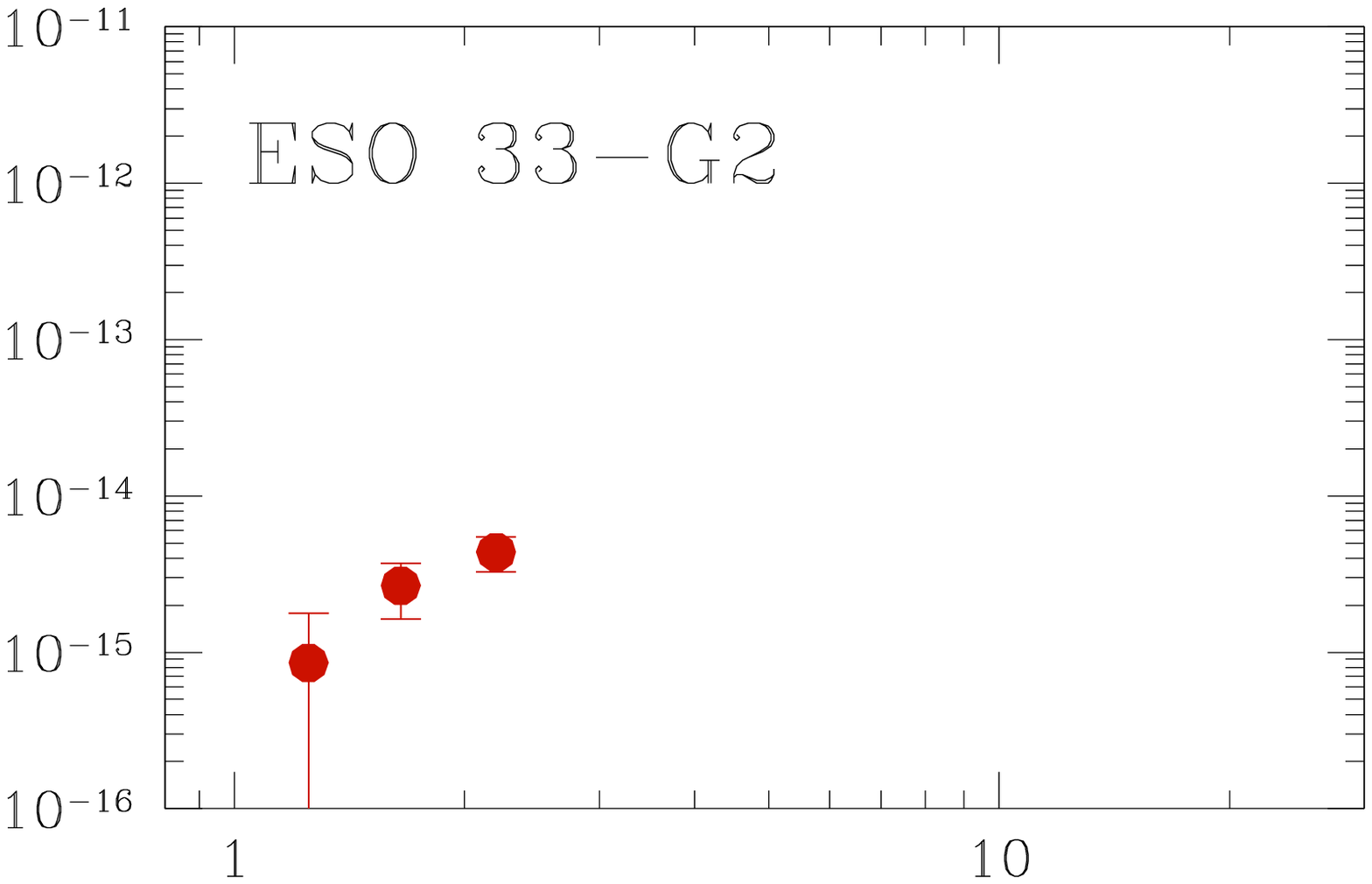}
    \includegraphics[scale=0.25,trim=100 100 0 100]{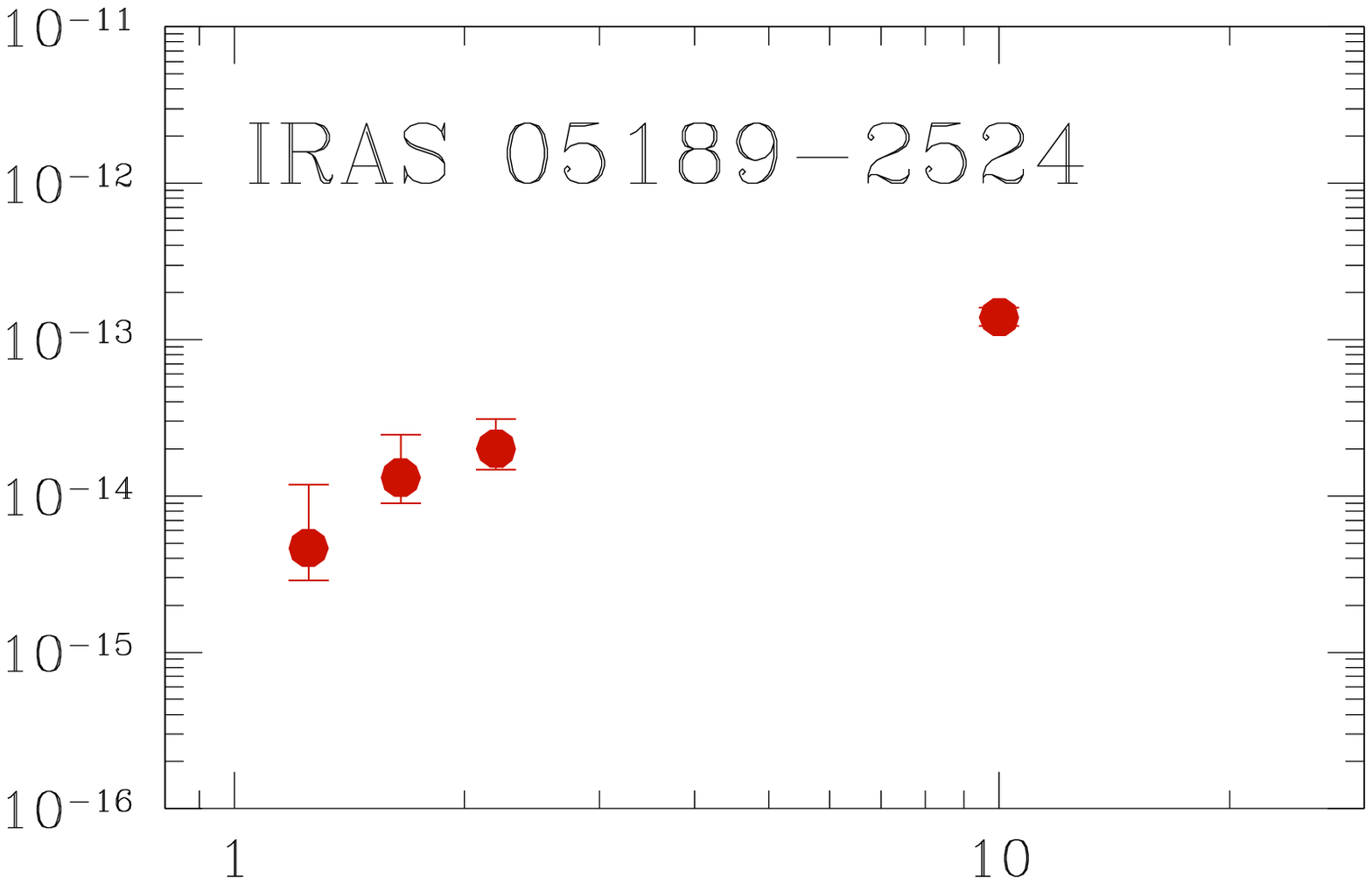}
    \includegraphics[scale=0.25,trim=100 100 0 100]{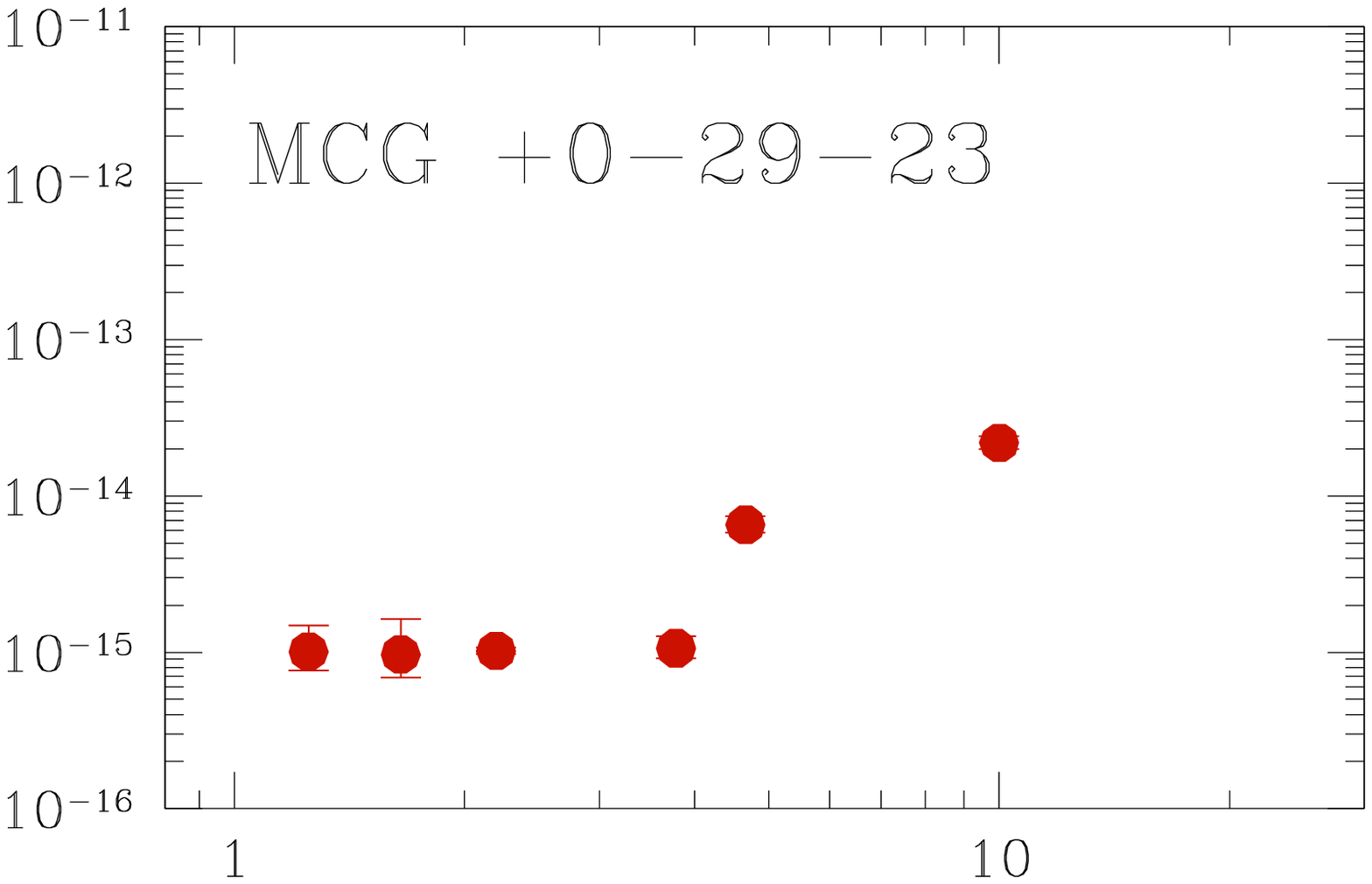}\\
    \includegraphics[scale=0.25,trim=0    30 0 100]{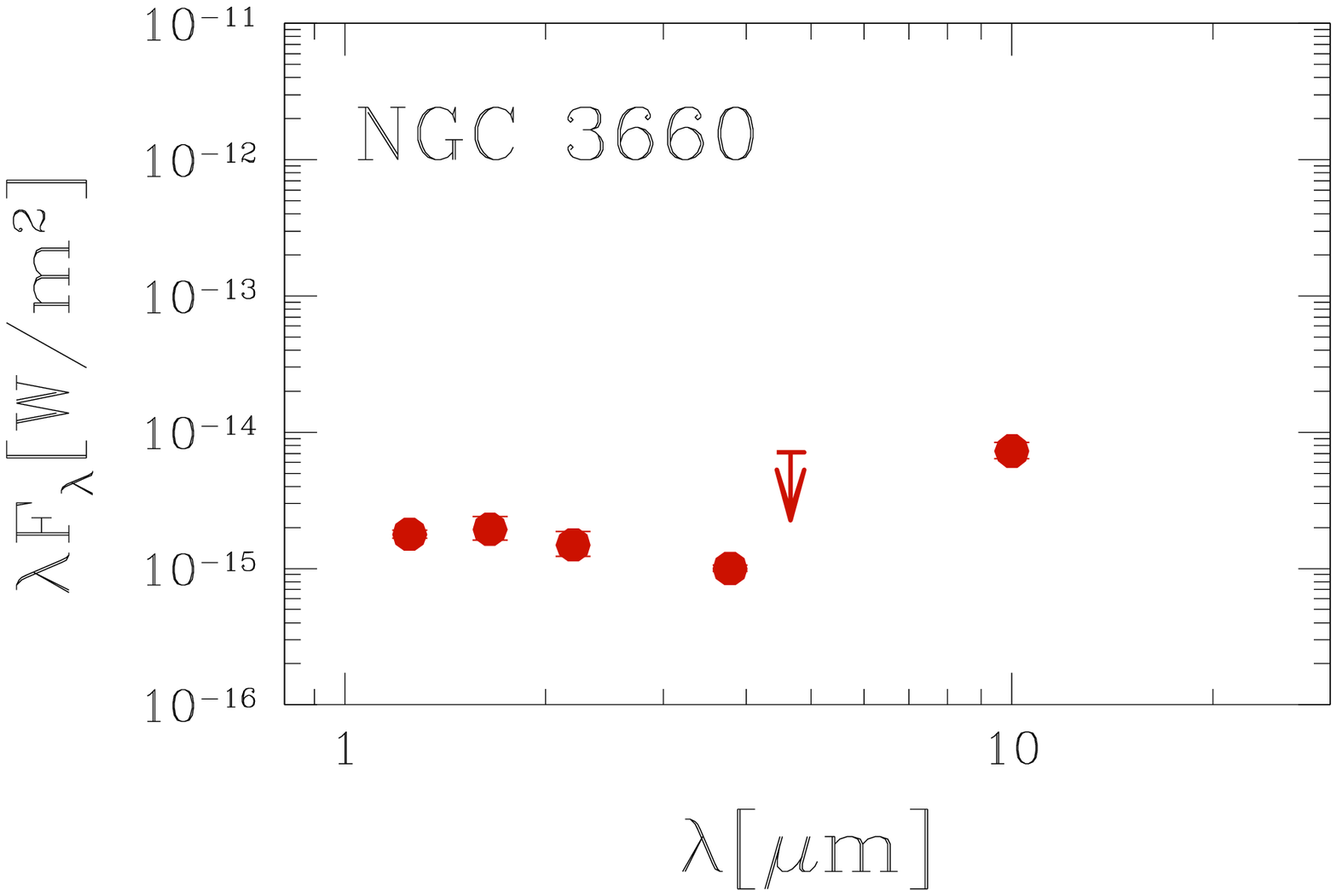}
    \includegraphics[scale=0.25,trim=100  30 0 100]{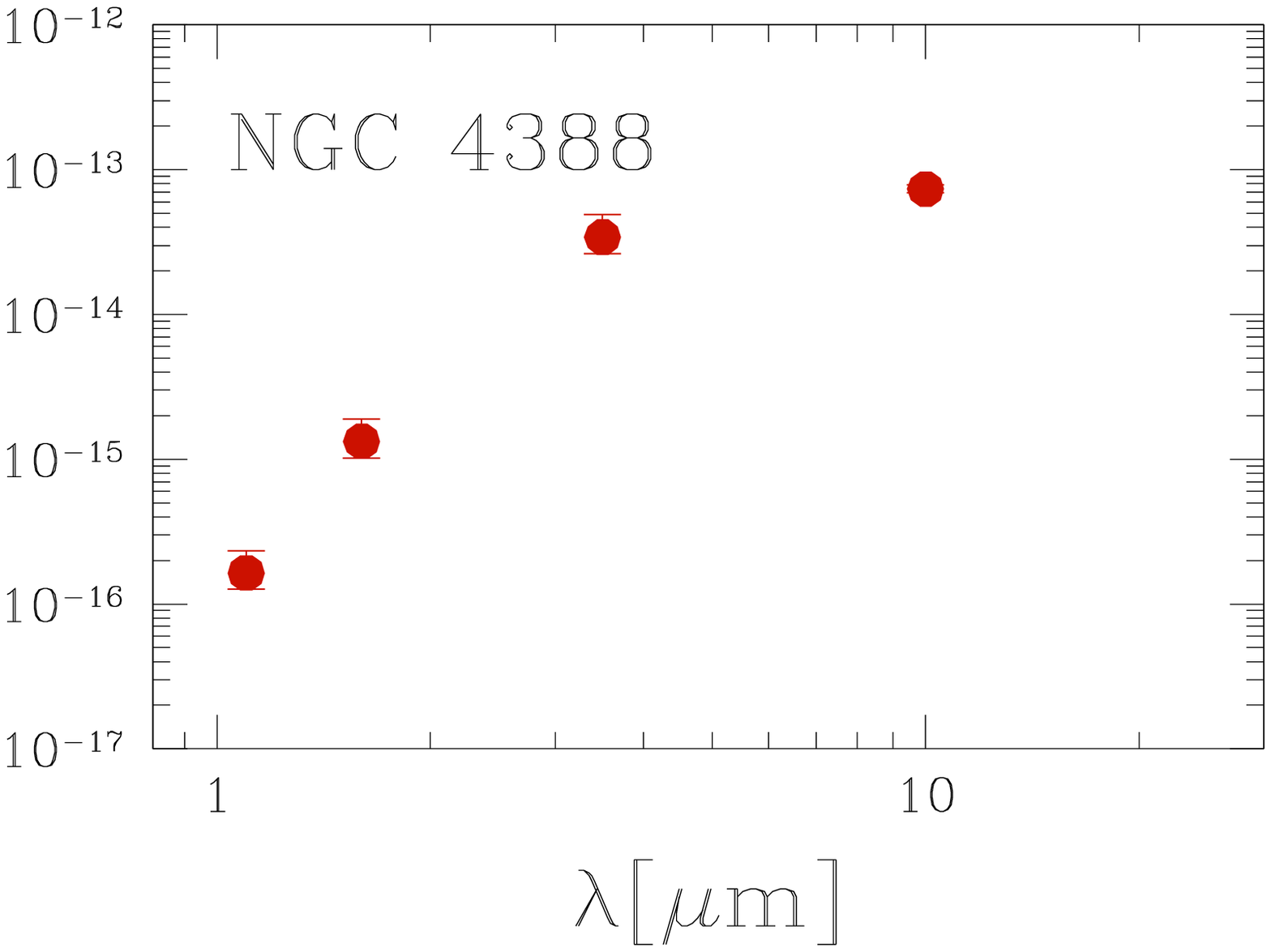}
    \includegraphics[scale=0.25,trim=100  30 0 100]{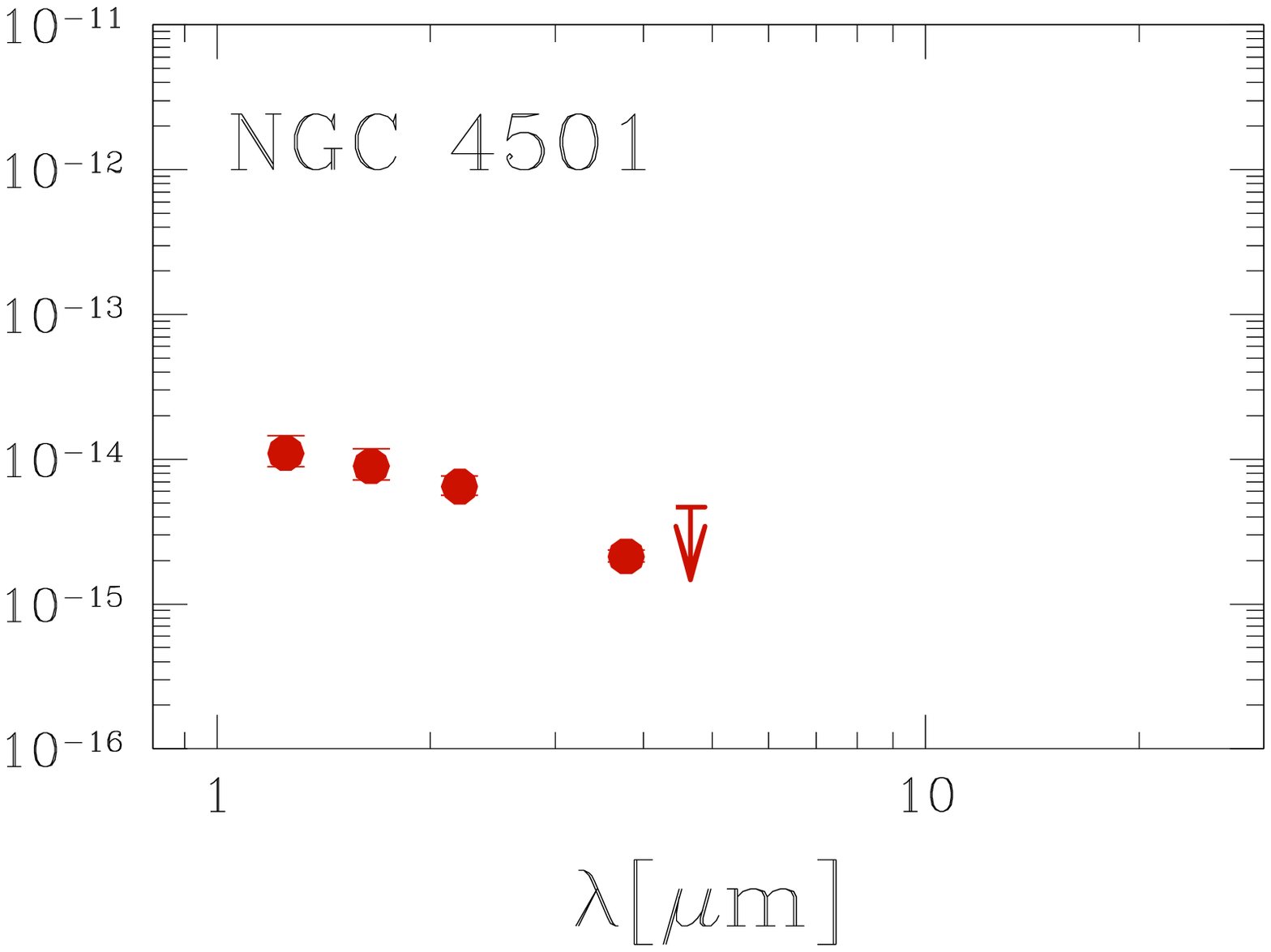}
    \includegraphics[scale=0.25,trim=100  30 0 100]{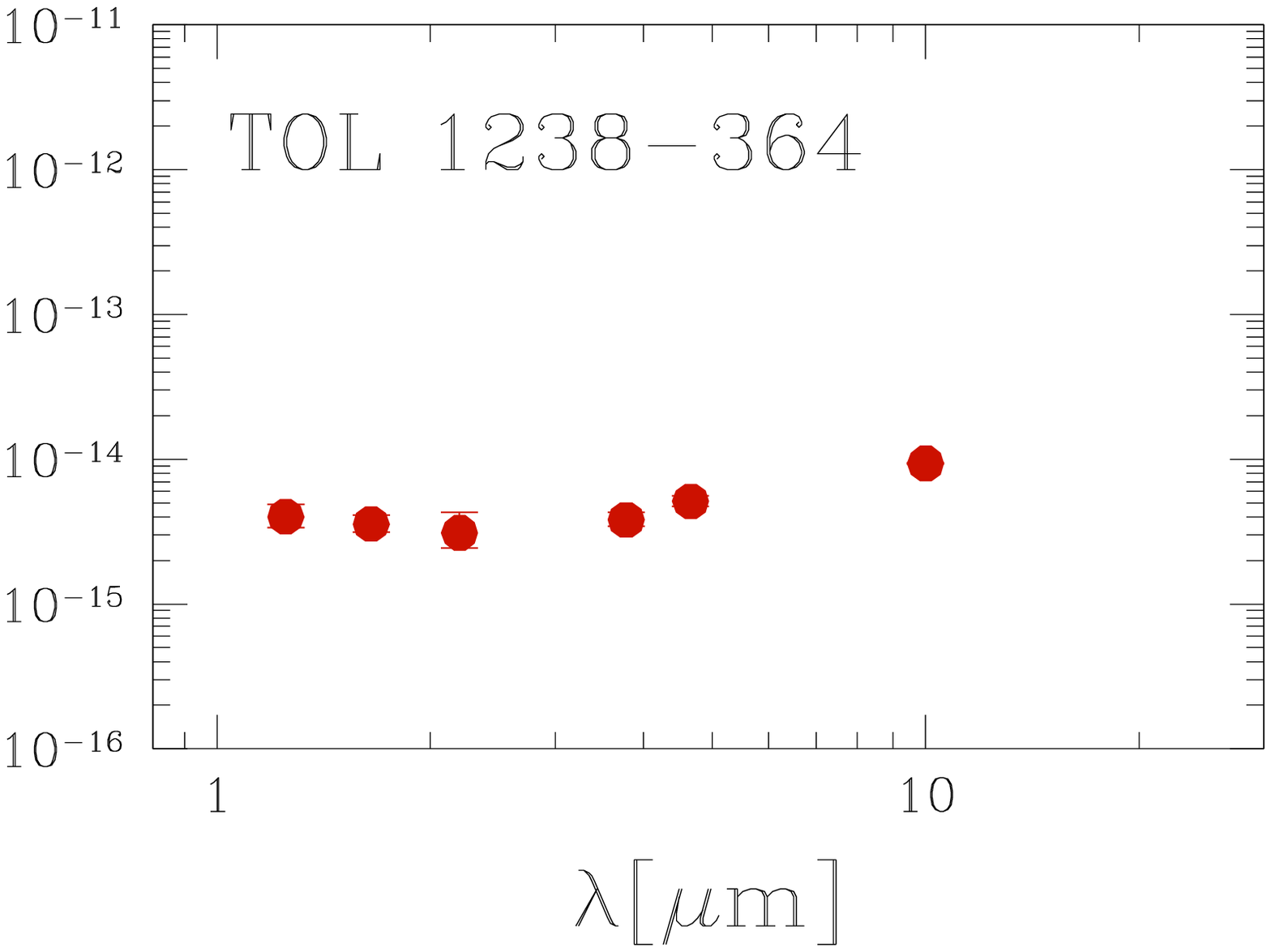}

  \end{center}
  \caption[SEDs]{Spectral Energy Distributions determined for our sample. Note
    the different scale in Y axis in galaxies NGC\,1068, MCG~-2-8-39,
    NGC\,1194, IRAS\,03362-1642, NGC\,5506, and NGC\,7590. 6 SEDs are from
    \citet{ah03}. \label{seds}}
\end{figure*}

\begin{figure*}
  \begin{center}
    \includegraphics[scale=0.25,trim=0   100 0 100]{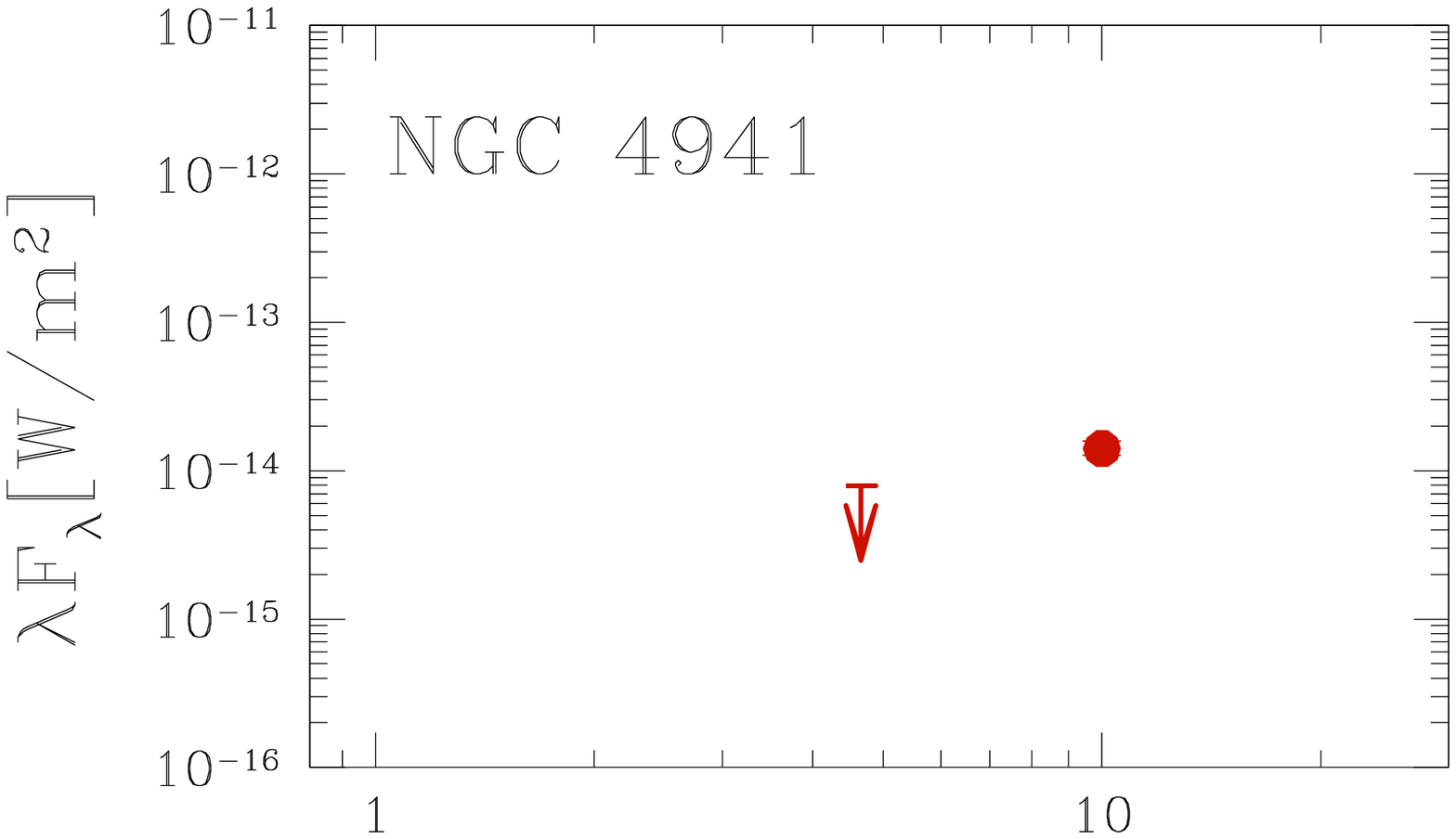}
    \includegraphics[scale=0.25,trim=100 100 0 100]{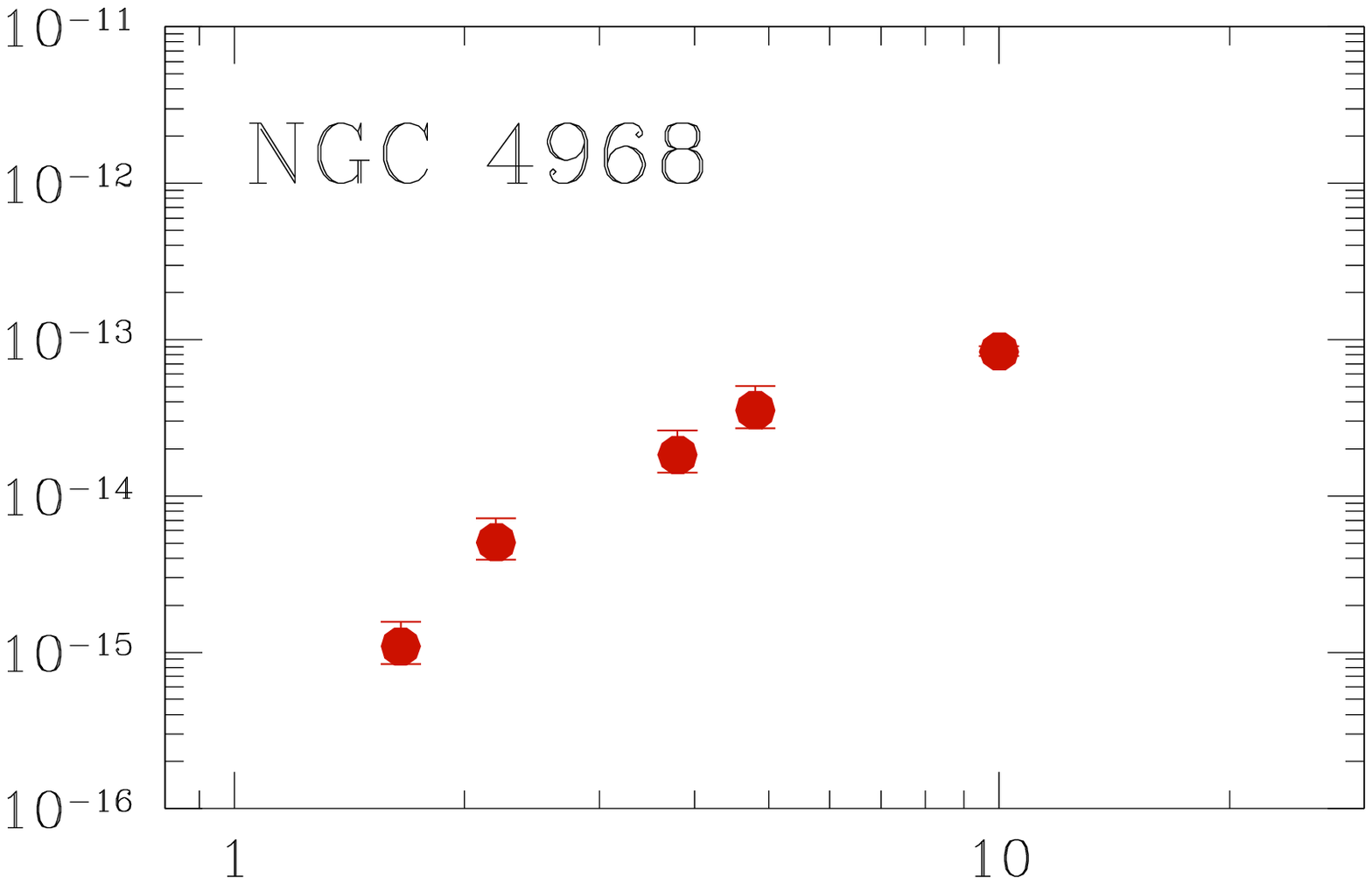}
    \includegraphics[scale=0.25,trim=100 100 0 100]{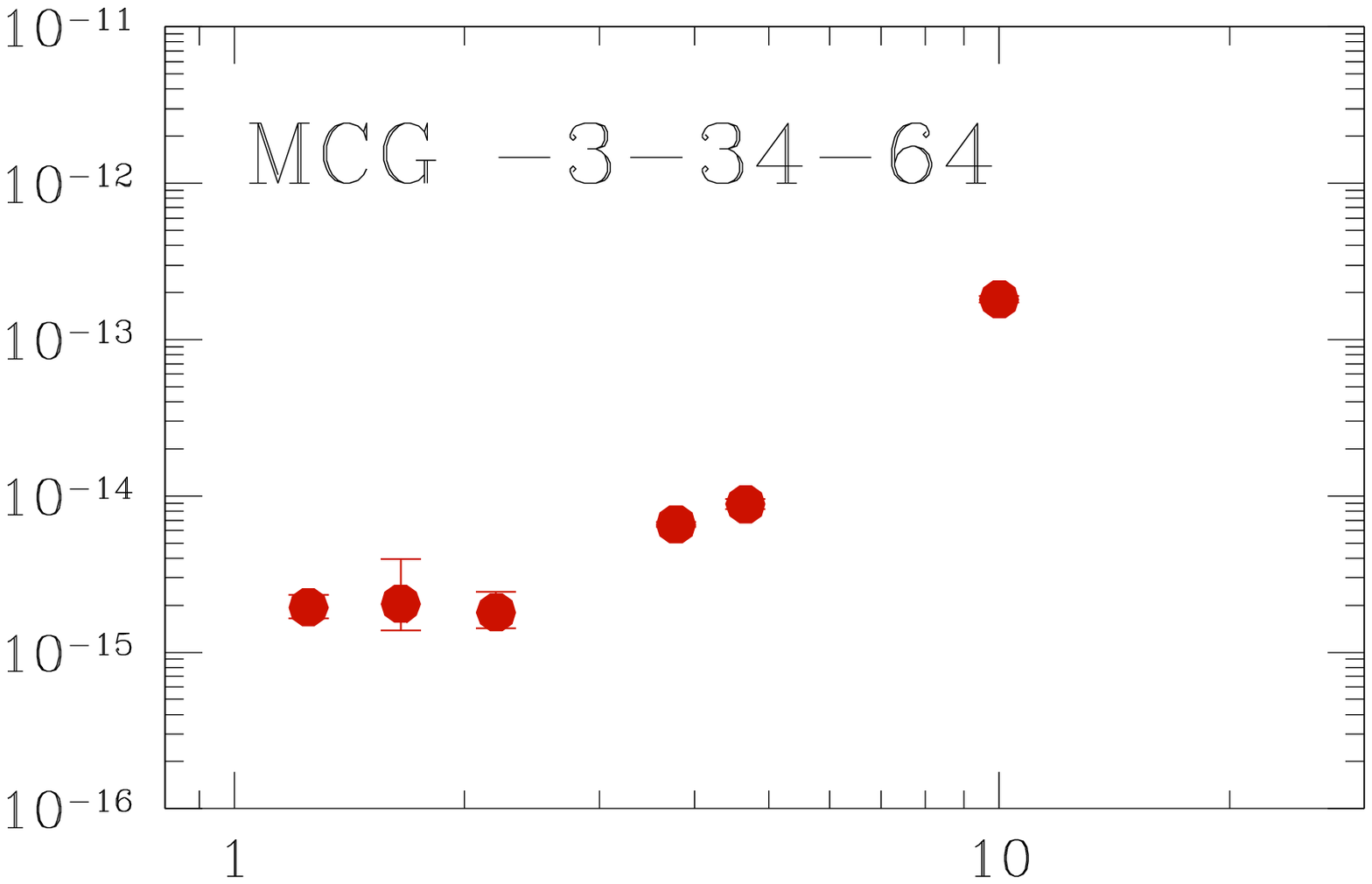}
    \includegraphics[scale=0.25,trim=100 100 0 100]{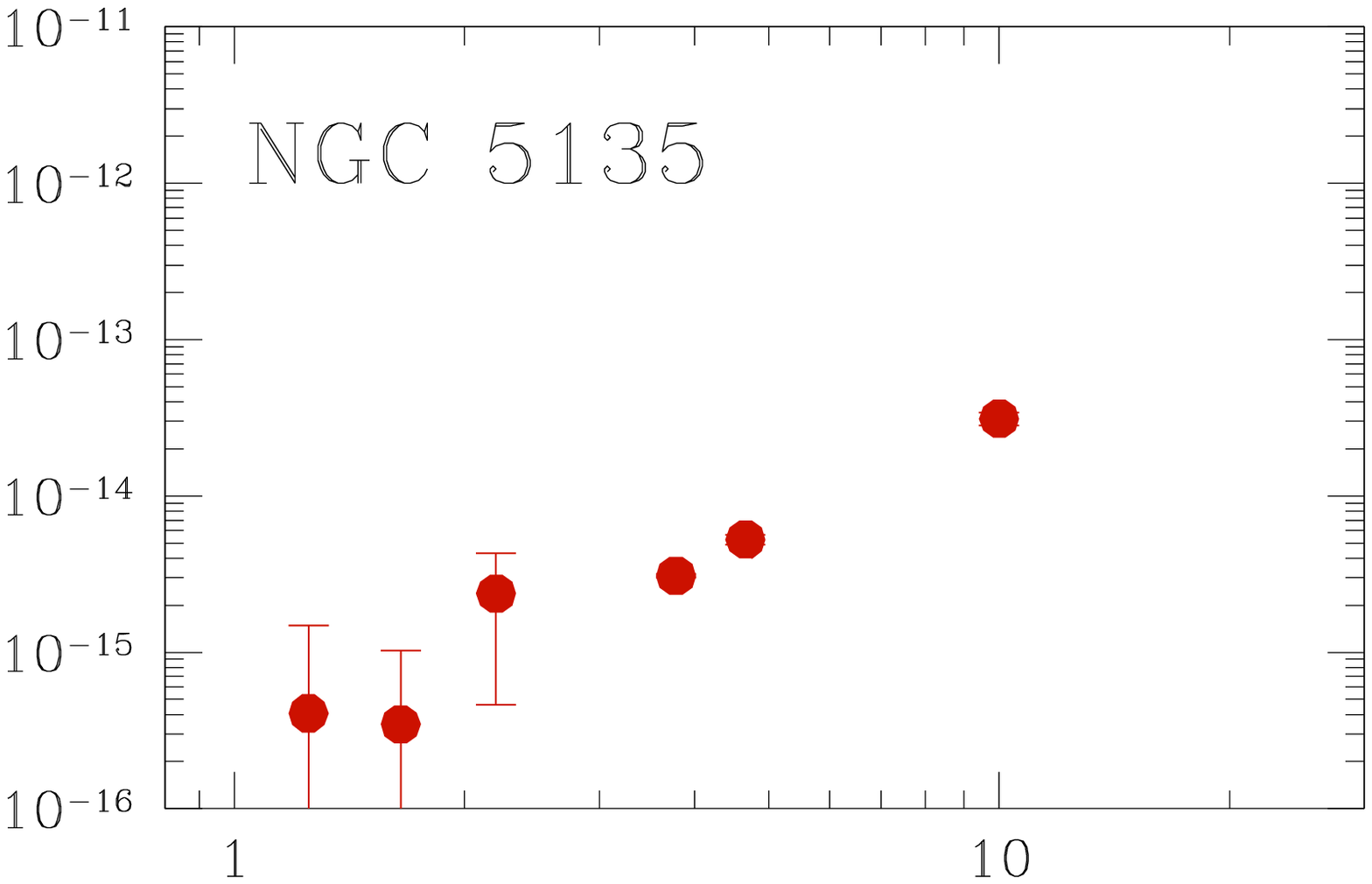}\\
    \includegraphics[scale=0.25,trim=0   100 0 100]{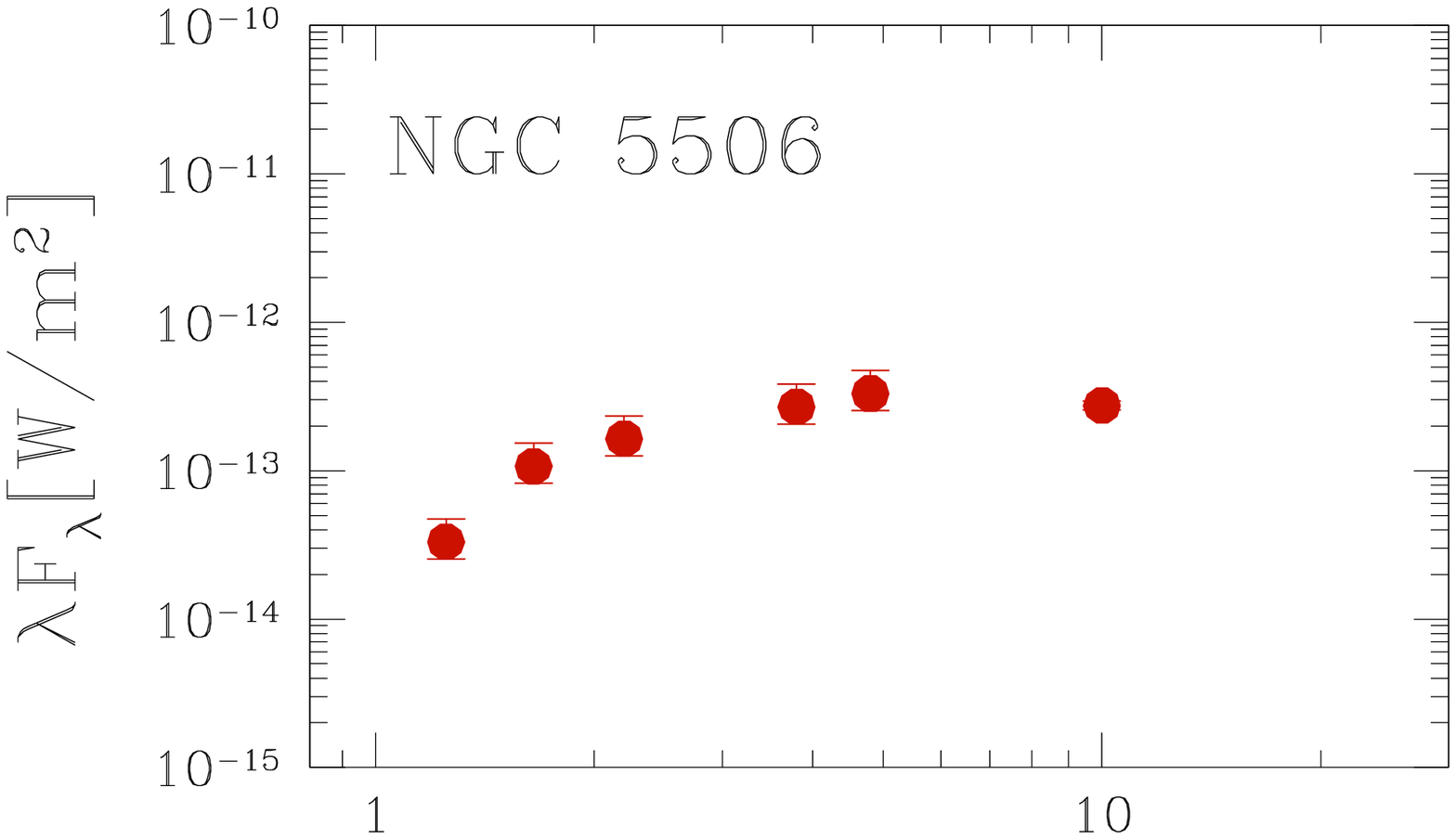}
    \includegraphics[scale=0.25,trim=100 100 0 100]{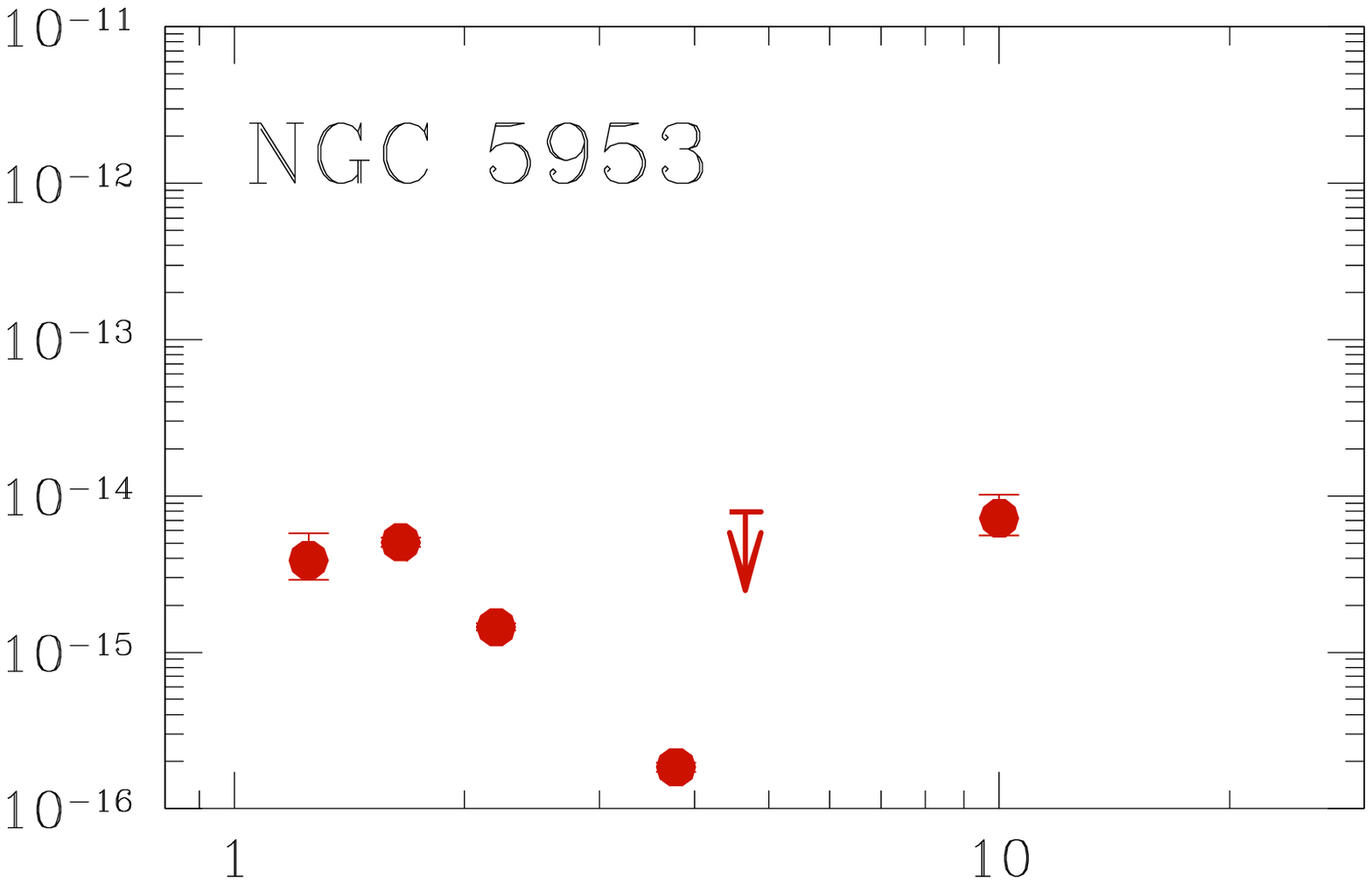}
    \includegraphics[scale=0.25,trim=100 100 0 100]{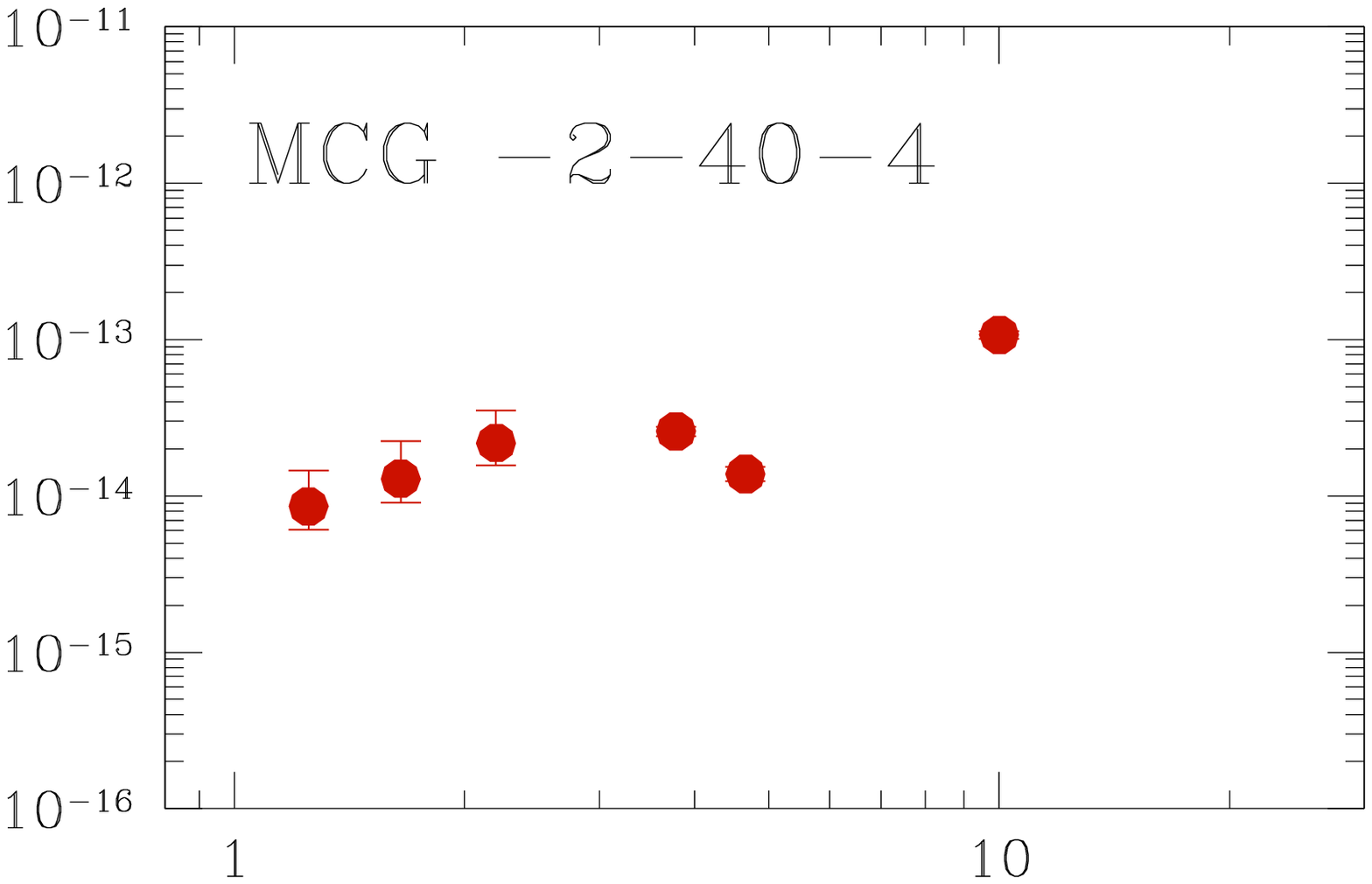}
    \includegraphics[scale=0.25,trim=100 100 0 100]{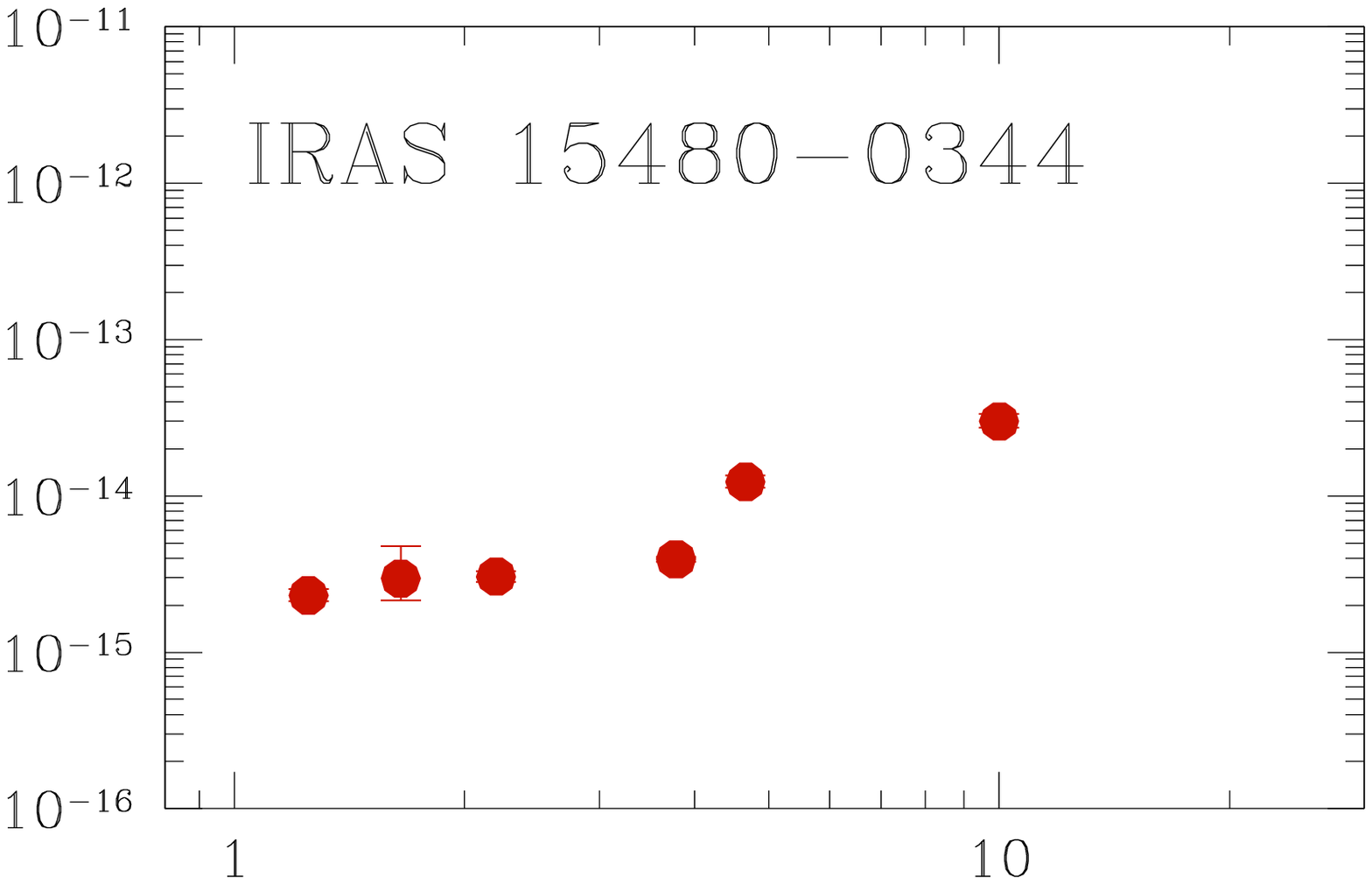}\\
    \includegraphics[scale=0.25,trim=0   100 0 100]{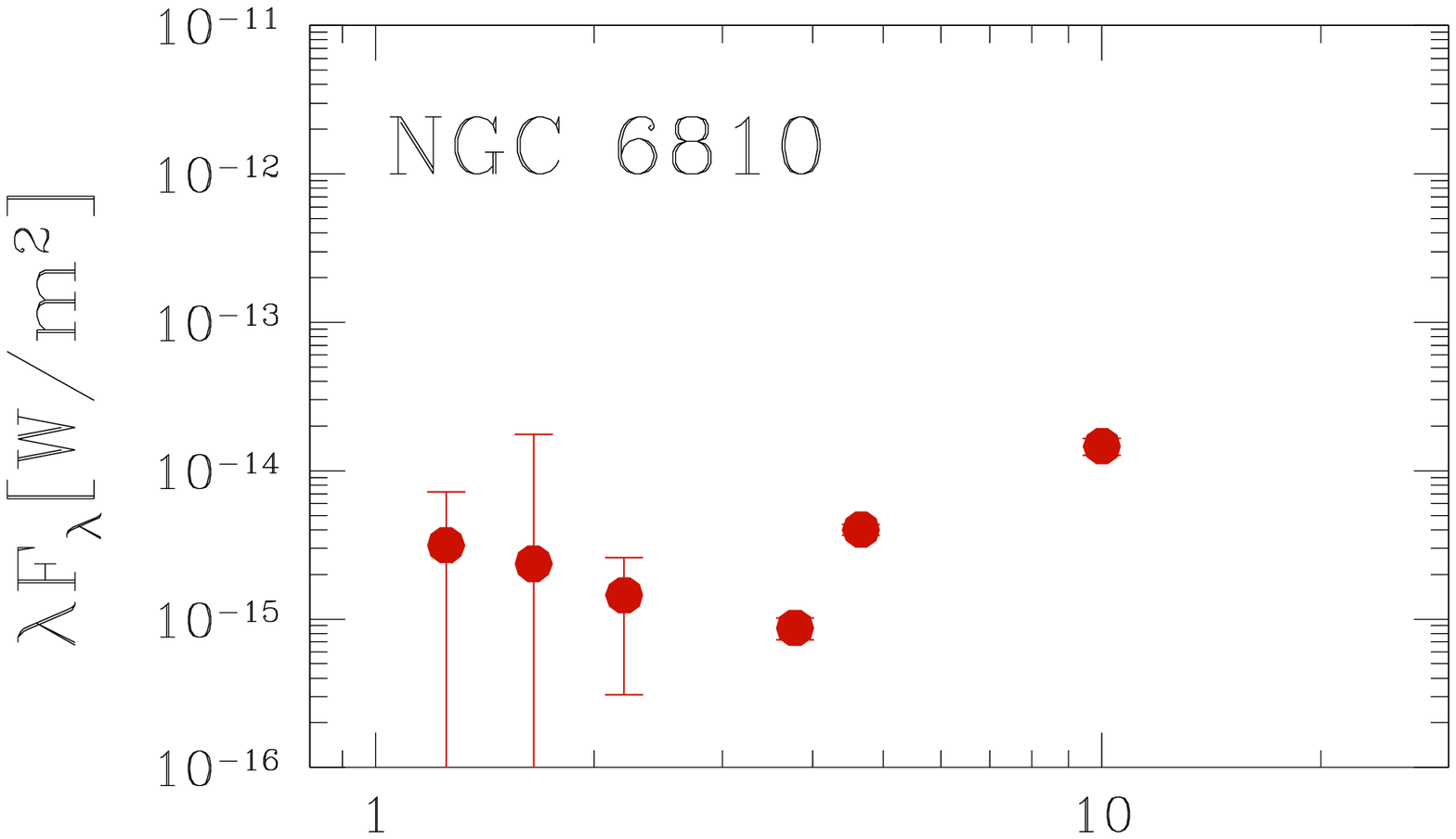}
    \includegraphics[scale=0.25,trim=100 100 0 100]{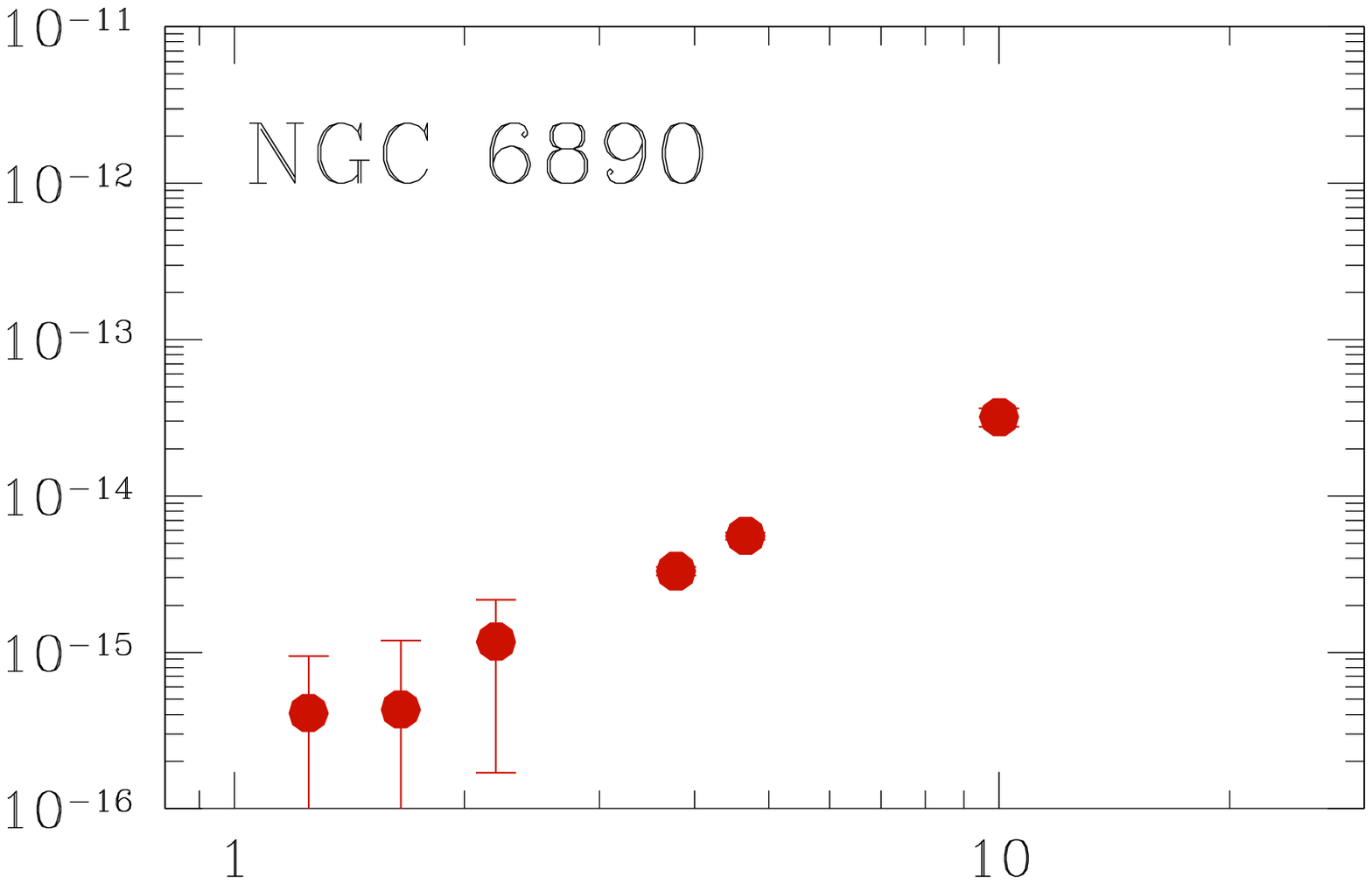}
    \includegraphics[scale=0.25,trim=100 100 0 100]{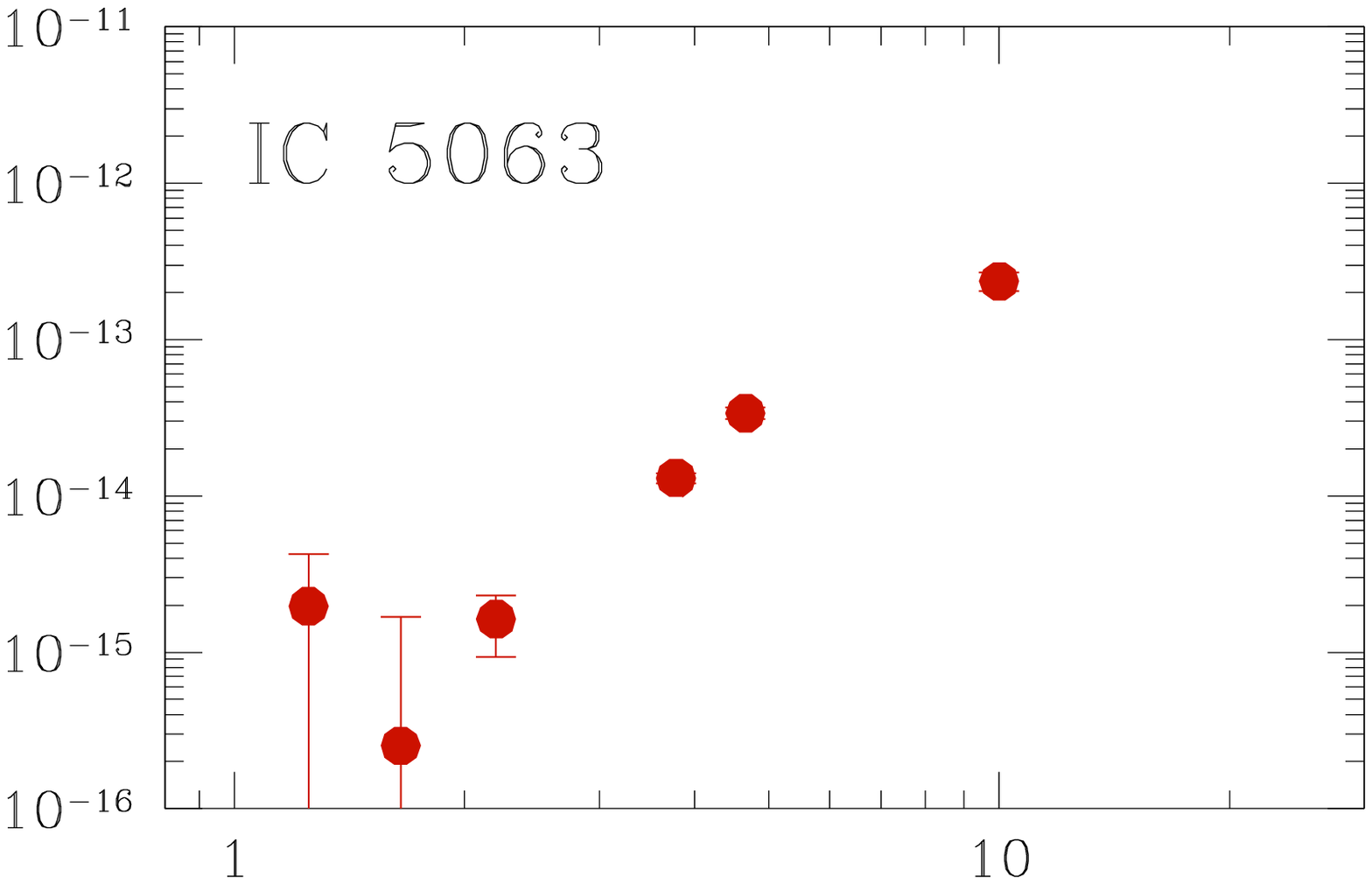}
    \includegraphics[scale=0.25,trim=100 100 0 100]{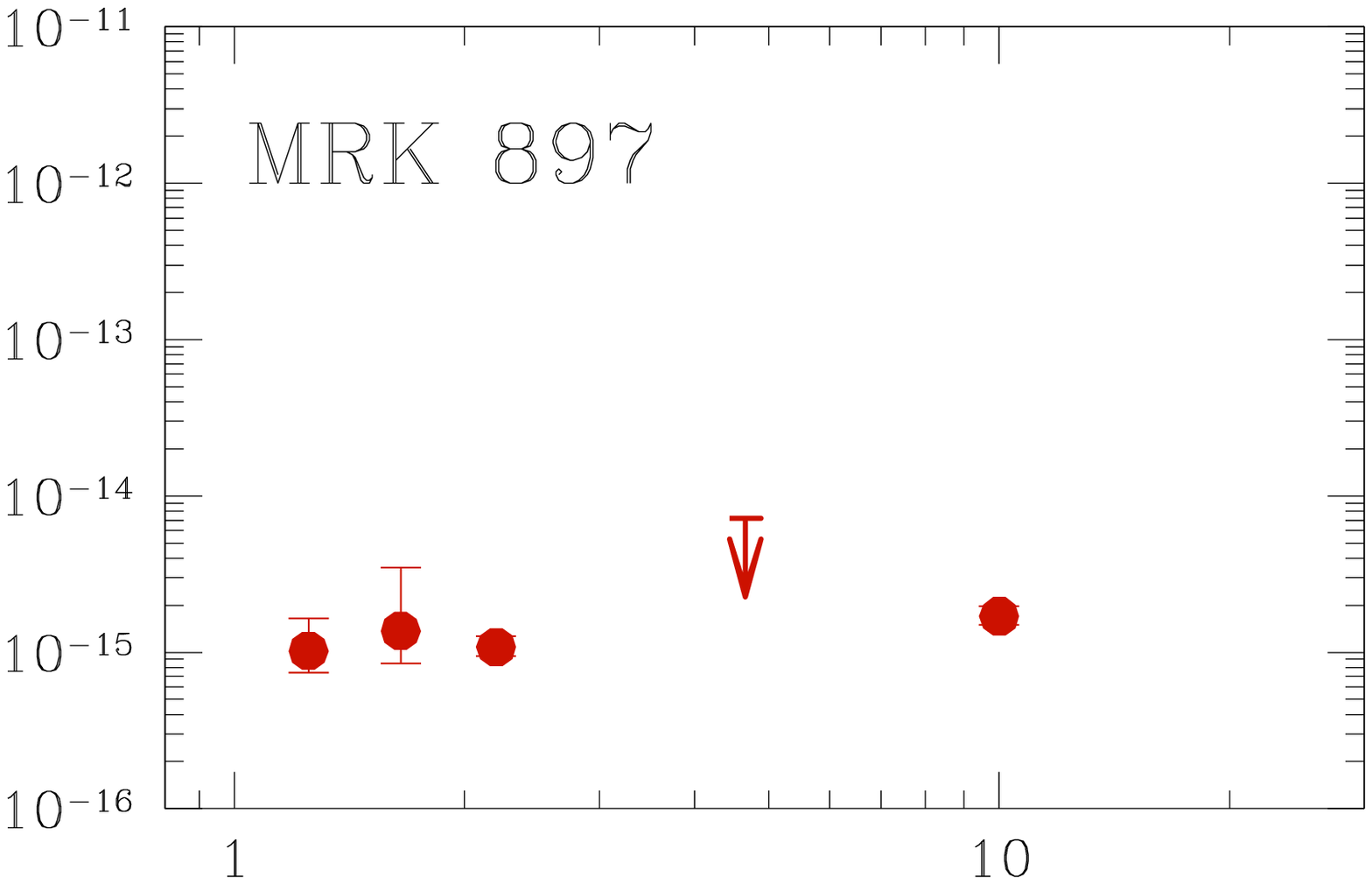}\\
    \includegraphics[scale=0.25,trim=0   100 0 100]{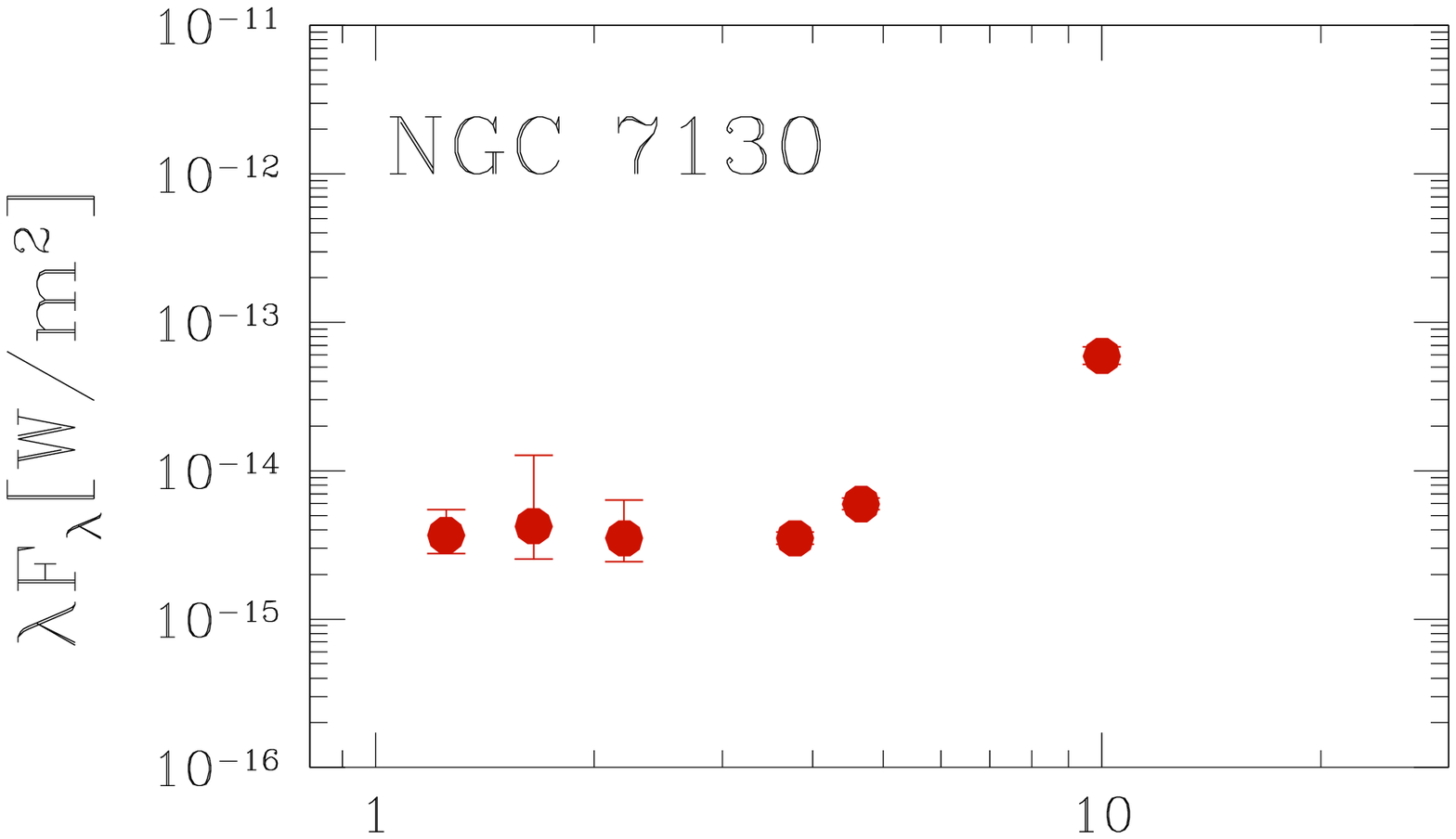}
    \includegraphics[scale=0.25,trim=100 100 0 100]{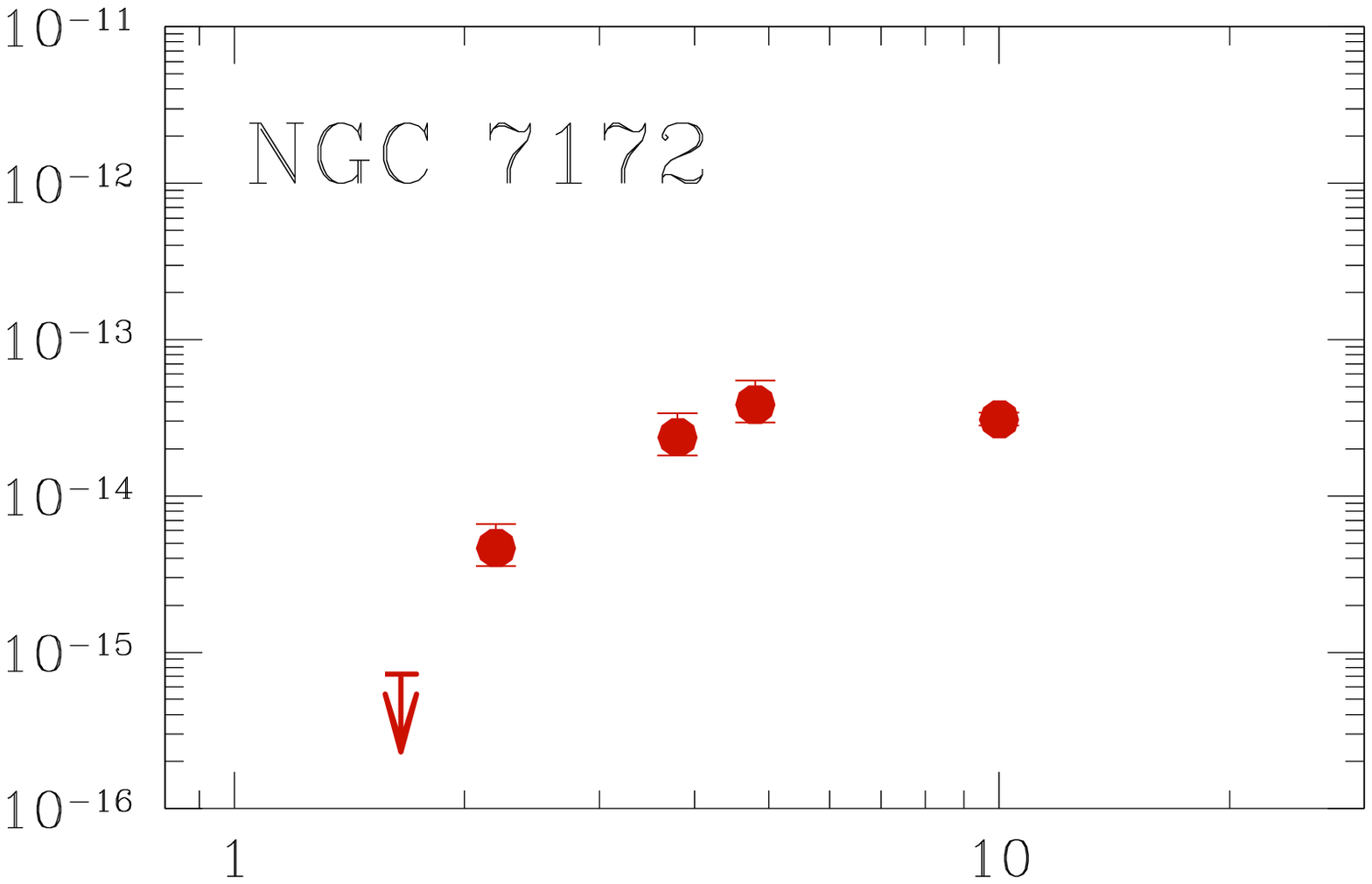}
    \includegraphics[scale=0.25,trim=100 100 0 100]{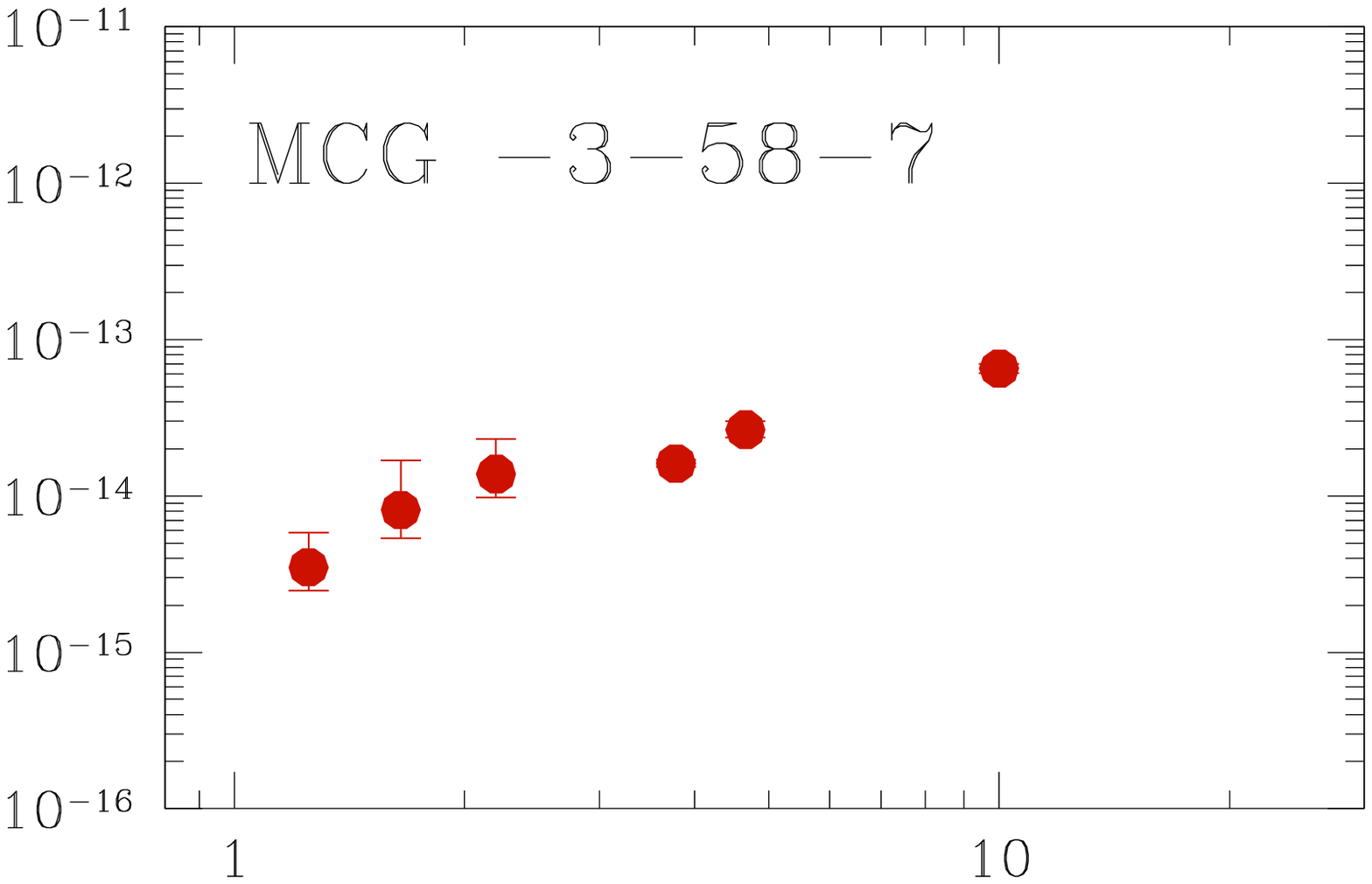}
    \includegraphics[scale=0.25,trim=100 100 0 100]{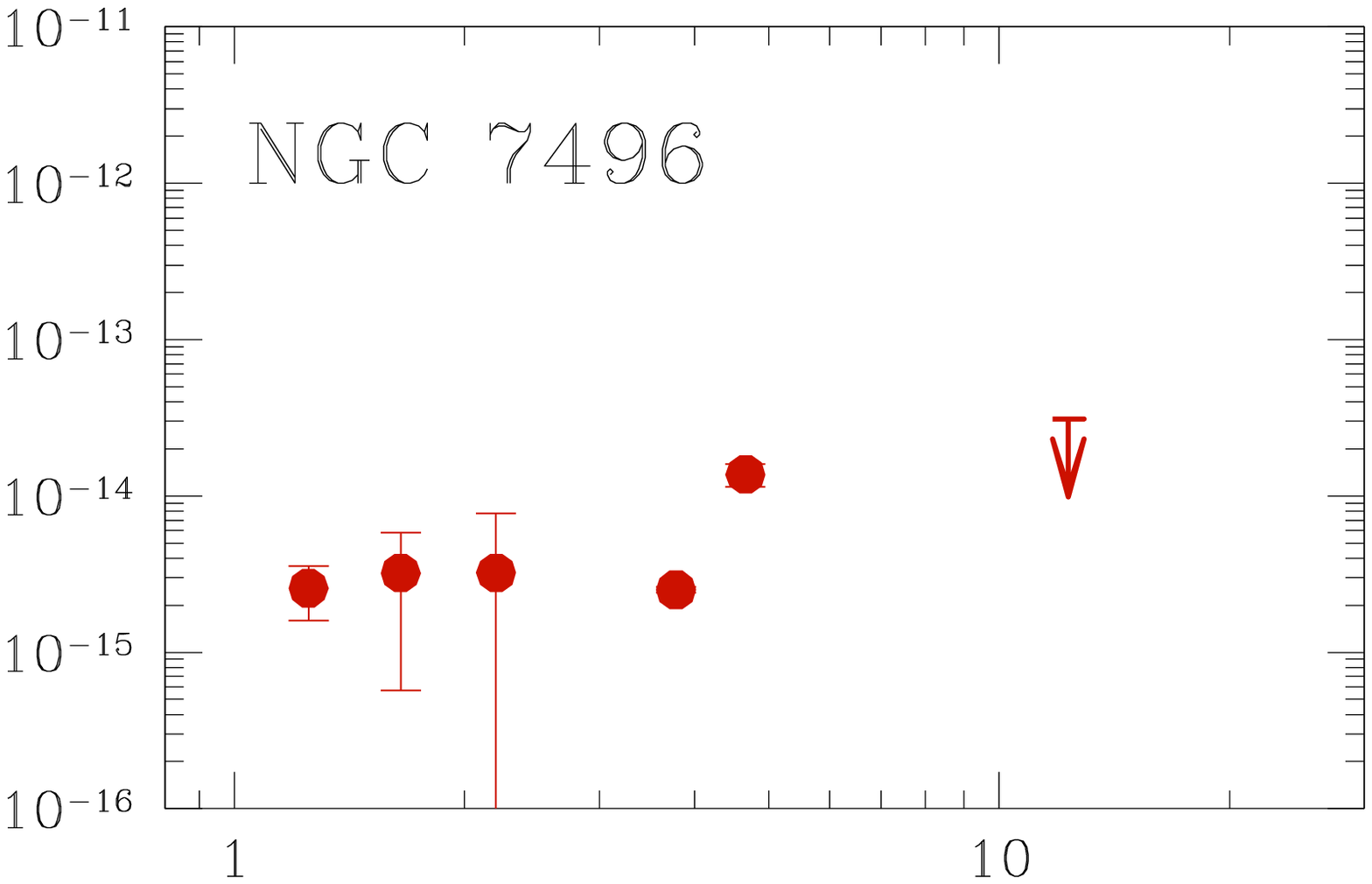}\\
    \includegraphics[scale=0.25,trim=0    30 0 100]{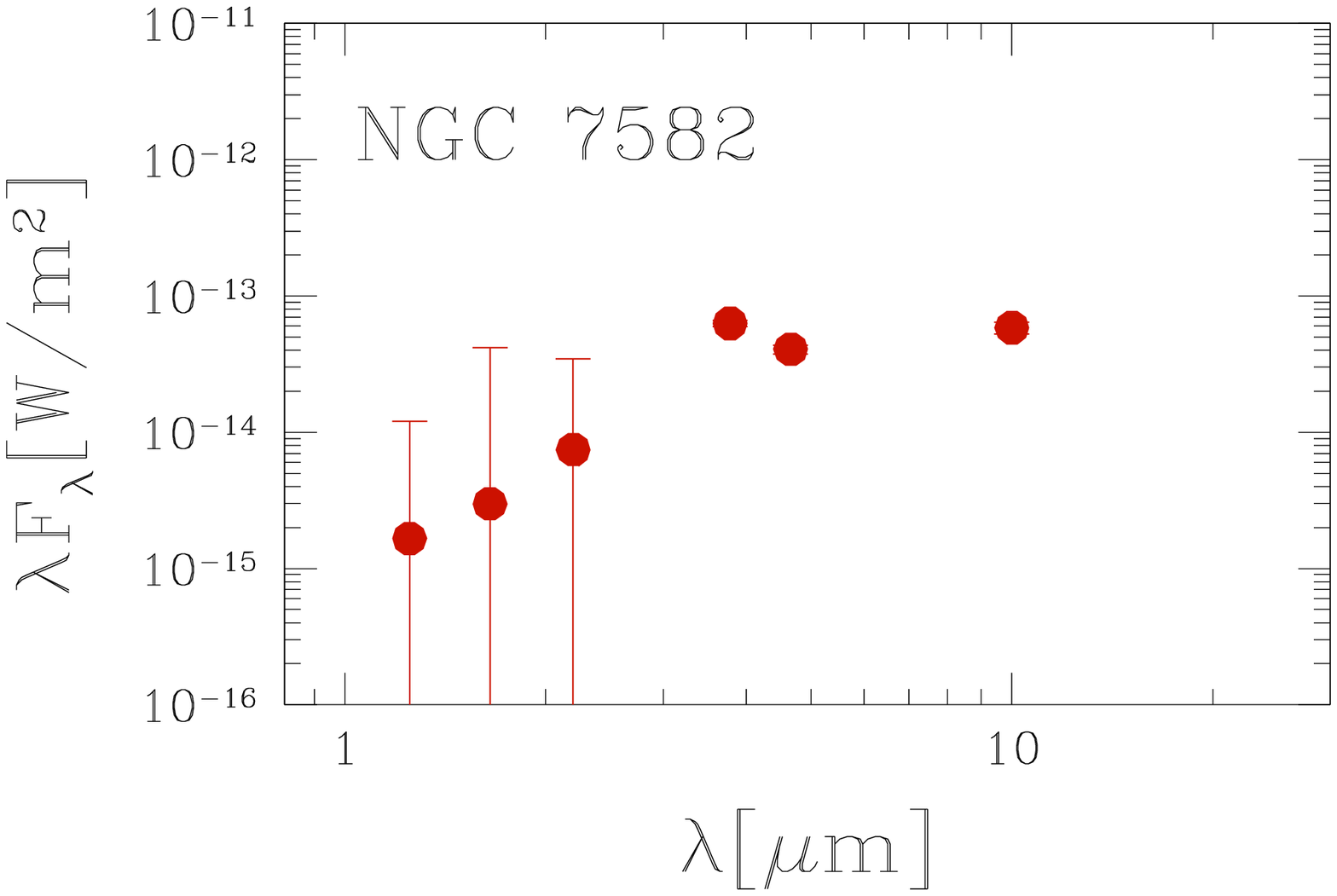}
    \includegraphics[scale=0.25,trim=100  30 0 100]{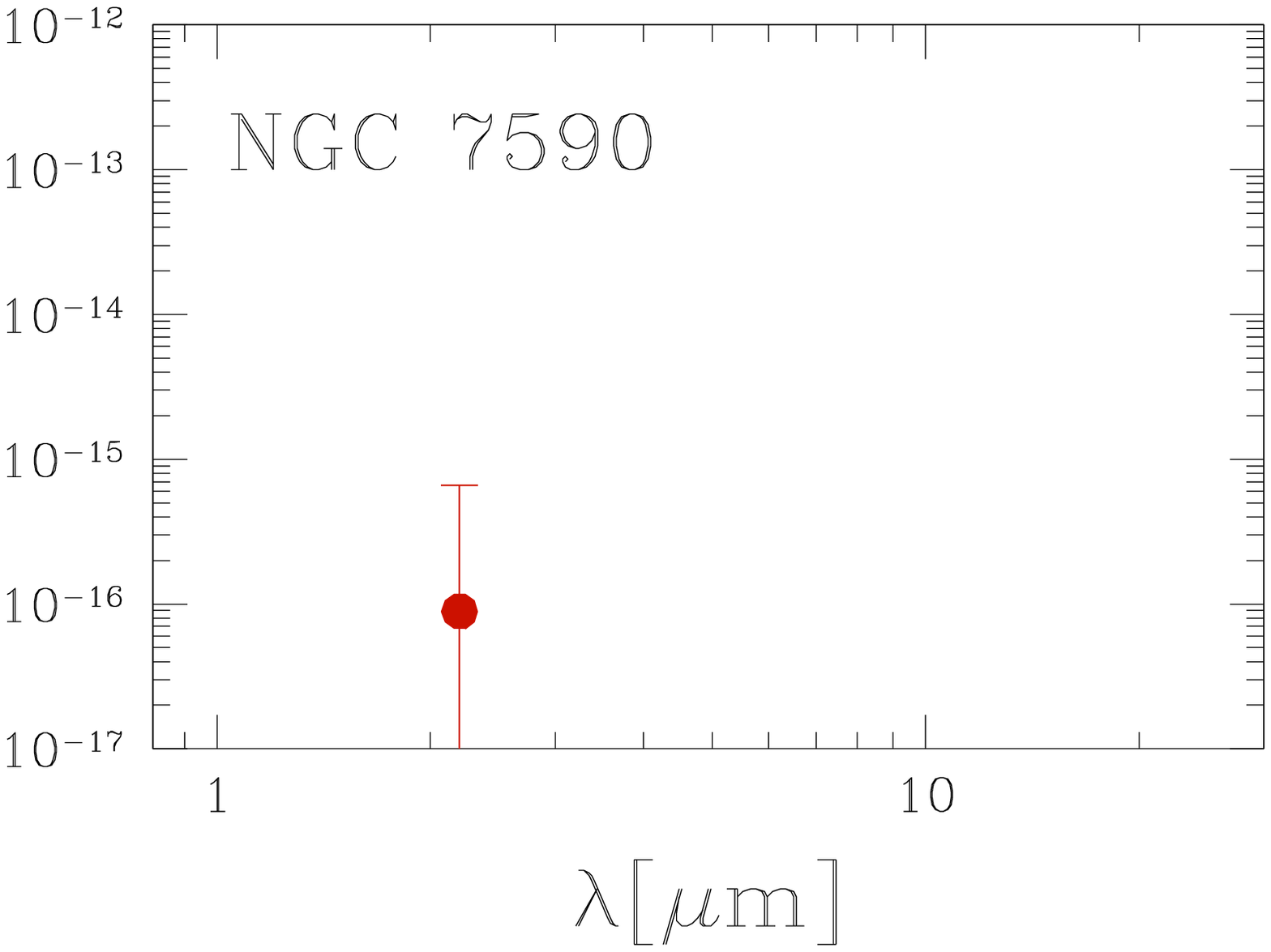}
    \includegraphics[scale=0.25,trim=100  30 0 100]{NGC7674_AH_sed.eps}
    \includegraphics[scale=0.25,trim=100  30 0 100]{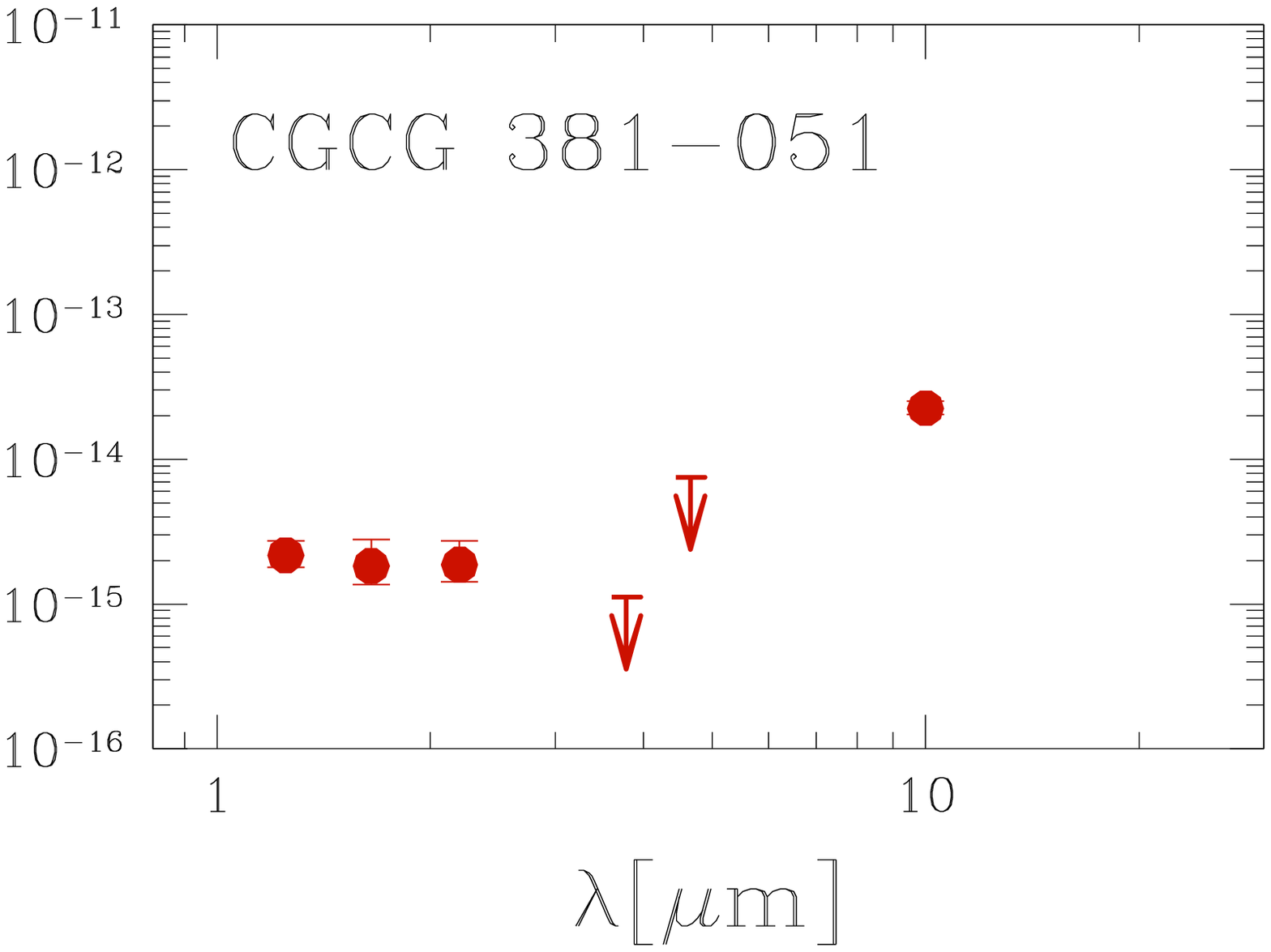}

  \end{center}
  \figurenum{5}
  \caption[SEDs]{Continuation.}
\end{figure*}

Fitted parameters are generally consistent between observing bands. The
following is a brief summary of the results, which are also presented in
Table~\ref{aju_contri}.

No disk or bar was detected in the M band, while the bulge was detected in
7/28 sources. In those objects the bulge contributed a mean brightness of 44\%
to the total flux. Only one bar was detected in the L band with a contribution
of 17\%. In the NIR bulge, disk and bar (if present) have similar
contributions of $\sim$ 30-50\%.

Seven out of 105 near--IR fits require a null nuclear contribution. The mean
percentage contribution of each galactic component is summarized in
Table~\ref{aju_contri}. A mean nuclear contribution of $\la$ 10\% is found in
the NIR bands (JHK), while a strong increase can be seen in the MIR, going
from $\sim$ 40\%\ in the M band to $\sim$ 100\%\ in the M and N bands.

The percentage contribution of the nuclear component in the J, K, L and M
bands is shown in Fig.~\ref{aju_contri_fig} as a function of the OIII
luminosities (tabulated in Paper II). It can be seen that, as expected, the
percentage nuclear contribution increases for more luminous AGN, as indicated
by the OIII measurements (Kotilainen et al.~1992; Alonso-Herrero et al.~2001,
2003).

The fitted parameters for the galactic structures are tabulated in
Table~\ref{fitted_param}. In Appendix~\ref{apendiceA} details of the
individual SBP fittings are given. The fits for the MIR profiles of all the
galaxies are shown at 2 different scales: a detailed one (bottom panels),
showing the central part of the galaxy, and a global one (middle panels),
where the whole galaxy is observable.

\subsection{Spectral Energy Distributions}

The constructed nuclear IR SEDs are shown in Fig.~\ref{seds}.  The nuclear
fluxes and standard deviations for each observed band are summarized in
Table~\ref{nuclear_flux}. Here we include the SEDs already determined by
\citet{ah01,ah03}. Upper limits when the galaxy was undetected in L, M, and N bands
are also shown.

There are some cases where a lack of photometric points prevents us from
obtaining a good SED, as with NGC\,1194, IRAS\,03362-1642, ESO\,33-G2,
NGC\,4941, and NGC\,7590. The SEDs of both nuclei in IRAS\,00198-7926 are
similar and do not determine whether the northern or southern nucleus (or
both) corresponds to the AGN.

\begin{figure*}
  \begin{center}
    \includegraphics[scale=0.35,trim=100 0 0 0]{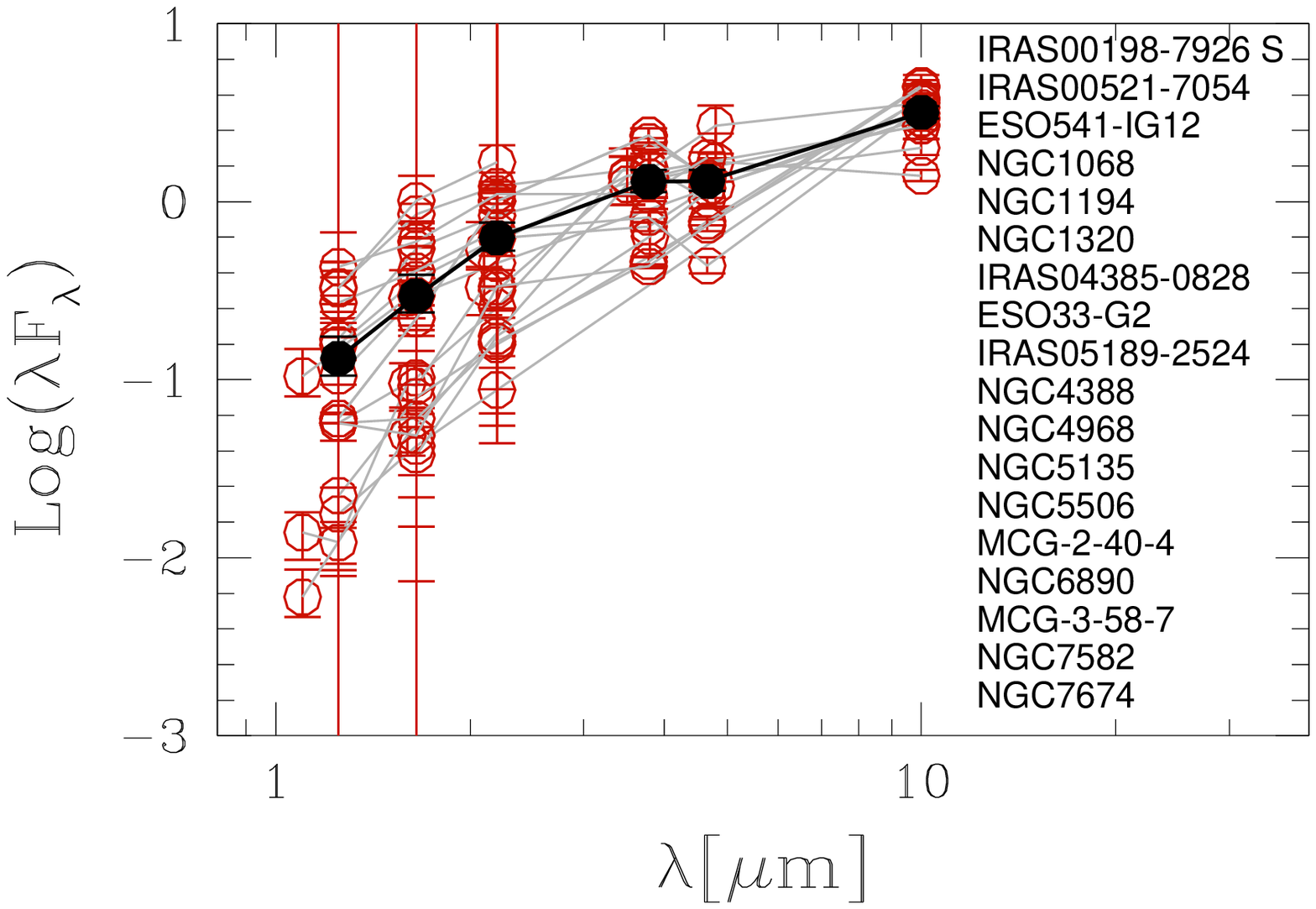}%
    \includegraphics[scale=0.35,trim=100 0 0 0]{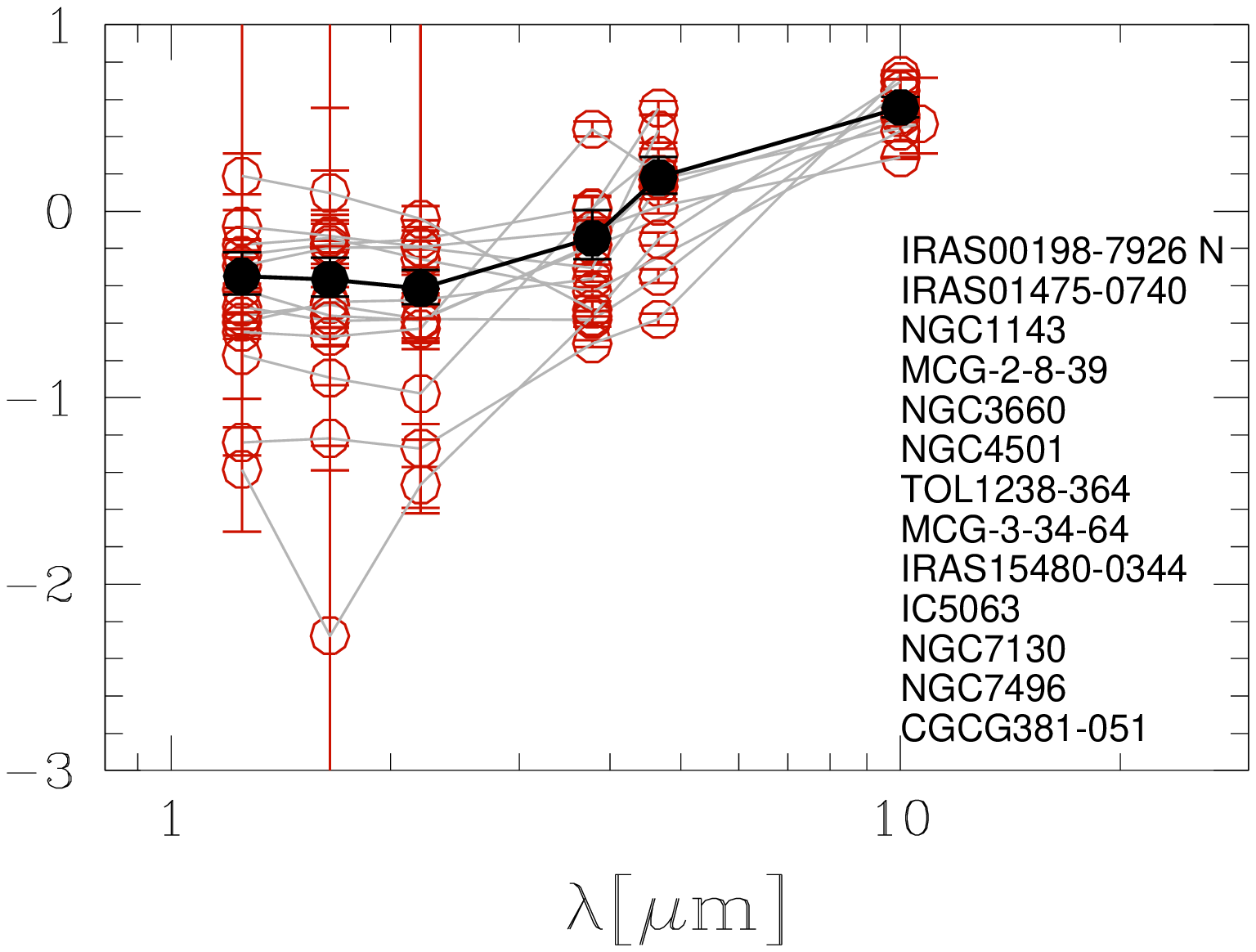}%
    \includegraphics[scale=0.35,trim=100 0 0 0]{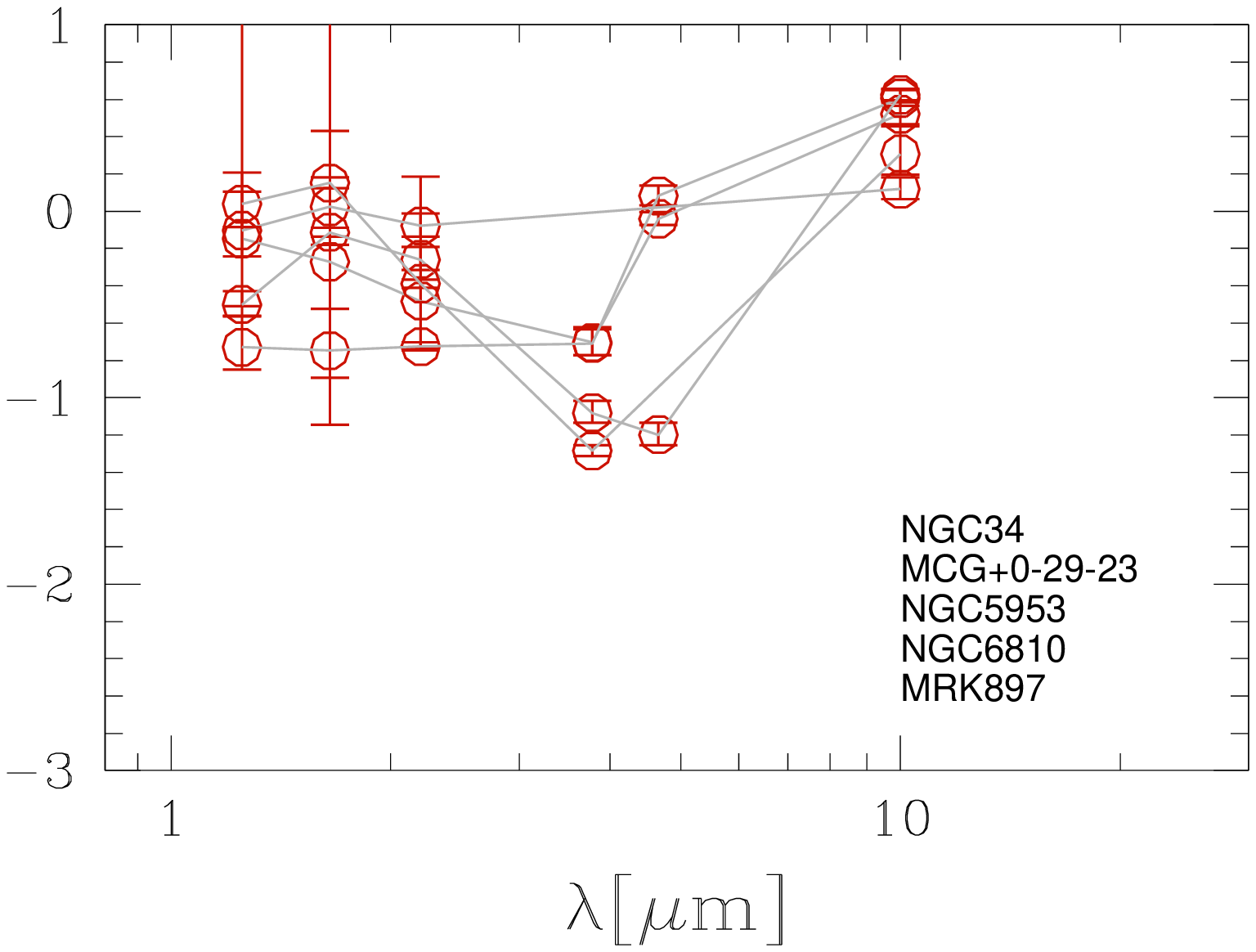}
  \end{center}
  \caption[SEDs]{Normalized Spectral Energy Distributions for three groups of
    sources. The left-hand-side panel shows nuclei with a NIR slope $>1$. The
    central panel shows those sources where a NIR excess is present, i.e.,
    with a NIR slope $<1$. Finally, the right-hand-side panel shows the SEDs
    of the three HII nuclei found in our sample and two nuclei which are
    likely to be contaminated by nearby starburst emission. Individual and
    average SEDs are shown for the first two cases, ignoring upper limits when
    deriving the averages.
    \label{avg_seds}}
\end{figure*}

Visual inspection of the SEDs reveals that we can broadly group most of them
into two types depending on the NIR behavior: one group shows a "classic''
bell-shaped SED, as expected from the emission of a dusty torus, while the
remaining SEDs show a much flatter NIR spectral distribution, with a clear
excess with respect to the first group. We have quantified this behavior by
measuring the NIR slope of the SEDs and classify them depending on whether the
slope $\alpha$ ($\lambda f_{\lambda} \propto \lambda^{\alpha}$) is larger or
smaller than 1. Individual and average SEDs for these two groups can be seen
in Figure \ref{avg_seds}. The tabulated values for the NIR slope are shown in
Table \ref{nuclear_flux}. Notice that some values have rather large errors and
therefore the classification can be ambiguous. The possible nature of the SEDs
presenting a NIR excess is discussed in the next section and further analysis
can be found in Paper II.

In the same figure we also include the SEDs of the three HII nuclei found in
our sample and of NGC\,34 and NGC\,5953 where contamination by a starburst
component is likely. It can be seen that the morphology of their SEDs clearly
departs from the bell-shaped spectra of the first group of Seyfert nuclei and
are somewhat closer to the second group of Seyfert galaxies, as it is also
noted by the flat value of their NIR slopes (see Table \ref{nuclear_flux}).

\begin{deluxetable*}{lccccccc}
\tabletypesize{\footnotesize}
\tablecolumns{8}

\tablecaption{Nuclear Fluxes \label{nuclear_flux}}

\tablehead{
\colhead{Galaxy} & \colhead{J band} & \colhead{H band} & \colhead{K band} &
\colhead{L band} & \colhead{M band} & \colhead{N band} & \colhead{NIR slope $\natural$}\\
\colhead{} & \colhead{mJy} & \colhead{mJy} & \colhead{mJy} & \colhead{mJy} & \colhead{mJy} & \colhead{mJy} &\\
}

\startdata
          NGC\,34       & 1.60$\pm$0.25      & 5.17$\pm$0.28       &  4.92  $\pm$0.70    &  1.27  $\pm$0.17 &   1.19  $\pm$0.16   &  171 $\pm$ 12  & 0.97$\pm$1.26 \\
 IRAS\,00198-7926 N     & 0.35$\pm$0.02      & 0.44$\pm$0.05       &  0.65  $\pm$0.08    &  13.10 $\pm$1.21 &   9.27  $\pm$0.75   &    --          & 0.10$\pm$0.15 \\
IRAS\,00198-7926 S      & 0.11$\pm$0.01      & 0.54$\pm$0.08       &  3.28  $\pm$0.43    &  13.10 $\pm$1.21 &   9.27  $\pm$0.75   &    --          & 5.02$\pm$0.16 \\
     IRAS\,00521-7054   & 2.57$\pm$0.14      & 4.70$\pm$0.29       &  11.66 $\pm$1.65    &  19.91 $\pm$1.84 &   39.31 $\pm$2.79   &    --          & 1.69$\pm$0.28 \\
       ESO\,541-IG12    & 1.35$\pm$0.16      & 4.56$\pm$0.56       &  8.73  $\pm$1.44    &  19.15 $\pm$3.18 &   15.71 $\pm$2.08   &  $<$140        & 2.28$\pm$0.62 \\
     IRAS\,01475-0740   & 0.66$\pm$0.02      & 1.00$\pm$0.04       &  1.41  $\pm$0.19    &  3.57  $\pm$0.45 &   8.47  $\pm$0.63   &    --          & 0.34$\pm$0.09 \\
      NGC\,1068\dag     & 9.8                & 97.6                &  449.6              &  3691	    &   8245      	  &  23000 	   & 5.77$\pm$0.86 \\
      NGC\,1144         & 0.35$\pm$0.11      & 0.34$\pm$0.15       &  0.42  $\pm$0.14    &  1.84  $\pm$0.04 &   5.10  $\pm$0.55   &  23$\pm$10 *   & -0.67$\pm$0.24\\
       MCG\,-2-8-39     & 0.07$\pm$0.05      & 0.07$\pm$0.09       &  0.07  $\pm$0.23    &  1.25  $\pm$0.19 &   5.39  $\pm$0.47   &  $<$114 *      & -1.00$\pm$0.01\\
         NGC\,1194      & 0.02$\pm$0.12      & 0.17$\pm$0.23       &  1.01  $\pm$0.90    &   --             &    --               &  133 $\pm$ 19  & 5.92$\pm$0.43 \\
         NGC\,1320      & 0.11$\pm$0.10      & 0.35$\pm$0.65       &  0.98  $\pm$0.97    &   --             &    --               &  195 $\pm$ 27  & 2.86$\pm$0.17 \\
     IRAS\,03362-1642   & 0.04$\pm$0.06      & no-det              &  no-det             &   --             &    --               &  970 $\pm$ 110 & --            \\
     IRAS\,04385-0828   & 0.19$\pm$0.10      & 0.90$\pm$0.14       &  2.36  $\pm$0.84    &   --             &    --               &  254 $\pm$ 46  & 3.43$\pm$0.64 \\
        ESO\,33-G2      & 0.36$\pm$0.38      & 1.47$\pm$0.57       &  3.21  $\pm$0.80    &   --             &    --               &    --          & 2.86$\pm$0.68 \\
     IRAS\,05189-2524   & 1.93$\pm$1.18      & 7.21$\pm$3.37       &  14.67 $\pm$5.23    &   --             &    --               &  463 $\pm$ 62  & 2.57$\pm$0.66 \\
       MCG\,+0-29-23    & 0.42$\pm$0.13      & 0.53$\pm$0.22       &  0.75  $\pm$0.04    &  1.33  $\pm$0.22 &   10.13 $\pm$1.22   &  73 $\pm$ 7    & 0.02$\pm$0.11 \\
         NGC\,3660      & 0.74$\pm$0.05      & 1.07$\pm$0.21       &  1.09  $\pm$0.23    &  1.26  $\pm$0.07 &   $<$11.07          &  24.2$\pm$3.3  & -0.32$\pm$0.37\\
         NGC\,4388\dag  & 0.06               & 0.71                &  --                 &  39.9            &   --                &  245$\pm$15    & 4.18$\pm$2.17 \\
         NGC\,4501\ddag & 4.58$\pm$1.12      & 4.93$\pm$1.18       &  4.76  $\pm$0.71    &  2.68  $\pm$0.25 &   $<$7.22           &  6.0$\pm$0.5 * & -0.93$\pm$0.11\\
       TOL\,1238-364    & 1.67$\pm$0.30      & 1.96$\pm$0.27       &  2.27  $\pm$0.64    &  4.81  $\pm$0.55 &   7.96  $\pm$0.68   &  31.2$\pm$0.4  & -0.46$\pm$0.02\\
         NGC\,4941      & no-det             & no-det              &  no-det             &   --             &   $<$12.23          &  47 $\pm$ 5    & --            \\
         NGC\,4968\dag  & no-det             & 0.6                 &  3.7                &  23.2            &   56.5		  &  280$\pm$20    & 5.32$\pm$0.01\\
       MCG\,-3-34-64    & 0.81$\pm$0.14      & 1.13$\pm$0.54       &  1.32  $\pm$0.34    &  8.29  $\pm$0.34 &   13.75 $\pm$0.96   &  603$\pm$30    & -0.14$\pm$0.19\\
         NGC\,5135      & 0.17$\pm$0.45      & 0.19$\pm$0.38       &  1.74  $\pm$1.41    &  3.88  $\pm$0.10 &   8.18  $\pm$0.62   &  104 $\pm$ 10  & 3.16$\pm$2.08 \\
         NGC\,5506\dag  & 13.8               & 59.0                &  120.4              &  340.1           &   530.0             &  100 $\pm$ 10  & 2.83$\pm$0.80 \\
         NGC\,5953\ddag & 1.61$\pm$0.53      & 2.77$\pm$0.18       &  1.06  $\pm$0.05    &  0.23  $\pm$0.02 &   $<$12.23          &  24$\pm$7      & -1.68$\pm$1.53\\
       MCG\,-2-40-4     & 3.57$\pm$1.47      & 7.10$\pm$3.00       &  15.99 $\pm$6.15    &  32.63 $\pm$2.21 &   21.43 $\pm$2.24   &  355$\pm$ 21   & 1.65$\pm$0.10 \\
     IRAS\,15480-0344   & 0.96$\pm$0.09      & 1.62$\pm$0.62       &  2.24  $\pm$0.18    &  4.95  $\pm$0.17 &   19.11 $\pm$1.68   &  100 $\pm$ 10  & 0.50$\pm$0.23 \\
         NGC\,6810      & 1.31$\pm$1.70      & 1.30$\pm$8.41       &  1.06  $\pm$0.84    &  1.10  $\pm$0.18 &   6.23  $\pm$0.52   &  49$\pm$ 6.5   & -1.37$\pm$0.21\\
         NGC\,6890      & 0.17$\pm$0.22      & 0.24$\pm$0.42       &  0.86  $\pm$0.73    &  4.16  $\pm$0.27 &   8.61  $\pm$0.48   &  107$\pm$ 14   & 1.89$\pm$0.92 \\
         IC\,5063       & 0.82$\pm$0.95      & 0.14$\pm$0.79       &  1.20  $\pm$0.50    &  16.38 $\pm$1.17 &   52.48 $\pm$4.39   &  787$\pm$105   & -0.37$\pm$4.00\\
         MRK\,897       & 0.43$\pm$0.16      & 0.75$\pm$0.45       &  0.79  $\pm$0.11    &                  &   $<$11.15          &  5.7$\pm$0.8   & 0.08$\pm$0.54\\         
         NGC\,7130      & 1.53$\pm$0.50      & 2.32$\pm$1.56       &  2.58  $\pm$1.15    &  4.41  $\pm$0.41 &   9.27  $\pm$0.79   &  197$\pm$ 26   & -0.08$\pm$0.33\\
         NGC\,7172\dag  & non-det            & $<$0.4              &  3.4                &  30.0            &   61.4              &  103 $\pm$ 10  & --            \\
      MCG\,-3-58-7      & 1.45$\pm$0.59      & 4.47$\pm$2.31       &  10.12 $\pm$4.11    &  20.29 $\pm$1.07 &   41.19 $\pm$4.94   &  218 $\pm$ 15  & 2.42$\pm$0.35 \\
         NGC\,7496      & 1.07$\pm$0.40      & 1.77$\pm$1.44       &  2.36  $\pm$3.31    &  3.16  $\pm$0.13 &   21.28 $\pm$3.55   &    --          & 0.39$\pm$0.24 \\
         NGC\,7582$\o$  & 0.69$\pm$4.33      & 1.64$\pm$21.24      &  5.45  $\pm$19.84   &  79.06 $\pm$3.88 &   63.19 $\pm$4.72   &  195           & 2.66$\pm$0.31 \\
         NGC\,7590      & no-det             & no-det              &  0.26  $\pm$0.55    &  no-det          &    --               &    --          & --            \\
         NGC\,7674\dag  & 1.25		     & 5.0		   & 12.3                &  53		    &    --		  & 344		   & 3.04$\pm$0.54 \\
       CGCG\,381-051    & 0.91$\pm$0.19      & 1.01$\pm$0.34       &  1.37  $\pm$0.43    &  $<$1.41         &   $<$11.65          &   75 $\pm$ 8   & -0.27$\pm$0.20\\

\enddata \tablecomments{\dag\ indicates when most of the SED measurements were
  determined by AH01 and AH03 (see Table 1 for more details); * indicates data
  from Joy \& Ghigo (1988) and Maiolino et al.\ (1995). Caution must be taken
  because of the very different dates of the observations when compared with
  our shorter-wavelength data. \ddag\ N-band observations from Gemini and G04
  only provided upper limits. $\natural$ NIR slope $\alpha$ ($\lambda
  f_{\lambda} \propto \lambda^{\alpha}$) was tabulated using least square fits
  to the data. $\o$ Photometry for NGC\,7582 will be complemented with
  observations from Prieto et al.~2010 and Ramos-Almeida et al.~2009 in paper
  II.}
\end{deluxetable*}

\begin{deluxetable}{lccc}
\tablecolumns{4}
\tablecaption{Near-IR Excess and Radio-loudness \label{excess}}
\tablehead{
\colhead{Galaxy} & \colhead{Excess} & \colhead{HII Nuclei} & \colhead{$R_L$}}
\startdata
IRAS\,00198-7926\,N & yes & no & --\\
IRAS\,01475-0740    & yes & no & 70 yes\\
NGC\,1144           & yes & no & 3.4 no\\
MCG\,-2-8-39        & yes & no & 0.2 no\\ 
NGC\,3660           & yes & no & 0.9 no\\
NGC\,4501           & yes & no & 95 yes\\
Tol\,1238-364       & yes & no & 0.4 no\\
MCG\,-3-34-64       & yes & no & 2.2 no\\
IRAS\,15480-0344    & yes & no & 0.6 no\\
NGC\,7130           & yes & no & 1.1 no\\
NGC\,7496           & yes & no& 32 yes\\
CGGC\,381-051       & yes & no & 2.4 no\\
\enddata
\end{deluxetable}

\subsection{Near-IR excess}

In a Type I Seyfert, where the inner region of the central source is directly
observed, the accretion disk can make a significant contribution to the total
near--IR emission. Recently Landt et al.~(2011) and Lira et al.~(2011) have
shown that this is necessary to explain the continuum around 1\micron\ in
these objects.

A ``typical'' Type II source, on the other hand, should have an SED that
decreases monotonically towards shorter wavelengths. But inspection of the
SEDs in this work shows several NIR SEDs with a clear excess which resembles
Type I systems (see Fig.~\ref{seds} and Table \ref{excess}).

All galaxies in our sample had an early classification as Type II systems
(with MCG\,-3-34-64 being classified as Seyfert 1.8, NGC\,5506 as a Seyfert
1.9, and NGC\,7130 as a LINER). However, as noted above, more recent
observations show that NGC\,6810, MRK\,897, and MCG\,+0-29-23 correspond to
HII nuclei, while NGC\,34 and NGC\,5953 might have contamination by a
starburst component. An IR excess is expected for these objects because, in
the absence of an active nucleus, the Rayleigh-Jeans emission tail from cold
stars dominates this spectral region for moderately old starbursts ($\sim$ a
few dozen Myrs). This is clearly observed in other starbursts and extensively
modeleded (e.g., Efstathiou et al.~2000; Dopita et al.~2005;
Rodr\'iguez-Merino et al.~2011).

The remaining Seyfert II objects that show an IR excess require alternative
explanations; we discuss various mechanisms below.

A traditional torus --- a spatially continuous structure with large optical
depth --- would completely obscure the accretion disk in Type II objects, but
a clumpy torus gives a non-zero probability of observing the central source at
any orientation; the latter scenario is increasingly accepted and could
explain some near--IR emission.

Alternatively, a nuclear jet could make a significant contribution to the
observed near--IR fluxes. In this case the emission would be due to
synchrotron instead of thermal processes. Jet emission is found to be
prominent in radio-loud AGN and in relativistically boosted radio-quiet AGN
(also known as radio-intermediate AGN; Falcke et al.~1996; Barvainis et
al.~2005). Radio-loud AGN are defined as those with $R_L > 100$ (where
$R_L=F_{5 GHz}/F_{B}$; Kellermann et al.~1989), radio-intermediate AGN are
those with $3-10 \la R_L \la 100$, while radio-quiet AGN have $R_L <<10$.

We have compiled 5 GHz measurements for our sample and determined $R_L$ by
computing $L^{\rm bol}$ from measured [OIII] fluxes and then turning them into
nuclear $B$ fluxes following Marconi et al.~(2004). The resulting $R_L$ values
are presented here for the SEDs with NIR excess and in Paper II for the
complete sample. We found that $\sim 30\%$ of the sources with NIR excess can
be classified as radio-loud (F\,01475-0740, NGC\,4501 and NGC\,7496); only two
additional radio-loud sources present a bell-shaped SED (F\,04385-0828 and
NGC\,7172), representing a 10\% of that subsample. NGC\,4501 presents a rather
peculiar SED, with a power-law distribution across the entire 1-10$\mu$m range
(see Paper II). In this case synchrotron emission might explain the spectral
shape.

In summary, jet emission might explain some observed features in the infrared
SEDs, but the link is not proven and does not explain all cases.

A very hot dust component has been proposed to explain near--IR emission in
Mrk\,1239 and Mrk\,766 (Rodr\'iguez-Ardila et al.~2005; Rodr\'iguez-Ardila \&
Mazzalay 2006). These are type I AGN, which provide an unimpaired view of the
hottest dust region, and emission from this dust --- possibly carbonaceous
grains surviving inside the silicate sublimation radius --- appears as an
excess superimposed on the disk component. Interestingly, the peak observed in
these two sources is clearly distinct from the putative torus component (which
is not observed in these sources as no data are available beyond 4\micron).

Similarly, a hot dust component was used in other works when modeling the
near-IR region of {\em Spitzer} spectra for powerful QSOs (Mor \&
Trakhtenbrot 2011, Mor et al.~2009, Schweitzer et al.~2008; Deo et
al.~2011). They argue that this component is required to fit the NIR part of
the spectrum and that disk emission can only contribute significantly at
1\micron, but not enough at longer wavelengths. Therefore very hot
pure-graphite dust emission is proposed.

A similar very hot dust component could explain some of our observations, but,
because of its location inside the torus sublimation radius, it again requires
a clumpy absorbing medium to explain its detection.

A very compact nuclear starburst might have a significant contribution to the
nuclear fluxes. To test this hypothesis we have looked at the positions of
Seyfert nuclei in the diagnostic diagrams presented in Figure \ref{diag_diag},
where nuclei with a NIR excess are shown with a square. It can be seen that
there is no clear segregation of these nuclei towards the HII location in the
plot. Also, these nuclei do not show preference for the presence of a very
young stellar continuum as shown in Table \ref{table_spec}.

Finally, a young and luminous nuclear stellar cluster could contribute to the
observed near-IR emission. However, with an absolute magnitude $z_{\rm
  AB}=-13$ for the most luminous examples (Cote et al.~2006) their flux is not
sufficient to make a significant contribution to the observed SEDs.

\section{Summary}\label{summary}

We have constructed {\emph{nuclear}} IR SEDs of 40 Type II Seyfert galaxies
from ground--based, high--resolution observations. The images were collected
between 2003 and 2004, to reduce variability distortion of the SEDs as much as
possible, and under good seeing conditions.

We also obtained optical spectroscopy for most of the sample. The spectra were
modeled using the STARLIGHT code to characterize their stellar continuum
properties and emission lines were used to classify the spectra as active or
HII nuclei.  Three objects from our sample most likely harbor a starburst
nucleus and not an AGN, in agreement with other findings.

The surface brightness profiles of each galaxy, in each observed band, were
constructed by fitting elliptical isophotes to the images. The profiles were
then modeled using a nucleus, bulge, disk and, where necessary, a bar. This
procedure was tested using synthetic data and IR observations of the Andromeda
galaxy, which showed that the scale length of the different galactic
structures does not change when observing at longer wavelengths. This allowed
us to fix the scale lengths in the MIR when analyzing undetected components of
distant galaxies. Careful modeling of the PSF was also included to take into
account seeing distortions.

The fitted SBP parameters for any particular object are generally consistent
between bands. In some cases, as in IRAS\,03362-1642, NGC\,4941, and
NGC\,7590, no nuclear component was detected by the deconvolution process in
the NIR. 

A variety of SED shapes is found in our sample, despite the common Type II
Seyfert classification. About 40\%\ of the Seyfert SEDs are characterized by
an upturn or excess in the near--IR. The 5 objects with strong starburst
emission (3 HII nuclei and 2 Seyfert nuclei with strong contamination from
nearby starburst regions) also show a similar excess. For genuine AGN this
component could be explained as emission from a jet, the accretion disk, or
from a very hot dust component, leaking out from the central region through a
clumpy obscuring structure. The presence of a very compact nuclear starburst
as the origin for this NIR excess emission is not favored by our data.

\acknowledgments 

We are forever indebted to Roberto Cid Fernandes for his help using the
STARLIGHT code. LV gratefully acknowledges fellowship support by project
MECESUP UCH0118, and partial support from Fondecyt project 1080603.  PL is
grateful for financial support from Fondecyt grant No 1080603. This
publication was also financed by the ALMA-Conicyt Fund, allocated to project
N° 31060003. A.A.H acknowledges support from the Spanish Plan Nacional de
Astronom\'{\i}a y Astrof\'{\i}sica under grant AYA2009-05705-E. Finally, PL
wants to express all her gratitude to A.~Cooke for his never-ending support.

\appendix

\section{Notes on the Surface Brightness Profile Deconvolution of Individual
  Sources}\label{apendiceA}

In the notes below, whenever a parameter is described as "constrained'' by
some other value, it means that the parameter was allowed to vary within a
limited range around the constraining value(s). Without this approach the
parameter would adopt a solution that was non--physical or at odds with values
found at other wavelengths.

\begin{figure}
\begin{center}
\epsscale{0.45}
\plotone{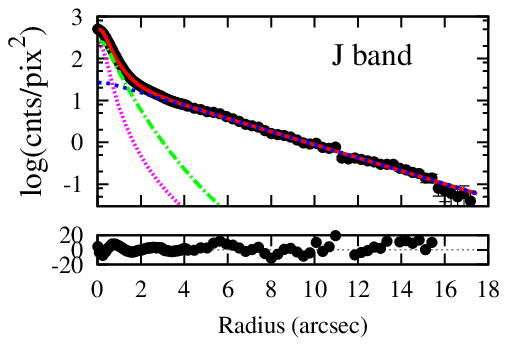}
\plotone{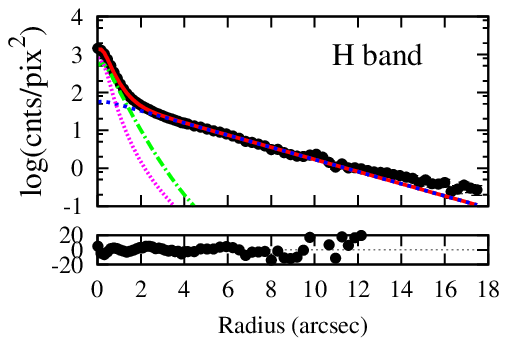}
\plotone{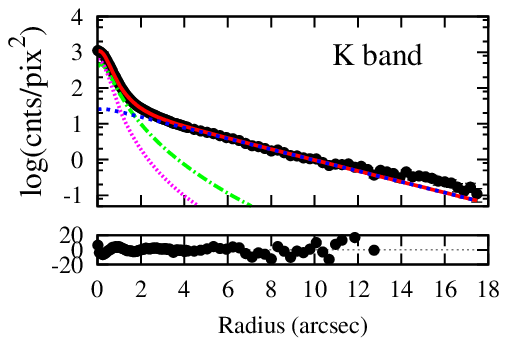}\\
\plotone{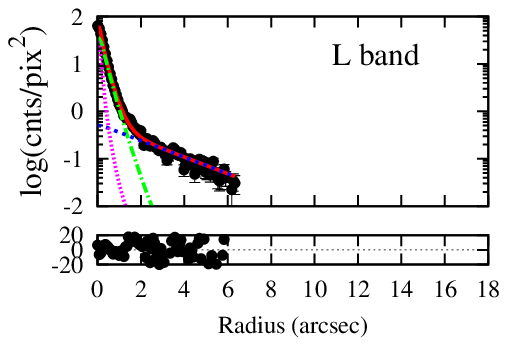}
\plotone{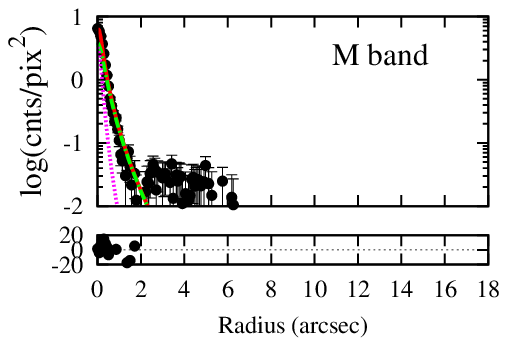}\\
\plotone{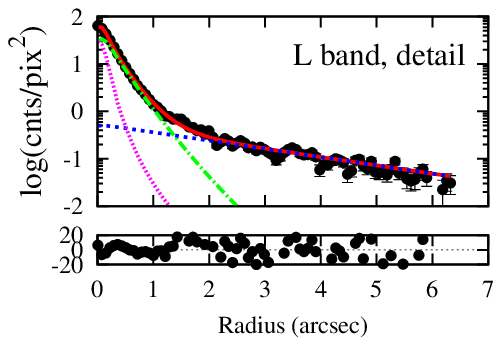}
\plotone{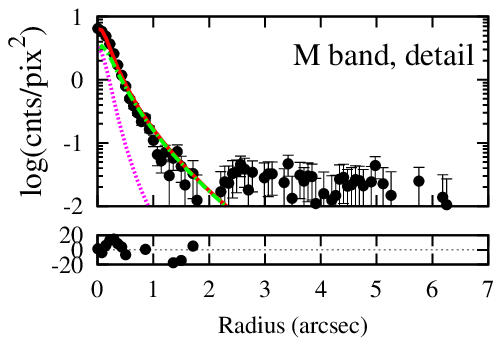}
\caption{Radial surface brightness profile of the J, H, K, L and M band
  observations of NGC\,34 (black dots with error bars). The y-axis corresponds
  to Log(counts/pix$^2$) and the ordinates show the distance from the center
  in arc-seconds. The best--fit model is shown with a thick continuous line.
  Residuals in percentage are shown in the bottom part of each panel. The
  different physical components of the model are: the nucleus (dotted--line),
  the bulge (dashed--dotted line), the disk (dashed line) and a bar whenever
  necessary (dashed--double--dotted line). The mid-IR profiles (L, M and N, if
  available) are plotted in two scales: a detailed scale showing the central
  part of the galaxy (bottom panels), and a global scale, showing the complete
  galaxy (middle panels). Color plots for this and the remaining galaxies are
  available in the online version of the Journal. \label{aj}}
\end{center}
\end{figure}

\paragraph{NGC\,34}\label{72}
In this object it is clear how the disk becomes weaker compared to the bulge
and nucleus from the J to L bands. The disk in the L band had to be
constrained to have a similar scale length to the J, H and K bands. In the
M-band the disk is not visible, but the bulge is still dominant. Here the
bulge effective radius had to be restricted to have similar values to those
derived at other wavelengths. The fits are shown in Fig.~\ref{aj}.

\paragraph{IRAS\,00198-7926} \label{71}
The NIR images for this galaxy show two very close compact objects. It is
therefore very difficult to determine which of the two corresponds to the
active nucleus. We label them as N and S. Elliptical photometry was performed
on both objects in each band. The fitting results are shown in Fig.~\ref{aj}
(online). The southern object has a strange behavior, where the effective
radius and the exponent of the Sersic's law of the bulge continuously
decreases from J to K. Only one object is detected in the L band image, and it
is included in both SEDs (north and south) because it is impossible to
determine its correspondence with any of the two nuclei. Also, the L-band
observations seem to suffer from seeing variations, as can be inferred from
the fitting results: constraining the parameters of the bulge to be similar to
the NIR values produces a very poor fit. As a result, all the emission
observed in this band was considered as coming from the nucleus.

\paragraph{IRAS\,00521-7054}\label{70}
This galaxy is classified as an E-S0 but, in order to obtain a good fit, the
disk component had to be excluded. Because the seeing changed between the
observation of the PSF star and the galaxy in the L band only the external
parts of the observed profile were considered when fitting the bulge (using
the values found in the NIR bands) and the nuclear flux was obtained assuming
$F_{nuc} = F_{gal}-F_{bul}$.

\paragraph{ESO\,541-IG12}\label{69}
This is a ``text--book'' active galaxy, with a bright dominant nuclear
source. However, in the L-band the S\'ersic's index was constrained by the
values obtained in the NIR fits, and in the M-band the effective radius of the
bulge was constrained to the values obtained in the NIR and L--band.

\paragraph{NGC\,424} \label{68}
The NIR observations could not be flux calibrated and the MIR observations
have no PSF star.  Hence, no SED was obtained.

\paragraph{IRAS\,01475-0740}\label{67}
In order to obtain a reasonable fit to the NIR profiles the scale length of
the disk was constrained to be larger than that of the effective radius of the
bulge. In the L band, the size and index of the bulge were restricted to have
similar values to those found in the NIR fits. 

\paragraph{NGC\,1125}\label{64}
This galaxy was only observed in the L and M bands. The disk and bulge
parameters could not be restricted and the parameter degeneration was too
large.  Hence, no SED was obtained.

\paragraph{NGC\,1144}\label{63}
This is an interacting system (NGC\,1143/44, Arp\,118) and the disk of the
Seyfert galaxy is very distorted. Comparing the J, H, and K band images it is
evident that the K band is less contaminated by the tidal tails, yielding well
behaved parameters for the disk. Using these values to constrain the disk
parameters in the J and H bands fitting results are consistent in all 3
bands. Results are plotted in Fig.~\ref{aj} (online) and clearly show excess
emission in the outer parts of the galaxy in the J and H bands, coming from
the streaming material. In the L band all 3 components are discernible and the
parameters are consistent with those of the NIR once the disk size is
constrained to be larger than the bulge size. The galaxy is point-like in the
M band, so all the emission is considered as nuclear and no fitting is needed.

\paragraph{MCG\,-2-8-39}\label{62}
This is a very nice example of a barred galaxy. The NIR fits were obtained
without any restriction, and the results are consistent. The L band profile
was fitted without disk or bar due to the bulge dominating the central region;
the size and index of the bulge had to be constrained using the NIR values.

\paragraph{NGC\,1194}\label{61}
This galaxy has a highly concentrated bulge ($n_B\approx3.3$). The results
were obtained constraining the index value of the K band profile to that found
in the J and H band fits.

\paragraph{NGC\,1320}\label{59}
This galaxy was fitted limiting the effective radius of the bulge in the K
band. 

\paragraph{IRAS\,03362-1642}\label{57}
The resulting fit of this galaxy does not include a nuclear component in the H
and K bands. In the J band a weak nucleus appears, with a contribution of $\la
1\%$, almost negligible. The bar is very dominant in all 3 bands.

\paragraph{IRAS\,04385-0828}\label{56}
To fit this galaxy it was necessary to constrain the bulge S\'ersic's index in
the J and H bands to the value found in K.

\paragraph{NGC\,1667}\label{55}
This galaxy has a very complex morphology, which prevents the determination of
a reliable profile fitting in the 3 bands observed.  Hence no SED was
obtained.

\paragraph{ESO\,33-G2}\label{54}
No restriction was needed in J and H bands to fit this galaxy. In the K band
the disk scale length was constrained to be larger than the bulge and bar
sizes, obtaining consistent results for all the parameters. 

\begin{figure*}[t]
  \begin{center}
    \includegraphics[scale=0.3,angle=270]{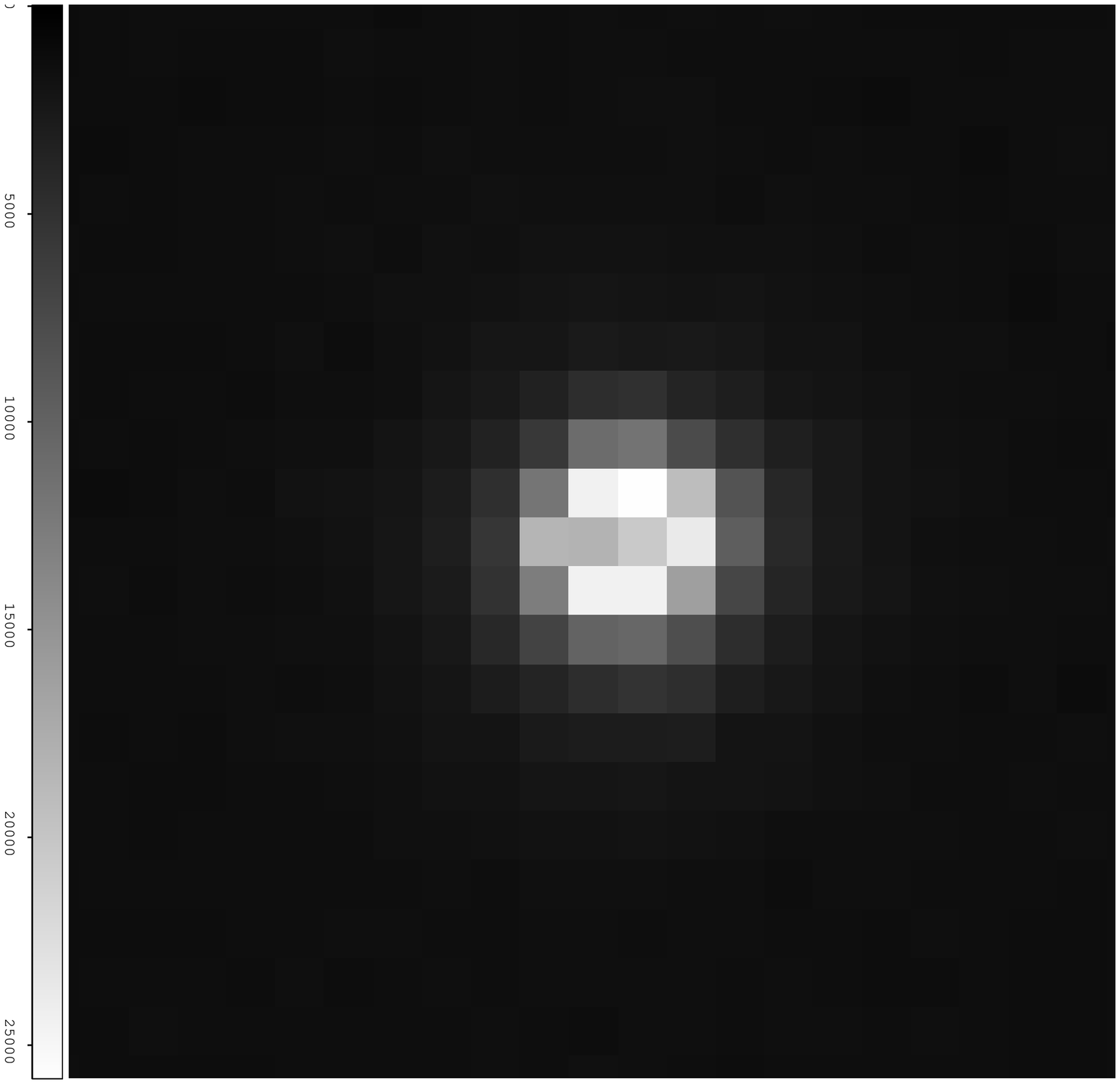}
    \includegraphics[scale=0.6,angle=270]{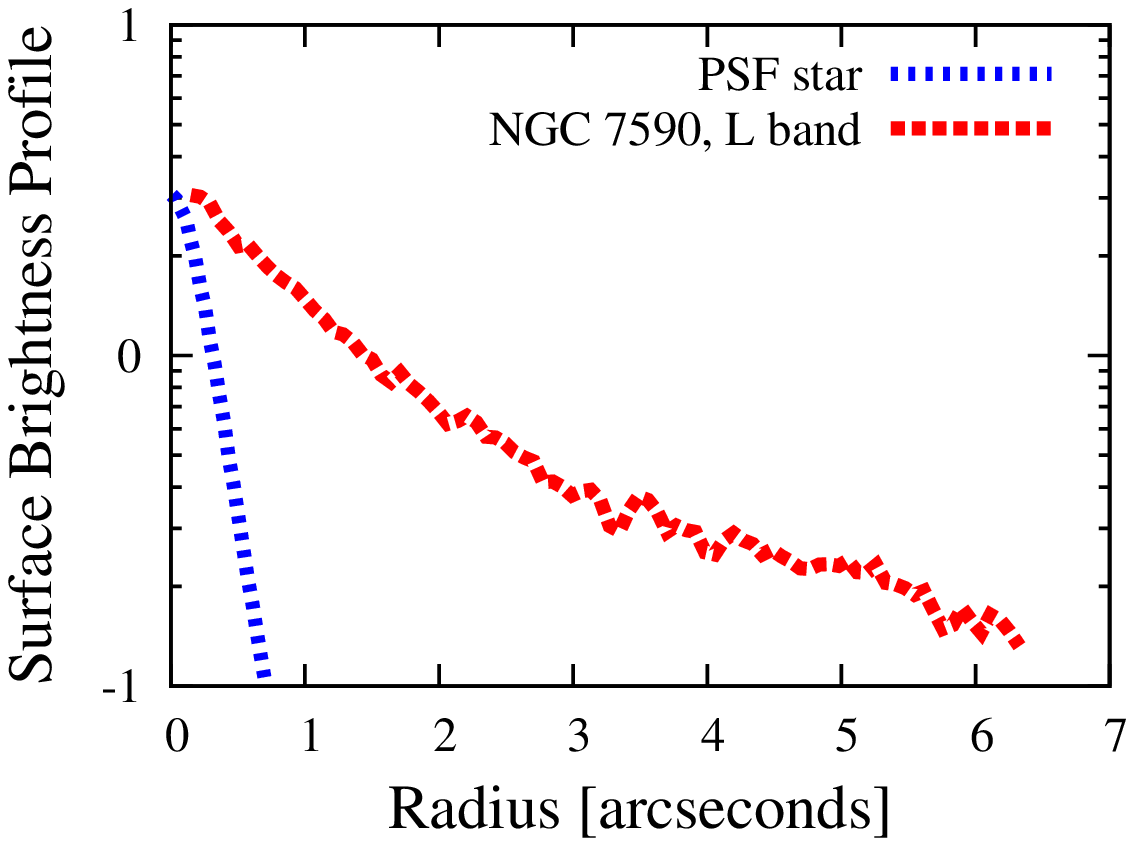}
  \end{center}
  \caption{{\emph{Left:}} Raw image in the K band of galaxy
    IRAS~05189-2524. The obscured central pixels are evident. {\emph{Right:}}
    Comparison of PSF star (dotted-blue line) and galaxy (dashed-red line)
    radial surface brightness profiles of L band of galaxy NGC~7590. It is
    evident that the galaxy is extended unless a strong variation in seeing
    conditions degraded the galaxy image to a level where no point source can
    be detected. Adding the fitting results in the near-IR and the resulting
    {\emph{Spitzer}} spectra after star-formation subtraction, it seems this
    galaxy has at most a very weak AGN.  \label{raw_53_K}}
\end{figure*}

\paragraph{IRAS\,05189-2524}\label{53}
In order to obtain a consistent fit to this galaxy the disk size in the H band
was constrained to be larger than the bulge size. In the K band the first
point of the radial profile is lower than the second and third points.  This
is due to obscuration that can be observed in all the raw images in this band,
as shown in the left panel of Fig.~\ref{raw_53_K}; this first point is not
included in the fitting process. In the N-band profile all the emission comes
from the nucleus.

\paragraph{ESO\,253-G3}\label{52}
This is an interacting system whose two nuclei are too close to fit
reliably. Hence, no SED was obtained.

\paragraph{MCG\,+0-29-23}\label{84}
Fig.~\ref{aj} (online) shows the fitting results for this very special
object. The results barely change from one band to another and the galaxy is
very extended even in the L band. It is not clear whether the central
spheroidal component corresponds to a bulge, since the images and the fitted
S\'ersic's index both resembles more closely a bar. In any case, bulge and/or
bar, the contribution to the central emission can be determined, even if it is
not possible to unveil the exact nature of the component.

\paragraph{NGC\,3660}\label{85}
This galaxy is very large, and due to the small size of the detector the
galaxy has been truncated in the J and H bands. In order to determine the real
value of the disk parameters the K band image was used and then these fitted
values were used to constrain the disk parameters in the J and H band fits,
excluding the profile for $r>45\arcsec$. In the L band, the profile with
$r>2\arcsec$ was skipped because it is mostly sky contribution. The parameters
of the bulge were restricted to those obtained in the J, H and K fits.

\paragraph{NGC\,4501}\label{86}
This is a huge galaxy compared to the detector, so the PSF was taken from the
standard star. The only restriction needed to obtain consistent results was
that the disk size was larger than the bulge size in the J and H bands. In the
L band the change in seeing conditions is evident, and to overcome this
problem the bulge was fitted for r$>3\arcsec$, constraining only the
S\'ersic's index; the nucleus is then computed as $F_{nuc} =
F_{gal}-F_{bul}$. The galaxy was not detected in the M or N bands. The upper
limit in the N band is that obtained by \citet{gorjian04}.

\paragraph{TOL\,1238-364}\label{87}
The fit to this galaxy in the NIR bands does not include the bump produced by
the ending of the flocculent arms between 16$\arcsec$ and 23$\arcsec$. In the
J and H bands a disk scale length larger than the bulge size was imposed. In
the K band the bulge index was constrained to have similar values to those
found in the J and H bands. In the L and M bands the bulge and disk parameters
were constrained to values found in the NIR fits. It seems that seeing
conditions changed between the observations in the M band, since the PSF star
did not produced an acceptable fit for the nucleus, so the bulge was fitted
and the difference between the galaxy and the bulge was considered as the
nuclear contribution, i.e., $F_{nuc} = F_{gal}-F_{bul}$.

\paragraph{NGC\,4941} \label{88}
This is a very big galaxy, which means that, again, the PSF is not contained
in the galaxy frame but comes from observations of the standard star.
Unfortunately the standard star is not a good approximation to the galaxy PSF,
resulting in a null detection of a nucleus in the 3 NIR band fits.  However,
the galaxy was also not detected in the L and M bands, so it may also be that
no detectable near-- and MIR emission is associated with the active nucleus.

\paragraph{MCG\,-3-34-64}\label{89}
Without special constraints, the fitted NIR parameters are consistent. In the
L and M band the seeing between the observation of the galaxy and the
observation of the PSF star seems to vary, resulting in a poor fit in the L
band. In order to fit both profiles, the bulge was fitted with the values
found in the NIR fits and the nuclear contribution was calculated as $F_{nuc}
= F_{gal}-F_{bul}$.

\paragraph{NGC\,5135}\label{90} 
The bar in this galaxy is very strong and the disk is not seen in the IR
images. Between 1$\arcsec$ and 3$\arcsec$ there is a bright star-forming
region, dominating the profile at all wavelengths; this region was excluded
from the fit in all the profiles. The galaxy profile in the M band looks
slightly broader than the PSF star, but the bulge component is not able to
account for it, so a change in the seeing conditions is assumed to be
responsible. The nuclear flux is calculated from the integration of the galaxy
profile for $r\leq 1\arcsec$.

\paragraph{MRK\,463}
This is an interacting system whose two nuclei are too close to fit
reliably. Hence, no SED was obtained.

\paragraph{NGC\,5953}\label{92}
The bump observed near r=6$\arcsec$ was excluded because it is produced by the
ending of the flocculent arms. In the L band, the seeing is worse in the
galaxy observation than the PSF star, as can be seen in Fig.~\ref{aj}
(online). In this case, the results of the J, H and K band fits were used to
fix the values of the bulge parameters and the flux of the nucleus was
estimated as $F_{nuc} = F_{gal}-F_{bul}$.

\paragraph{MCG\,-2-40-4}\label{93}
The bar and the disk are indistinguishable for $r\gtrsim 2\arcsec$ so the
fitted component in the L-band is a mixture of both structures. The bulge
parameters in this band were constrained to values found in the NIR to
reliably separate it from the mixed bar-disk component. The M band profile was
fitted with the bulge effective radius and S\'ersic's index constrained to the
values found in the NIR.

\paragraph{IRAS\,15480-0344}\label{66}
Barely no restriction was needed to obtain a good fit in NIR bands. In the L
band only a nucleus and a bulge were fitted, and the index of the S\'ersic's
law was limited by the values found in the NIR bands.

\paragraph{IRAS\,19254-7245}\label{83}
This is an interacting system whose two nuclei are too close to fit
reliably. Hence, no SED was obtained.

\paragraph{NGC\,6810}\label{82}
To obtain a good fit for this galaxy bumps in the radial profile produced by
two large star forming regions were not considered. In the L, M and N bands,
the seeing was better in the observations of the PSF star than in the
observations of the galaxy. It is impossible for the bulge to account for the
difference at $r \lesssim 0.5\arcsec$. Fixed values for the bulge parameters
were used, similar to the values found at shorter wavelengths, and the nuclear
flux was estimated as $ F_{nuc} = F_{gal}-F_{bul}$. The extended emission of
this galaxy in the N band is an exceptional case.

\paragraph{NGC\,6890}\label{81}
To fit this galaxy the bumps at $r\approx 15 \arcsec$ and $r\approx 30
\arcsec$ were excluded from the fitting process because they correspond to the
ending of the flocculent arms. In the L band the signal beyond $r\approx 4
\arcsec$ corresponds to sky emission. The galaxy in the M and N bands is
consistent with being point-like, so no fit is performed. 

\paragraph{IC\,5063}\label{80}
This galaxy is classified as SA0, but the best fit uses only bulge and nucleus
components (no disk), which get stronger from shorter to longer
wavelengths. In the L band we restricted the values of the parameters of the
bulge to be close to the values found in the J, H and K fits. In the M band a
change in seeing is evident, causing the PSF star to be sharper and narrower
than the galaxy. The bulge parameters were frozen and only the nucleus was
allowed to vary. The galaxy is point-like in the N band, so no fit is needed.

\paragraph{MRK\,897}\label{79}
In the L band the object fell too close to the edge of the image and it was
not possible to obtain the profile necessary to accurately determine the bulge
contribution, so the bulge parameters were restricted to have values similar
to the J H and K ones.

\paragraph{NGC\,7130}\label{78}
The only restriction needed to fit the NIR bands was that the disk scale
length was larger than the bulge effective radius. In the L band the seeing
apparently varied between observations of the galaxy and the PSF star. In
order to obtain a reliable nuclear flux, the bulge component was fitted
between 2$\arcsec$ and 4$\arcsec$ and the parameters constrained to be as
close as possible to the J, H and K values. However, the fit to the galaxy was
not adequate (mainly in the innermost region), so the nuclear flux is
estimated as $ F_{nuc} = F_{gal}-F_{bul}$. In the M band, the galaxy looks
almost point-like so no fit is needed.

\paragraph{MCG\,-3-58-7}\label{77}
The disk size in the K band image was constrained to be larger than the bulge
size. In the L band, only a nucleus and a bulge were included, and the index
of the bulge was constrained to have a value similar to those obtained at
shorter wavelengths.

\paragraph{NGC\,7496} \label{76}
While fitting this galaxy, the K band bulge index had to be restricted by the
values obtained in the J and H bands. Also, the disk scale length was
restricted to be larger than the bulge effective radius. In the L and M bands
only a bulge and a nucleus were included.  In L the bulge size was restricted
by the values found in the NIR bands; in M the size and index of the bulge
component were constrained by values found at shorter wavelengths.

\paragraph{NGC\,7582} \label{75}
The NIR images show a prominent galaxy, orientated nearly face-on, and two
very weak arms with a different apparent inclination angle. In the literature
this system has been characterized by an inclination angle of $\sim60$ degrees
\citep{morris85}, corresponding to the disk traced by the weak arms. The
galaxy has a strong star-forming region near the nucleus, which is also
observable in X-rays \citep{bianchi07}, so the region near $r\approx
2.5\arcsec$ was not included in the fitting process.

\paragraph{NGC\,7590} \label{74}
This galaxy was observed twice in the NIR, once in service mode and once in
visitor mode, with a different detector each time. The two fits gave
consistent results. The point source detected in the NIR bands contributes at
most 0.3\% to the total flux. In the L and M bands, no point source was
detected, again suggesting a very weak AGN.  After subtracting the star
forming component (see Paper II) a {\emph{Spitzer}}-IRS spectrum of NGC\,7590
showed only a weak AGN continuum.

\paragraph{CGCG\,381-051} \label{73}
This object was not detected in the L and M bands. In the J, H and K bands the
fit did not include the region between $r\approx 15 ``$ and $r\approx 20 ``$,
which correspond to the end of the galactic arms.  The H band results were
used to contains the bulge index in the K band and the bulge size in the J
band.  Although the bulge is small it can be distinguished from the bar in all
images.

\end{document}